\documentclass{vldb}

\usepackage{xspace,subfigure,multirow}
\usepackage{graphicx}
\usepackage{balance}  
\usepackage{courier}
\usepackage{enumitem}
\usepackage{wrapfig}
\usepackage{mathtools}
\usepackage{bbm}
\usepackage{bm}
\usepackage{mathrsfs}

\usepackage{graphicx}
\usepackage{subfigure}
\usepackage{balance}
\usepackage{xspace,colortbl,subfigure,multirow}
\usepackage{amsmath}
\usepackage{cancel}
\usepackage{array}
\usepackage{verbatim}
\usepackage{color}
\usepackage{bbm}
\usepackage{multirow}
\usepackage{diagbox}
\usepackage{graphicx}
\usepackage{subfigure}
\usepackage[labelfont=bf,textfont={bf}]{caption}
\usepackage{epstopdf}
\usepackage{amsfonts}

\usepackage{listings}
\usepackage{framed}
\usepackage{xcolor}
\colorlet{shadecolor}{gray!20}

\makeatletter
\newif\if@restonecol
\makeatother

\usepackage[lined,boxed,vlined,ruled]{algorithm2e}
\usepackage{hyperref}

\newcommand{\lgl}[1]{ {\mbox{$<$}} ( { \bf \textcolor{blue} {#1} } ) {\mbox{$>$}}}

\vldbTitle{Relational Data Synthesis using Generative Adversarial Networks: 
	A Design Space Exploration}
\vldbAuthors{Ju Fan, Tongyu Liu, Guoliang Li, Junyou Chen, Yuwei Shen, Xiaoyong Du}
\vldbDOI{https://doi.org/10.14778/3407790.3407802}
\vldbVolume{13}
\vldbNumber{11}
\vldbYear{2020}

\subtitle{Technical Report}

\begin{document}

\pagestyle{plain}
	
\newtheorem{definition}{Definition}
\newtheorem{example}{Example}
\newtheorem{lemma}{Lemma}
\newtheorem{theorem}{Theorem}

\pagenumbering{roman}


\newcommand{\T}{\mathcal{T}\xspace}
\newcommand{\TX}{X\xspace}
\newcommand{\Tfake}{{\T}^{\prime}\xspace}
\newcommand{\TXfake}{{\TX}^\prime\xspace}

\newcommand{\ST}{\mathcal{ST}\xspace}
\newcommand{\STfake}{{\ST}^{\prime}\xspace}

\newcommand{\Ttrain}{\T_{\tt train}\xspace}
\newcommand{\Ttest}{\T_{\tt test}\xspace}
\newcommand{\TXtest}{\TX_{\tt test}\xspace}
\newcommand{\Tvalidation}{\T_{\tt valid}\xspace}
\newcommand{\Tsyn}{\T_{\tt syn}\xspace}

\newcommand{\tu}{t}
\newcommand{\tuvec}{\bm{\tu}\xspace}
\newcommand{\tux}{x\xspace}
\newcommand{\noise}{\bm{z}\xspace}
\newcommand{\condt}{\bm{c}\xspace}

\newcommand{\La}{Y\xspace}
\newcommand{\LaDom}{\Omega}
\newcommand{\la}{y\xspace}
\newcommand{\Lafake}{{Y}^{\prime}\xspace}
\newcommand{\Latest}{Y_{\tt test}\xspace}
\newcommand{\lafake}{\tilde{y}\xspace}

\newcommand{\eval}{{\tt Eval}\xspace}
\newcommand{\diff}{{\tt Diff}\xspace}

\newcommand{\Nn}{N_n\xspace}
\newcommand{\Ac}{A^{\tt c}}
\newcommand{\Ad}{A^{\tt d}}

\newcommand{\Gen}{G\xspace}
\newcommand{\Dis}{D\xspace}

\newcommand{\dshtru}{${\tt HTRU2}$\xspace}
\newcommand{\dsadult}{${\tt Adult}$\xspace}
\newcommand{\dsct}{${\tt CovType}$\xspace}
\newcommand{\dspendigits}{${\tt Digits}$\xspace}
\newcommand{\dsanuran}{${\tt Anuran}$\xspace}
\newcommand{\dscensus}{${\tt Census}$\xspace}
\newcommand{\dscredit}{${\tt Credit}$\xspace}
\newcommand{\dssat}{${\tt SAT}$\xspace}
\newcommand{\dsbing}{${\tt Bing}$\xspace}
\newcommand{\dssda}{${\tt SDataNum}$\xspace}
\newcommand{\dssdb}{${\tt SDataCat}$\xspace}

\newcommand{\tanhh}{{\tt tanh}\xspace}
\newcommand{\sigmoid}{{\tt sigmoid}\xspace}
\newcommand{\softmax}{{\tt softmax}\xspace}
\newcommand{\relu}{{\tt ReLU}\xspace}
\newcommand{\lrelu}{{\tt LeakyReLU}\xspace}
\newcommand{\bn}{{\tt BN}\xspace}
\newcommand{\fc}{{\tt FC}\xspace}
\newcommand{\deconv}{{\tt DeConv}\xspace}
\newcommand{\conv}{{\tt Conv}\xspace}
\newcommand{\lstm}{{\tt LSTMCell}\xspace}
\newcommand{\adam}{{\tt Adam}\xspace}
\newcommand{\rms}{{\tt RMSProp}\xspace}
\newcommand{\clip}{{\tt clip}\xspace}

\newcommand{\AlgoVTrain}{\textsc{VTrain}\xspace}
\newcommand{\AlgoWTrain}{\textsc{WTrain}\xspace}
\newcommand{\AlgoSimD}{\textsc{Simplified}\xspace}
\newcommand{\AlgoCTrain}{\textsc{CTrain}\xspace}
\newcommand{\AlgoCTrainPlus}{\textsc{CTrain}\xspace}
\newcommand{\AlgoDPTrain}{\textsc{DPTrain}\xspace}

\newcommand{\bsgan}{$\term{GAN}$\xspace}
\newcommand{\bsvgan}{$\term{VGAN}$\xspace}
\newcommand{\bscgan}{$\term{CGAN}$\xspace}
\newcommand{\bsctrain}{$\term{CTrain}$\xspace}
\newcommand{\bsvtrain}{$\term{VTrain}$\xspace}
\newcommand{\bsvae}{$\term{VAE}$\xspace}
\newcommand{\bsbn}{$\term{BN}$\xspace}
\newcommand{\bsbayes}{$\term{PB}$\xspace}

\newcommand{\reminder}[1]{ {\mbox{$<=$}} [ { \bf \textcolor{blue} {#1} } ] {\mbox{$=>$}}}
\newcommand{\term}[1]{{\tt #1}\xspace}

\newcommand{\norm}{${\tt sn}$\xspace}
\newcommand{\gmm}{${\tt gn}$\xspace}
\newcommand{\ordinal}{${\tt od}$\xspace}
\newcommand{\onehot}{${\tt ht}$\xspace}

\newcommand{\fanj}[1]{{#1}\xspace}
\newcommand{\revise}[1]{\textcolor{blue}{#1}}

\title{Relational Data Synthesis using Generative Adversarial Networks: 
A Design Space Exploration}

\title{Relational Data Synthesis using Generative Adversarial Networks: 
	A Design Space Exploration}

\numberofauthors{6}
\author{
	\alignauthor
	Ju Fan\\
	\affaddr{Renmin University of China}\\
	\email{fanj@ruc.edu.cn}
	\alignauthor
	Tongyu Liu\\
	\affaddr{Renmin University of China}\\
	\email{ltyzzz@ruc.edu.cn}
	\alignauthor Guoliang Li\\
	\affaddr{Tsinghua University}\\
	\email{liguoliang@tsinghua.edu.cn}
	\and  
	\alignauthor Junyou Chen\\
	\affaddr{Renmin University of China}\\
	\email{kanamemadoka@ruc.edu.cn}
	\alignauthor Yuwei Shen\\
	\affaddr{Renmin University of China}\\
	\email{rmdxsyw@ruc.edu.cn}
	\alignauthor Xiaoyong Du\\
	\affaddr{Renmin University of China}\\
	\email{duyong@ruc.edu.cn}
}

\pagenumbering{arabic}

\setcounter{page}{1}

\maketitle

\begin{abstract}
	The proliferation of big data has brought an urgent demand for privacy-preserving data publishing. Traditional solutions to this demand have limitations on effectively balancing the tradeoff between privacy and utility of the released data. Thus, the database community and machine learning community have recently studied a new problem of relational data synthesis using generative adversarial networks (GAN) and proposed various algorithms. However, these algorithms are not compared under the same framework and thus it is hard for practitioners to understand GAN's benefits and limitations.
	To bridge the gaps, we conduct so far the most comprehensive experimental study that investigates applying GAN to relational data synthesis.
	We introduce a unified GAN-based framework and define a space of design solutions for each component in the framework, including neural network architectures and training strategies.
	We conduct extensive experiments to explore the design space and compare with traditional data synthesis approaches. 
	Through extensive experiments, we find that GAN is very promising for relational data synthesis, and provide guidance for selecting appropriate design solutions.
	We also point out limitations of GAN and identify future research directions.
\end{abstract}

\section{Introduction}\label{sec:intro}
%
The tremendous amount of big data does not automatically lead to be easily accessed. The difficulty in data access is still one of the top barriers of many data scientists, according to a recent survey~\cite{kaggle/survey}.
In fact, organizations, such as governments and companies, have intention to publish data to the public or share data to partners in many cases, but they are usually restricted by regulations and privacy concerns. For example, a hospital wants to share its electronic health records (EHR) to a university for research purpose.
However, the data sharing must be carefully reviewed to avoid disclosure of patient privacy, which usually takes several months without guarantee of approval~\cite{Journals/Jama/JGJ}.

To address the difficulties, \emph{privacy-preserving data publishing} has been extensively studied recently to provide a safer way for data sharing~\cite{DBLP:conf/kdd/BrickellS08,DBLP:conf/icde/LiLV07,DBLP:series/ads/Domingo-Ferrer08,DBLP:conf/pods/AgrawalA01,DBLP:conf/sigmod/ZhangCPSX14,DBLP:journals/tods/ZhangCPSX17}. However, the existing solutions suffer from the limitations on effectively balancing privacy and utility of the released data~\cite{DBLP:journals/pvldb/ParkMGJPK18}. Therefore, efforts have been made recently in the database and machine learning communities to apply \emph{generative adversarial networks (GAN)} to relational data synthesis~\cite{DBLP:conf/ijcai/ChenJLPSS19,DBLP:journals/corr/abs-1907-00503,DBLP:journals/corr/abs-1811-11264,DBLP:journals/pvldb/ParkMGJPK18,DBLP:journals/jamia/BaowalyLLC19,DBLP:journals/corr/ChoiBMDSS17,DBLP:conf/wims/LuWY19}.
The main advantages of GAN are as follows. First, different from the conventional methods~\cite{DBLP:conf/kdd/BrickellS08,DBLP:conf/icde/LiLV07,DBLP:conf/pods/AgrawalA01,DBLP:conf/sigmod/ZhangCPSX14} that inject noise to the original data, GAN utilizes neural networks to generate ``fake'' data directly from noise. Thus, there is no \emph{one-to-one} relationship between real and synthetic data, which reduces the risk of re-identification attacks~\cite{DBLP:journals/pvldb/ParkMGJPK18}.
Moreover, the adversarial learning mechanism of GAN enables the synthetic data to effectively preserve utility of the original data for supporting down-streaming applications, such as classification and {clustering}.


However, compared with the success of using GAN for image generation~\cite{DBLP:conf/nips/LucicKMGB18}, GAN-based {relational} data synthesis is still in its infancy stage. Despite some very recent attempts~\cite{DBLP:conf/ijcai/ChenJLPSS19,DBLP:journals/corr/abs-1907-00503,DBLP:journals/corr/abs-1811-11264,DBLP:journals/pvldb/ParkMGJPK18,DBLP:journals/jamia/BaowalyLLC19,DBLP:journals/corr/ChoiBMDSS17,DBLP:conf/wims/LuWY19}, as far as we know, the proposed methods are not compared under the same framework and thus it is hard for practitioners to understand GAN's benefits and limitations. 
To bridge the gaps, in this paper, we provide a comprehensive experimental study that examines applying GAN to relational data synthesis.
We introduce a general framework that can unify the existing solutions for GAN-based data synthesis. Based on the framework, we conduct extensive experiments to systemically investigate the following two key questions.


Firstly, it remains an unresolved question on how to effectively apply GAN to relational data synthesis. It is worth noting that relational data has its own characteristics that make the adoption very challenging. $(i)$ Relational data has mixed data types, including categorical and numerical attributes. $(ii)$ Different attributes have correlations. $(iii)$ Many real-world datasets have highly imbalanced data distribution. Thus, the state-of-the-art GAN design for image synthesis (e.g., DCGAN~\cite{DBLP:journals/corr/RadfordMC15}) may not perform well for relational data. 
We review the existing solutions that realize GAN, including neural network design and training strategies, and define a \emph{design space} by providing a categorization of the solutions. Through exploring the design space, we systemically evaluate the solutions on datasets with various types and provide insightful experimental findings.

The second question is whether GAN is more helpful than the existing approaches to relational data synthesis. To answer this, this paper considers various baseline approaches, including a representative deep generative model, variational auto-encoder (VAE)~\cite{DBLP:journals/corr/KingmaW13,DBLP:conf/icml/RezendeMW14}, and the state-of-the-art data synthesis approach using statistical models~\cite{DBLP:conf/sigmod/ZhangCPSX14,DBLP:journals/tods/ZhangCPSX17}. To provide a comprehensive comparison, we evaluate their performance on both privacy and the utility of the synthetic data. 
Moreover, we also examine whether GAN can support provable privacy protection, i.e., differential privacy~\cite{DBLP:journals/fttcs/DworkR14}.
Based on the comparison, we analyze the benefits and limitations of applying GAN to relational data synthesis.

To summarize, we make the following contributions.

(1) We conduct so far the most comprehensive experimental study for applying GAN to relational data synthesis. We formally define the problem and review the existing approaches (Section~\ref{sec:overview}). We introduce a unified framework and define a design space that summarizes the solutions for realizing GAN (Sections~\ref{sec:methods}, \ref{sec:preproc} and \ref{sec:design-choices}), which can help practitioners to easily understand how to apply GAN.

(2) We empirically conduct a thorough evaluation to explore the design space and compare with the baseline approaches (Section~\ref{sec:exp-method}). We make all codes and datasets in our experiments public at Github\footnote{https://github.com/ruclty/Daisy}.
We provide extensive experimental findings and reveal insights on strength and robustness of various solutions, which provide guidance for an effective design of GAN.

(3) We find that GAN is highly promising for relational data synthesis, as it empirically provides better tradeoff between synthetic data utility and privacy. We point out its limitations and identify research directions (Section~\ref{sec:conclusion}).

\section{Relational Data Synthesis}\label{sec:overview}
\vspace{-.5em}
\subsection{Problem Formalization} \label{subsec:problem}

This paper focuses on a relational table $\T$ of $n$ records, i.e., $\T = \{\tu_{1}, \tu_{2}, \ldots, \tu_n\}$. 
We use $\T[j]$ to denote the $j$-th attribute (column) of table $\T$ and $\tu[j]$ to denote the value of record $\tu$'s $j$-th attribute. 
In particular, we consider both categorical (nominal) and numerical (either discrete or continuous) attributes in this paper.
%
%
%
%
%
We study the problem of synthesizing a ``fake'' table $\Tfake$ from the original $\T$, with the objective of preserving {data utility} and protecting {privacy}.

(1) \emph{Data utility} is highly dependent to the specific need of the synthetic data for down-streaming applications.
This paper focuses on the specific need on using the fake table to train machine learning (ML) models, which is commonly considered by recent works~\cite{DBLP:conf/ijcai/ChenJLPSS19,DBLP:journals/corr/abs-1907-00503,DBLP:journals/corr/abs-1811-11264,DBLP:journals/pvldb/ParkMGJPK18,DBLP:journals/jamia/BaowalyLLC19,DBLP:journals/corr/ChoiBMDSS17,DBLP:conf/wims/LuWY19}. This means that an ML model trained on the fake table should achieve similar performance as that trained on $\T$. For simplicity, this paper considers classification models.
%
We represent the original table as $\T=[\TX;\La]$, where each $\tux_{i} \in \TX$ and each $\la_{i} \in \La$ respectively represent \emph{features} and \emph{label} of the corresponding record $\tu_{i}$.
We use $\T$ to train a classifier $f:\TX \rightarrow \La$ that maps $\tux_{i} \in \TX$ to its predicted label $f(\tux_{i})$. Then, we evaluate the performance of $f$ on a test set $\Ttest=[\TXtest; \Latest]$ using a specific metric $\eval(f|\Ttest)$. Some representative metrics include F1 score and Area Under the ROC Curve (AUC).
Similarly, we can train a classifier $f^{\prime}$ on the synthetic table $\Tfake$ and evaluate the classifier on the same $\Ttest$ to obtain its performance $\eval(f^\prime|\Ttest)$.
The utility of $\Tfake$ is measured by the difference between these two classifiers' performance metrics, i.e.,
\begin{equation}\label{eq:diff}
\diff(\T, \Tfake) = \mid \eval(f|\Ttest)-\eval(f^{\prime}|\Ttest) \mid.
\end{equation}

\begin{figure}[!t]
	\begin{center} 
		\epsfig{figure=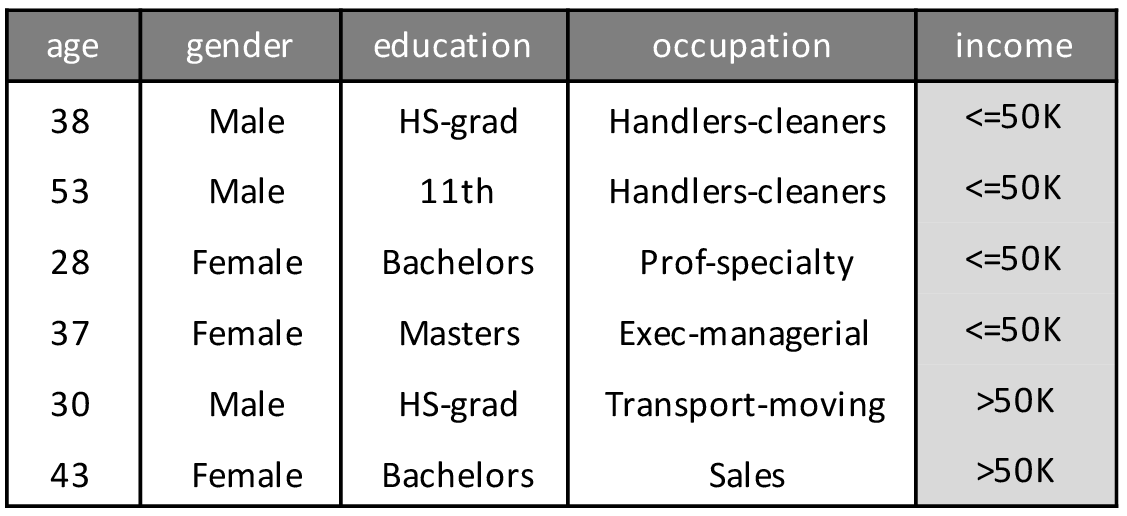,width=0.45\textwidth}
		\vspace{-1em}
		\caption{An example relational table.
		}
		\label{fig:example}
		\vspace{-2em}
	\end{center}
			\vspace{-1em}
\end{figure}

%
%

\vspace{-1em}
\begin{example}[Synthetic Data Utility]
Consider an example table $\T$ in Figure~\ref{fig:example}, where label $\term{income}$ has two unique values: $0$ ($\term{income} \leq 50K$) and $1$ ($\term{income} > 50K$). We use $\T$ to train a synthesizer $\Gen$ and generate a fake table $\Tfake$ via $\Gen$. We train models $f$ and $f^{\prime}$ to predict $\term{income}$ on $\T$ and $\Tfake$ respectively, and evaluate these models on a test table $\Ttest$. We measure the performance difference of these two models as $\diff(\T, \Tfake)$ between the original $\T$ and synthetic table $\Tfake$. 
Intuitively, the lower the difference $\diff(\T, \Tfake)$ is, the better the synthetic table preserves the data utility.
\end{example}
\vspace{-.5em}

(2) \emph{Privacy risk} evaluation for synthetic table $\Tfake$ is also an independent research problem.
This paper adopts two commonly-used metrics in the existing works~\cite{DBLP:journals/tdp/ParkG14,DBLP:conf/wims/LuWY19,DBLP:conf/psd/Mateo-SanzSD04}, namely hitting rate and distance to the closest record (DCR). Intuitively, the metrics measure the likelihood that the original data records can be re-identified by an attacker. 

%

\vspace{1mm}
\noindent \textbf{Synthetic data generation for Clustering.}
%
For example, suppose that a hospital wants to ask a CS team to develop a clustering algorithm that discovers groups of similar patients. It can first share the synthetic data to the team for ease of algorithm development. Then, it deploys the developed algorithm in the hospital to discover groups on the original data. In this case, the data utility is that a clustering algorithm should achieve similar performance on both original table $\T$ and fake table $\Tfake$.
Let $\mathcal{C}=\{c_1, c_2, \ldots,c_k\}$ and $\mathcal{C}^\prime=\{c_1^\prime, c_2^\prime, \ldots,c_k^\prime \}$ respectively denote the sets of clusters discovered by a clustering algorithm on $\T$ and $\Tfake$. We can use a standard evaluation metric for clustering, such as normalized mutual information (NMI), to examine the quality of $\mathcal{C}$ and $\mathcal{C}^\prime$. Then, the utility of $\Tfake$ for clustering is measured by the difference between these two metrics, i.e., 
$\diff_{\tt CST}(\T, \Tfake) = \mid \eval(\mathcal{C}|\T)-\eval(\mathcal{C}^\prime|\Tfake) \mid$, where $\eval(\mathcal{C}|\T)$ ($\eval(\mathcal{C}^\prime|\Tfake)$) is the evaluation metric for clusters $\mathcal{C}$ ($\mathcal{C}^\prime$) from original table $\T$ (fake table $\Tfake$).
Intuitively, we prefer a smaller $\diff_{\tt CST}$ for preserving the utility.


\vspace{1mm}
\noindent \textbf{Synthetic data generation for \emph{approximate query processing (AQP)}~\cite{DBLP:conf/sigmod/ChaudhuriDK17,DBLP:journals/corr/abs-1903-10000}.}
For example, suppose that a user wants to perform data exploration or visualization on a large dataset. To reduce latency, some work~\cite{DBLP:journals/corr/abs-1903-10000} introduces a \emph{lightweight} approach that utilizes synthetic data in the client to quickly answer aggregate queries, without communicating with the server. To support this, the synthetic data should preserve the utility that answers aggregate queries as accurate as possible to the original data $\T$.
To formally measure the data utility, we adopt the \emph{relative error difference}~\cite{DBLP:journals/corr/abs-1903-10000}, as defined as below.
For each aggregate query $q$, we compute the relative error $e^{\prime}$ over the synthetic table $\Tfake$,
and the relative error $e$ over a fixed size sample obtained from $\T$.
Then, we compute the relative error difference as the absolute difference between these two errors, 
$\diff_{\tt AQP}(\T, \Tfake|q) = \mid e -e^\prime \mid$.
Given a workload with a set $Q$ of queries, we compute the average $\diff_{\tt AQP}(\T, \Tfake) = \sum_{q \in Q}{\diff_{\tt AQP}(\T, \Tfake|q)}/|Q|$.

\subsection{Related Works for Data Synthesis} \label{subsec:rw}
%
Data synthesis has been extensively studied in the last decades, and the existing approaches can be broadly classified into \emph{statistical model} and \emph{neural model}. The statistical approach aims at modeling a joint multivariate distribution for a dataset and then generating fake data by sampling from the distribution. To effectively capture dependence between variates, existing works utilize copulas~\cite{DBLP:journals/pvldb/LiXZJ14,DBLP:conf/dsaa/PatkiWV16}, Bayesian networks~\cite{DBLP:conf/sigmod/ZhangCPSX14,DBLP:journals/tods/ZhangCPSX17}, Gibbs sampling~\cite{DBLP:journals/tdp/ParkG14} and Fourier decompositions~\cite{DBLP:conf/pods/BarakCDKMT07}. 
Synopses-based approaches, such as wavelets and multi-dimensional sketches, build compact {data summary} for massive data~\cite{DBLP:journals/ftdb/CormodeGHJ12,DBLP:journals/tkde/XiaoWG11}, which can be then used for estimating joint distribution.
As the statistical models may have limitations on effectively balancing privacy and data utility,
neural models have been recently emerging to synthesize relational data. Existing works aim to use deep generative models to approximate the distribution of an original dataset. To this end, some studies devise deep de-noising autoencoders~\cite{DBLP:conf/pakdd/GondaraW18} and variational autoencoders (VAE)~\cite{DBLP:journals/corr/abs-1903-10000}, while more attentions are paid on generative adversarial networks (GAN)~\cite{DBLP:conf/ijcai/ChenJLPSS19,DBLP:journals/corr/abs-1907-00503,DBLP:journals/corr/abs-1811-11264,DBLP:journals/pvldb/ParkMGJPK18,DBLP:journals/jamia/BaowalyLLC19,DBLP:journals/corr/ChoiBMDSS17,DBLP:conf/wims/LuWY19,DBLP:journals/pvldb/YangFWLLD18,DBLP:conf/sigmod/LiuYFWLD19}. 

However, despite the aforementioned attempts on GAN-based relational data synthesis, existing works have not systemically explored the design space, as mentioned previously. Thus, this paper conducts an experimental study to systemically investigate the design choices and compare with the state-of-the-art statistical approaches for data synthesis. Note that, besides data synthesis, private data release can also be achieved by anonymization~\cite{DBLP:conf/kdd/BrickellS08,DBLP:conf/icde/LiLV07,DBLP:journals/dase/SinghS18} and perturbation~\cite{DBLP:series/ads/Domingo-Ferrer08,DBLP:conf/pods/AgrawalA01}. However, the existing study~\cite{DBLP:journals/pvldb/ParkMGJPK18} has shown that GAN-based data synthesis outperforms these techniques.

%
%
%

%

\vspace{-.25em}
\subsection{Generative Adversarial Networks (GAN)} \label{subsec:gan}
\vspace{-.25em}

Generative adversarial networks (GAN)~\cite{DBLP:conf/nips/GoodfellowPMXWOCB14,DBLP:conf/nips/LucicKMGB18}, are a kind of deep generative models, which have achieved breakthroughs in many areas, such as image generation~\cite{DBLP:journals/corr/RadfordMC15,DBLP:journals/corr/MirzaO14,DBLP:conf/nips/ChenCDHSSA16}, and sequence generation~\cite{DBLP:conf/aaai/YuZWY17,DBLP:conf/ismir/YangCY17}. 
Typically, GAN consists of a generator $\Gen$ and a discriminator $\Dis$, which are competing in an adversarial process. The generator $\Gen(\mathbf{z};\theta_{g})$ takes as input a random noise $\mathbf{z} \in \mathbb{R}^{z}$ and generates synthetic samples $\Gen(\mathbf{z}) \in \mathbb{R}^{d}$, while the discriminator $\Dis(\tuvec;\theta_{d})$ determines the probability that a given sample comes from the real data instead of being generated by $\Gen$.
Intuitively, the optimal $\Dis$ could distinguish real samples from fake ones, and the optimal $\Gen$ could generate indistinguishable fake samples which make $\Dis$ to randomly guess. 
Formally, $\Gen$ and $\Dis$ play a minimax game with value function $V(\Gen, \Dis)$, i.e., 
$
\min_{\Gen}\max_{\Dis}{V(\Gen,\Dis)} = \mathbb{E}_{\tuvec \in p_{data}(\tuvec)}\big[\log{\Dis(\tuvec)}\big] + \mathbb{E}_{\mathbf{z} \in p_{z}(\mathbf{z})}\big[1-\log{\Dis(\Gen(\mathbf{z}))}\big],
$
%
%
%
where $p_{data}$ is the distribution of the real samples transformed from our relational table $\T$ and $p_{z}$ is the distribution of the input noise $\mathbf{z}$. 

\vspace{-.75em}

\section{GAN-based Synthesis Overview} \label{sec:methods}

\vspace{-.25em}

Relational data has its own characteristics that make the adoption of GAN to data synthesis challenging. First, relational data has \emph{mixed} data types, and thus it is non-trivial to transform a record into the input of GAN.
Second, different attributes in relational data usually have correlations. It remains challenging to enable the generator to capture such correlations.
Third, most real-world data has highly \emph{imbalanced} label distribution. 
This increases the difficulty of relational data synthesis, especially for records with minority labels. To address these challenges, we introduce a framework that unifies the existing solutions for applying GAN to relational data synthesis.

%
\vspace{-.5em}
\subsection{Framework of GAN-based Synthesis} \label{subsec:framework}
\vspace{-.25em}

\begin{figure}[!t]\vspace{-1.5em}
	\begin{center} 
		\epsfig{figure=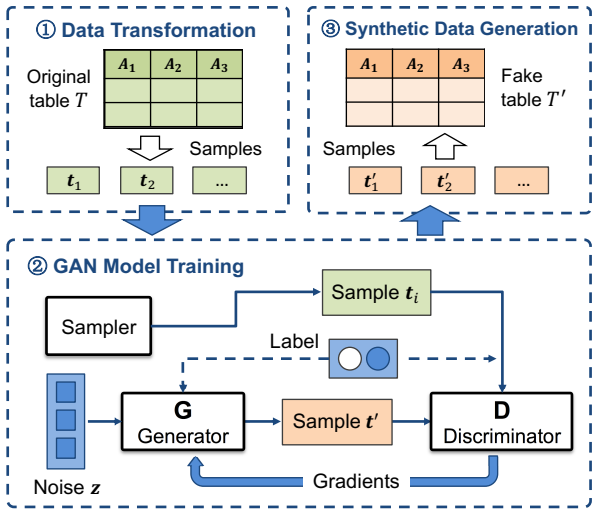,width=0.44\textwidth}
		\vspace{-0.5em}
		\caption{Overview of data synthesis using GAN. (1) It transforms each record in a relational table into a sample $\tuvec \in \mathbb{R}^{d}$. (2) It takes the samples as input to train a deep generative model $\Gen$ using the adversarial training framework in GAN. (3) It utilizes the trained $\Gen$ to generate a set of synthetic samples, which are then transformed back into fake records.}
		\label{fig:archi}
		\vspace{-3em}
	\end{center}
\end{figure}

Figure~\ref{fig:archi} shows a unified framework of GAN-based relational data synthesis. 
It takes a relational table $\T$ as input and generates a table $\Tfake$ of synthetic data in three phases. 

\noindent \textbf{Phase I - Data Transformation.}
This phase aims at preparing \emph{input data} for the subsequent GAN model training. Specifically, it transforms each record $\tu \in \T$ with mixed attribute types into a sample $\tuvec \in \mathbb{R}^{d}$ of numerical values, which can be then fed into neural networks in GAN.

\noindent \textbf{Phase II - GAN Model Training.}
This phase aims at training a deep generative model $\Gen$. Specifically, $\Gen$ takes as input a random noise $\mathbf{z} \in \mathbb{R}^{z}$ and generates synthetic sample $\tuvec^{\prime} = \Gen(\mathbf{z}) \in \mathbb{R}^{d}$. Meanwhile, a \textsc{Sampler} picks a sample $\tuvec_{i}$ from the data prepared by the previous phase. Then, fed with both real and synthetic samples, our discriminator $\Dis$ determines the probability that a given sample is real. By iteratively applying minibath stochastic gradient descent, parameters of both $\Gen$ and $\Dis$ are optimized, and thus $\Gen$ could be improved towards generating indistinguishable samples that fool $\Dis$. 
One key technical issue here is to design effective neural networks for $\Gen$ that can capture correlations among attributes. Moreover, considering {imbalanced} label distribution of real datasets, our framework also supports {conditional GAN}~\cite{DBLP:journals/corr/MirzaO14} that also feeds a target label to both generator and discriminator as their input, so as to ``guide'' them for generating records with the label.

\noindent \textbf{Phase III - Synthetic Data Generation.}
This phase utilizes $\Gen$, which is well trained in the previous phrase, to generate a synthetic table $\Tfake$. It repeatedly feeds $\Gen$ with the prior noise $\noise$ (as well as target label), which generates a set of synthetic samples $\{\tuvec^\prime\}$. Next, it adopts the same data transformation scheme used in Phase I to convert the samples back into records that then compose $\Tfake$.
\vspace{-1mm}
\begin{example}[Framework]
	Considering our example in Figure~\ref{fig:example}, the framework transforms each record into a sample. Suppose that we adopt ordinal encoding for categorical attributes.
	The first record is transformed to $[38,0,0,0,0]$. Then, it uses the transformed samples to train the GAN model for obtaining an optimized generator $\Gen$. It leverages $\Gen$ to generate samples, e.g., $[40, 1, 2, 1, 1]$, and transforms the samples back to synthetic records, e.g., $(40$, $\term{Female}$, $\term{Bachelors}$, $\term{Prof}$-$\term{specialty}$, $>{50K})$.
\end{example}

\vspace{-.5em}
\subsection{Categorization of Design Choices} \label{subsec:design-space}

\begin{figure}[!t]\vspace{-1.5em}
	\begin{center} 
		\epsfig{figure=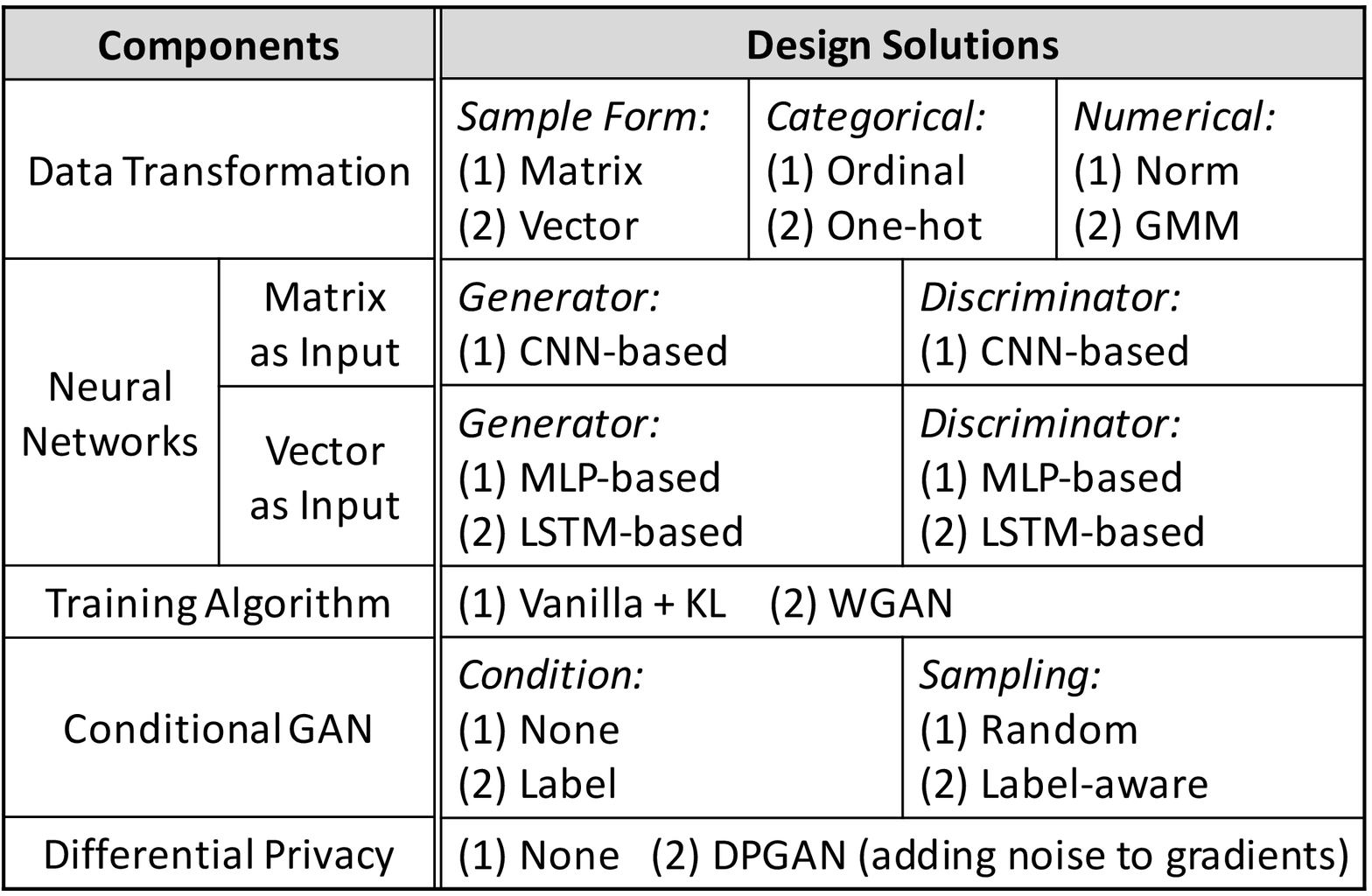,width=0.46\textwidth}
		\vspace{-1em}
		\caption{A categorization of design solutions.}
		\label{fig:design-space}
		\vspace{-2.5em}
	\end{center}
\end{figure}

We provide a categorization of design solutions for each component in our framework, as summarized in Figure~\ref{fig:design-space}. 

\noindent \textbf{Data transformation.}
We examine how to encode categorical attributes to numerical values, and normalize numerical attributes to appropriate ranges that fit neural networks. We consider widely-used encoding schemes for categorical attributes, i.e., ordinal or one-hot encoding, and normalization schemes for numerical attributes, i.e., simple normalization or normalization using Gaussian Mixture Model (GMM). We will take exploration of more sophisticated transformation schemes as a future work.
Moreover, as different neural networks have different requirements for the input, sample $\tuvec$ can be in the form of either \emph{matrix} or \emph{vector}. More details of data transformation are in Section~\ref{sec:preproc}. 

%
%

\noindent \textbf{Neural networks.}
Existing works for relational data synthesis consider three representative neural networks.
(1) Inspired by the success of image synthesis, some apply DCGAN~\cite{DBLP:journals/corr/RadfordMC15}, and use Convolutional Neural Networks (CNN) for $\Gen$ and $\Dis$, in which $\Gen$ is a deconvolution process and $\Dis$ is a convolution process~\cite{DBLP:conf/ijcai/ChenJLPSS19,DBLP:journals/pvldb/ParkMGJPK18}.
(2) Following the original GAN~\cite{DBLP:conf/nips/GoodfellowPMXWOCB14}, some studies~\cite{DBLP:journals/corr/ChoiBMDSS17,DBLP:journals/corr/abs-1907-00503} use multilayer perceptron (MLP) consisting of multiple fully-connected layers.
(3) Some approaches utilize a \emph{sequence generation} mechanism that generates attributes separately in sequential time-steps~\cite{DBLP:journals/corr/abs-1811-11264}, and use recurrent neural networks, such as long short-term memory (LSTM) networks~\cite{DBLP:journals/neco/HochreiterS97} for $\Gen$.
Note that different neural networks have different forms of input: CNN takes matrix-formed samples, while MLP and LSTM uses vector-formed samples. This paper focuses on comparing the aforementioned representative neural networks under the same framework. We will take an exploration of more sophisticated models, such as Bidirectional LSTM~\cite{DBLP:journals/corr/abs-1303-5778}, as a future work. More details can be referred to Section~\ref{subsec:nn}.

%
%

\noindent \textbf{Training algorithm.} 
Minibatch-based stochastic gradient descent (SGD) strategy is applied for GAN training.
This paper focuses on investigating \emph{mode collapse}~\cite{DBLP:conf/nips/SalimansGZCRCC16,DBLP:journals/corr/MetzPPS16}, a well-recognized challenge in GAN training.
To this end, we evaluate different training algorithms with various loss functions and variants of SGD optimizer, such as $\adam$ and $\rms$. This paper investigates two alternatives to train GAN: (1) the vanilla training algorithm~\cite{DBLP:conf/nips/GoodfellowPMXWOCB14} with an improved loss function and (2) Wasserstein GAN (WGAN) training~\cite{DBLP:journals/corr/ArjovskyCB17}.
See Section~\ref{subsec:train} for more details of these algorithms.

\noindent \textbf{Conditional GAN.}
The imbalanced label distribution in real-world data may result in \emph{insufficient training} for records with minority labels~\cite{DBLP:journals/corr/abs-1907-00503}. Thus, some studies~\cite{DBLP:journals/corr/abs-1907-00503} apply {conditional GAN}~\cite{DBLP:journals/corr/MirzaO14} to data synthesis. We examine the adoption of {conditional GAN} that encodes a label as a condition vector $\condt$ to guide $\Gen$ ($\Dis$) to generate (discriminate) samples with the label. We evaluate the performance of GAN with/without label as a condition. Moreover, we also investigate different sampling strategies (i.e., \textsc{Sampler} in Figure~\ref{fig:archi}): random sampling as commonly used in GAN training, and label-aware sampling that gives fair opportunity for samples with different labels. See Section~\ref{subsec:cgan} for more details.

\vspace{1mm}
\noindent \textbf{Differential privacy.}
We consider differential privacy~\cite{DBLP:journals/fttcs/DworkR14}, a well-adopted formalization of data privacy, to evaluate whether GAN can still be effective to preserve data utility while providing provable privacy protection.
%
Intuitively, although $\Gen$ does not access the real data $\T$ (only $\Dis$ accesses $\T$ via \textsc{Sampler}), $\Gen$ may still implicitly disclose privacy information as the gradients for optimizing $\Gen$ is computed based on $\Dis$. Thus, we adopt the DPGAN model~\cite{DBLP:journals/corr/abs-1802-06739} in the GAN training process, as elaborated in Section~\ref{subsec:dpgan}.
\section{Data Transformation}\label{sec:preproc}
Data transformation converts a record $\tu$ in $\T$ into a sample $\tuvec \in \mathbb{R}^{d}$. To this end, it processes each attribute $\tu[j]$ in $\tu$ independently to transform $\tu[j]$ into a vector $\tuvec_{j}$. Then, it generates $\tuvec$ by combining all the attribute vectors.
Note that the transformation is reversible: after generating synthetic sample $\tuvec^\prime$ using $\Gen$, we can apply these methods to reversely convert $\tuvec^\prime$ to a fake record. 

%

\vspace{1mm}
\noindent \textbf{Categorical attribute transformation.}
We consider two commonly-used encoding schemes.

\emph{1) Ordinal encoding} assigns an ordinal integer to each category of categorical attribute $\T[j]$, e.g., starting from $0$ to $|\T[j]|-1$ ($|\T[j]|$ is domain size of $\T[j]$). After ordinal encoding, $\T[j]$ is equivalent to a discrete numeric attribute. 

\emph{2) One-hot encoding} first assigns each category of categorical attribute $\T[j]$ with an integer, starting from $0$ to $|\T[j]|-1$. Then, it represents each category as a \emph{binary vector} with all zero values, except that the index of the integer corresponding to the category is set as one. 


\noindent \textbf{Numerical attribute transformation.}
We normalize values in a numerical attribute to $[-1,1]$, to enable neural networks in $\Gen$ to generate values in the attribute using $\tanhh$ as an activation function.

\emph{1) Simple normalization} uses $\T[j].\max$ and $\T[j].\min$ to respectively denote the maximum and minimum values of attribute $\T[j]$. Given an original value $v$ in $\T[j]$, it normalizes the value as
$v_{\tt norm}= -1 + 2 \cdot \frac{v-\T[j].\min}{\T[j].\max-\T[j].\min}.$

\emph{2) GMM-based normalization.}
Some studies~\cite{DBLP:journals/corr/abs-1907-00503,DBLP:journals/corr/abs-1811-11264} propose to consider the \emph{multi-modal} distribution of a numerical attribute $\T[j]$, to avoid limitations of simple normalization, such as gradient saturation. They utilize a Gaussian Mixture model (GMM) to cluster values of $\T[j]$, and normalize a value by the cluster it belongs to. 
They first train a GMM with $s$ components over the values of $\T[j]$, where the mean and standard deviation of each component $i$ are denoted by $\mu^{(i)}$ and $\sigma^{(i)}$. Then, given a specific value $v$, they compute the probability distribution $(\pi^{(1)}, \pi^{(2)}, \ldots, \pi^{(s)})$ where $\pi^{(i)}$ indicates the probability that $v$ comes from component $i$, and normalize $v$ as
$
v_{\tt gmm} = \frac{v - \mu^{(k)}}{2 \sigma^{(k)}}, where~ k=\arg\max_{i}{\pi^{(i)}}.
$
For example, suppose that the records in our example table can be clustered into two modes, i.e., ``young generation'' and ``old generation'' with Gaussian distributions $G(20, 10)$ and $G(50,5)$ respectively. Then, given an $\term{age}$ value $43$, we first determine that it is more likely to belong to the old generation, and then normalize it into a vector $(-0.7, 0,1)$ where $(0,1)$ indicates the second mode and $-0.7$ is $v_{\tt gmm}$.

\vspace{1mm}
\noindent \textbf{Combination of multiple attributes.}
Once all attributes in $\tu$ are transformed by the above schemes, we need to combine them together to generate sample $\tuvec$. 

\emph{1) Matrix-formed samples.} For CNN-based neural networks, we follow the method in~\cite{DBLP:journals/pvldb/ParkMGJPK18} to convert attributes into a square matrix.
For example, a record with $8$ attributes is converted into a $3\times3$ square matrix after padding one zero. 
Note that this method requires each attribute is transformed into one value instead of a vector (otherwise, the vector of an attribute may be split in the matrix). Thus, one-hot encoding and GMM-based normalization are not applicable.

\emph{2) Vector-formed samples.} For MLP-based and LSTM-based neural networks, we concatenate all the attribute vectors to generate a sample vector, i.e., $\tuvec=\tuvec_{1} \oplus \tuvec_{2} \oplus \ldots \oplus \tuvec_{m}$. Obviously, this method is compatible to all the attribute transform schemes described above.

\vspace{-.5em}
\begin{example}[Data Transformation]
Let's consider the last record shown in Figure~\ref{fig:example}. When transforming the record into a matrix-formed sample, we can only apply ordinal encoding and simple normalization and obtain a square matrix $((0.2,1,2),(4,1,0),(0,0,0))$. 
In contrast, when transforming the record into a vector-formed sample, we may choose to use one-hot encoding and GMM-based normalization, and obtain $(\underline{-0.7,0,1},\underline{0,1}, \underline{0,0,1,0}, \underline{0,0,0,0,1}, \underline{0,1})$, where the underlines indicate different attributes.
\end{example}

\section{GAN Model Design}\label{sec:design-choices}

\vspace{-.5em}


\subsection{Neural Network Architectures} \label{subsec:nn}
We describe the basic idea of neural networks evaluated in this paper, and leave more details in our report~\cite{daisy}. 

%


\noindent \textbf{CNN: convolutional neural networks.}
CNN is utilized in the existing works for data synthesis~\cite{DBLP:conf/ijcai/ChenJLPSS19,DBLP:journals/pvldb/ParkMGJPK18}.
Generator $\Gen$ takes as input a prior noise $\noise$, which is also denoted by $\bm{h}_{g}^{0}$. It then uses $L$ de-convolution layers $\{\bm{h}_{g}^{l}\}$ (i.e., fractionally strided convolution) to transform $\noise$ to a synthetic sample in the form of matrix, where 
$\bm{h}_{g}^{l+1} = \relu(\bn(\deconv(\bm{h}_{g}^{l})))$ and $\tuvec = \tanhh(\deconv(\bm{h}_{g}^{L}))$.
%
%
Discriminator $\Dis$ takes as input a real/fake sample $\tuvec$ in matrix form, which is also denoted by $\bm{h}_{d}^{0}$. It applies $L$ convolution layers $\{\bm{h}_{d}^{l}\}$ where $\bm{h}_{d}^{l+1} = \lrelu(\bn(\conv(\bm{h}_{d}^{l})))$ and $\conv$ is a convolution function. Finally, $\Dis$ outputs a probability indicating how likely $\tuvec$ is real, i.e., $f = \sigmoid(\bn(\conv(\bm{h}_{d}^{L})))$.

\noindent \textbf{MLP: fully connected neural networks.}
MLP is used in the existing works~\cite{DBLP:journals/corr/ChoiBMDSS17,DBLP:journals/corr/abs-1907-00503}. 
%
%
In this model, $\Gen$ takes as input noise $\noise$, which is also denoted by $\bm{h}^{(0)}$, and utilizes $L$ fully-connected layers. Each layer is computed by
$
\bm{h}^{l+1} = \phi\big(
		\bn(
			\fc_{|\bm{h}^{l}| \rightarrow |\bm{h}^{l+1}|}(\bm{h}^{l})
		)
	\big),
$
where $\fc_{|\bm{h}^{l}| \rightarrow |\bm{h}^{l+1}|}(\bm{h}^{l}) = \bm{W}^{l}  \bm{h}^{l} + \bm{b}^{l}$ with weights $\bm{W}^{l}$ and bias $\bm{W}^{l}$, $\phi$ is the activation function (we use $\relu$ in our experiments), and $\bn$ is the batch normalization~\cite{DBLP:conf/icml/IoffeS15}.
Discriminator $\Dis$ is an MLP that takes a sample $\tuvec$ as input, and utilizes multiple fully-connected layers and a $\sigmoid$ output layer to classify whether $\tuvec$ is real or fake. 

One issue here is how to make the output layer in $\Gen$ \emph{attribute-aware}. We propose to generate each attribute vector $\tuvec_{j}$ depending on the transformation method on the corresponding attribute $\T[j]$, e.g., using $\tanhh$ and $\softmax$ for simple normalization and one-hot encoding respectively.
In particular, for GMM-based normalization, we adopt the following method in~\cite{DBLP:journals/corr/abs-1907-00503}. We first use $\tanhh(\fc_{|\bm{h}^{L}| \rightarrow 1}(\bm{h}^{L}) $ to generate $v_{\tt gmm}$ and then use $\softmax(\fc_{|\bm{h}^{L}| \rightarrow s}(\bm{h}^{L}))$ to generate a one-hot vector indicating which component $v_{\tt gmm}$ belongs to.
After generating $\{\tuvec_{j}\}$ for all attributes, we concatenate them to obtain $\tuvec$ as a synthetic sample.

\vspace{1mm}
\noindent \textbf{LSTM: recurrent neural networks.}
%
%
The basic idea is to formalize record synthesis as a \emph{sequence generation} process~\cite{DBLP:journals/corr/abs-1811-11264}: it models a record $\tuvec$ as a {sequence} and each element of the sequence is an attribute $\tuvec_{j}$. It uses LSTM to generate $\tuvec$ at multiple timesteps, where the $j$-th timestep is used to generate $\tuvec_{j}$.
Let $\bm{h}^{j}$ and $\bm{f}^{j}$ respectively denote the hidden state and output of the LSTM at the $j$-th timestep. Then, we have $\bm{h}^{j+1} = \lstm(\noise, \bm{f}^{j}, \bm{h}^{j})$ and $\bm{f}^{j+1} = \tanhh( \fc_{|\bm{h}^{j+1})| \rightarrow |\bm{f}^{j+1})|}(\bm{h}^{j+1}))$, where $\bm{h}^{0}$ and $\bm{f}^{0}$ are initialized with random values.
To realize discriminator $\Dis$, we use a typical \emph{sequence-to-one} LSTM~\cite{DBLP:conf/nips/SutskeverVL14}.

Similar to MLP, we make the output layer in $\Gen$ attribute aware by considering transformation method for each attribute.
%
%
In particular, 
for $\tu[j]$ transformed by GMM based normalization, we use two timesteps to generate its sample $\tuvec_{j}$, and concatenate these two parts. 



%
%

\subsection{GAN Training and Mode Collapse} \label{subsec:train}

We use the vanilla GAN training algorithm~\cite{DBLP:conf/nips/GoodfellowPMXWOCB14} (\AlgoVTrain) to iteratively optimize parameters $\theta_{d}$ in $\Dis$ and $\theta_{g}$ in $\Gen$.
%
%
In each iteration, it trains $\Dis$ and $\Dis$ alternately. 
%
%
%
As the algorithm may not provide sufficient gradient to train $\Gen$ in the early iterations~\cite{DBLP:conf/nips/GoodfellowPMXWOCB14}, existing work~\cite{DBLP:journals/corr/abs-1811-11264} introduces the KL divergence between real and synthetic data to warm up model training. Let ${\tt KL}(\T[j], \Tfake[j])$ denote the KL divergence regarding attribute $\T[j]$ between the sampled real examples $\{\tuvec^{(i)}\}^{m}_{i=1}$ and synthetic samples $\{\Gen(\noise^{(i)})\}^{m}_{i=1}$. Based on this, we optimize $\Gen$ by considering the original loss and KL divergences regarding all attributes, i.e.,
\begin{align}\label{eq:vtrain-loss}
\mathcal{L}_{\Gen}=\mathbb{E}_{\noise \sim p(\noise)}[\log(1 - D(G(\noise)))] +\sum_{j=1}^{|\T|}{\tt KL}(\T[j], \Tfake[j]),
\end{align}
%

\noindent \textbf{Mode Collapse.}
We investigate \emph{mode collapse}~\cite{DBLP:conf/nips/SalimansGZCRCC16,DBLP:journals/corr/MetzPPS16}, a well-recognized challenge in GAN training.
Mode collapse would result in similar, or even nearly duplicated records in synthetic table $\Tfake$. The reason is that generator $\Gen$ would generate a limited diversity of samples, regardless of the input noise. Then, as synthetic records are transformed from the samples (see Section~\ref{sec:preproc}), many records will have the same values for most of the attributes as well as the labels.
As a result, the synthetic table $\Tfake$ would fail to preserve the data utility of original table $\T$. For example, a classifier trained on $\Tfake$ may perform badly, and it sometimes achieves very low F1 scores on the test set. 
A deep investigation further shows that, when mode collapse happens, $\Gen$ cannot get sufficient gradient in training iterations and the training algorithm fails to decrease the loss of $\Gen$. In this case, $\Gen$ won't converge and may overfit to a few training records.

We study how to avoid mode collapse by examining the following two strategies.
The first one is to utilize the training algorithm of Wasserstein GAN (\AlgoWTrain)~\cite{DBLP:journals/corr/ArjovskyCB17}, which is commonly used for addressing mode collapse in image synthesis.  
Different from \AlgoVTrain, \AlgoWTrain removes the $\sigmoid$ function of $\Dis$ and changes the gradient optimizer from $\adam$ to $\rms$. It uses the loss functions as
\begin{align} \label{eq:wtrain-loss}
& \mathcal{L}_{D} = -\mathbb{E}_{\tuvec \sim p_{data}(\tuvec) }[D(\tuvec)]+\mathbb{E}_{\noise \sim p(\noise)}[D(G(\noise))] \nonumber \\
& L_{G}=-\mathbb{E}_{\noise\sim p(\noise)}[D(G(\noise))].
\end{align}\vspace{-2mm}
%

The second strategy is to still use the vanilla GAN training algorithm \AlgoVTrain, but simplify the neural network of discriminator $\Dis$. The idea is to make $\Dis$ not trained too well, and thus avoid the chance of gradient disappearance of generator $\Gen$. Specifically, based on some theoretical analysis~\cite{DBLP:conf/nips/GoodfellowPMXWOCB14,DBLP:conf/iclr/ArjovskyB17}, if $\Dis$ is perfectly trained, the loss of $\Gen$ would become a constant and the gradient of $\Gen$ will be vanishing.
	%
In order to avoid such circumstance, we may choose to use a relatively simple neural network to realize $\Dis$, e.g., reducing the numbers of layers or neurons in the neural network.

\begin{table}[!t]
	\centering
	\caption{Comparison of training algorithms.}\label{table:algo-compare} \vspace{-1mm}
		\begin{tabular}{|c||c|c|c|c|}
			\hline
			\textbf{Algorithm} & \textbf{Loss} & \textbf{Optimizer} & \textbf{Sampling} & \textbf{DP}  \\  
			\hline		
			\hline
			\AlgoVTrain & Eq.(\ref{eq:vtrain-loss}) & $\adam$ & random & $\times$ \\ \hline
			\AlgoWTrain & Eq.(\ref{eq:wtrain-loss}) & $\rms$ & random & $\times$ \\ \hline
			\AlgoCTrainPlus & Eq.(\ref{eq:cgan}) & $\adam$ & label-aware & $\times$ \\ \hline
			\AlgoDPTrain & 	Eq.(\ref{eq:wtrain-loss}) & $\rms$ & random & $\surd$ \\
			\hline		
	\end{tabular}
\vspace{-1em}
\end{table}

\subsection{Conditional GAN} \label{subsec:cgan}
%
%
The basic idea of conditional GAN is to encode label as a \emph{condition vector} $\condt \in \mathbb{R}^{c}$ and feed $\condt$ to both generator and discriminator as an additional input.
We respectively represent the generator and the discriminator as $\Gen(\noise | \condt; \theta_{g}) \in \mathbb{R}^{d}$ and $\Dis(\tuvec | \condt; \theta_{d})$. Then, generator $\Gen$ would like to generate samples conditioned on $\condt$ which can perfectly fool discriminator $\Dis$, while $\Dis$ wants to distinguish real samples with condition $\condt$ from synthetic ones, i.e.,
%
\begin{align}\label{eq:cgan}
\min_{\Gen}\max_{\Dis}{V(\Gen,\Dis)} = & \mathbb{E}_{\tuvec \in p_{data}(\tuvec)}\big[\log{\Dis(\tuvec|\condt)}\big] \nonumber \\
& + \mathbb{E}_{\mathbf{z} \in p_{z}(\mathbf{z})}\big[1-\log{\Dis(\Gen(\mathbf{z}|\condt))}\big].
\end{align}
%

One obstacle is that, due to the highly imbalanced label distribution, the minority label may not have sufficient training opportunities. 
To overcome the obstacle, we introduce \emph{label-aware} data sampling in model training (\AlgoCTrainPlus).
The idea is to sample minibatches of real examples by considering labels as a condition, instead of uniformly sampling data. Specifically, in each iteration, the algorithm considers every label in the real data, and for each label, it samples records with corresponding label for the following training of $\Dis$ and $\Gen$. Using this method, we can ensure that data with minority labels also have sufficient training opportunities.

\subsection{Differential Privacy Preserving GAN} \label{subsec:dpgan}
We apply DPGAN~\cite{DBLP:journals/corr/abs-1802-06739} to enable our data synthesizer to support differential privacy. The basic idea is to add noise to the gradients used to
update parameters $\theta_{d}$ to make discriminator $\Dis$ differentially private, since $\Dis$ accesses the real data and has the risk of disclosing privacy information. Then, according to the \emph{post-processing} property of differential privacy, a differentially private $\Dis$ will also enable $\Gen$ differentially private, as parameters $\theta_{g}$ are updated based on the output of $\Dis$.
Overall, DPGAN follows the framework of Wasserstein GAN training with minor modifications (\AlgoDPTrain). 
See the original paper~\cite{DBLP:journals/corr/abs-1802-06739} for more details.

\vspace{1mm}
\noindent \textbf{Algorithm comparison.}
All the algorithms in Sections~\ref{subsec:train} - \ref{subsec:dpgan} share the same optimization framework, i.e., minibatch stochastic gradient descent, but use different strategies. Table~\ref{table:algo-compare} compares the algorithms in loss function, gradient optimizer, sampling and differential privacy (DP) supporting.
We also present the pseudo-codes of all training algorithms in our technical report~\cite{daisy} due to the space limit.

\vspace{-.5em}

\section{Evaluation Methodology} \label{sec:exp-method}

\vspace{-.5em}

\subsection{Datasets}\label{subsec:exp-datasets}

\begin{table}[!t]
	\centering
	\caption{Real datasets for our evaluation: \#Rec, \#C, \#N, and \#L are respectively numbers of records, numerical attributes, categorical attributes, and unique labels.}
	\label{table:data-info}
	\resizebox{0.48\textwidth}{!}{%
	\begin{tabular}{|c|c|c|c|c|c|c|} 
		\hline
		\textbf{Dataset} & \textbf{Domain} & \textbf{\#Rec} 
		&\textbf{\#N} &\textbf{\#C} &\textbf{\#L} & \textbf{Skewness} \\ 
		\hline \hline
		\multicolumn{7}{|c|}{\textbf{low-dimensional (\#Attr$\le20$)}} \\
		\hline
		\dshtru~\cite{HTRU2} & Physical & 17,898 &8 &0 &2  & skew \\		 
		\hline
		\dspendigits~\cite{Pendigits} & Computer & 10,992 &16 &0 &10 &  balanced \\ 
		\hline
		\dsadult~\cite{Adult} & Social &41,292 &6 &8 &2 & skew \\ 
		\hline
		\dsct~\cite{Covertype} & Life &116,204 &10 &2 &7   & skew \\ \hline \hline
		\multicolumn{7}{|c|}{\textbf{high-dimensional (\#Attr$>20$)}} \\
		\hline
		\dssat~\cite{SAT} & Physical &6,435 &36 &0 &6  &  balanced \\  \hline
		\dsanuran~\cite{Anuran} & Life &7,195 &22 &0 &10  & skew \\  \hline
		\dscensus~\cite{Census} & Social &142,522 &9 &30 &2 & skew \\  \hline
		\dsbing~\cite{DBLP:journals/tkde/LiZLTY19} & Web &500,000 &7 &23 &- & - \\ \hline
	\end{tabular}
	}\vspace{-1.5em}
\end{table}

To consider various characteristics of relational data, e.g., mixed data types, attribute correlation and label skewness, we use $8$ real datasets from diverse domains, such as Physical and Social. The datasets are representative that capture different data characteristics, as summarized in Table~\ref{table:data-info}.
First, they have different numbers of attributes ($\term{\#Attr}$), which may affect the performance of data synthesis. For simplicity, we consider low-dimensional ($\term{\#Attr}\le20$) and high-dimensional ($\term{\#Attr}>20$).
Second, they have different attribute types. For both high- and low-dimensional datasets, we differentiate them into \emph{numerical} with only numerical attributes and \emph{mixed} with both numerical and categorical attributes.
Third, they have different label skewness. We consider a dataset is \emph{skew} if the ratio between numbers of records with the most popular and the rarest labels is larger than $9$.
We use four low-dimensional datasets, \dshtru, \dspendigits, \dsadult and \dsct. Among them, \dshtru and \dspendigits only contain numerical attributes,
while \dsadult and \dsct have both numerical and categorical attributes. Moreover, the datasets, except \dspendigits, are skew in label distribution.
For high-dimensional datasets, we use two numerical datasets, \dssat and \dsanuran and two mixed datasets, \dscensus and \dsbing, and both balanced and skew cases are considered.
In particular, \dsbing is a Microsoft production workload dataset, which is used for evaluating AQP in the existing work~\cite{DBLP:journals/tkde/LiZLTY19}. Thus, we only use the \dsbing dataset for AQP in our experiments.
Due to the space limit, we leave more details of the datasets in our technical report~\cite{daisy}.

%
%
	



To provide in-depth analysis on synthesis performance by varying degrees of attribute correlation and label skewness, we also use two sets of simulated datasets.

%
%

\noindent \textbf{(1) \dssda datasets} are used to evaluate data synthesis for records with purely \emph{numerical} attributes.
We follow the simulation method in~\cite{DBLP:journals/corr/abs-1907-00503} to first generate $25$ two-dimensional variables, each of which follows Gaussian distribution $f(x,y)=\mathcal{N}(\mu_{x},\mu_{y};\sigma_{x},\sigma_{y})$ where the means $\mu_{x},\mu_{y}$ are randomly picked from the set of two-dimensional points, $(u, v)|u,v \in \{-4,-2,0,2,4\}$ and standard deviation $\sigma_x$ ($\sigma_y$) $\sim \term{uniform}(0.5,1)$.
Then, we iteratively generate records, in which each record is randomly sampled from one of the 25 Gaussian variables, and assigned with a binary label.

In the simulation, we control \emph{attribute correlation} by varying correlation coefficient $\rho_{xy}={cov(x,y)}/{\sqrt{\sigma_{x}\sigma_{y}}})$ in each Gaussian distribution. 
We consider two degrees of attribute correlation by 
setting the coefficients to $0.5$ and $0.9$ respectively.
We also control \emph{label skewness} by varying the ratio between positive and negative labels assigned to the records. We consider two settings on skewness: $\term{balanced}$ with ratio $1:1$ and $\term{skew}$ with ratio $1:9$.
%

%

\noindent
\textbf{(2) \dssdb datasets} are used to evaluate the synthesis of records with purely \emph{categorical} attributes. 
We generate $5$ categorical attributes as follows.
We first construct a chain Bayesian network with $5$ nodes linked in a sequence, each of which corresponds to a random variable. Then, we generate each record by sampling from the joint distribution modeled by the network, and assign it with a binary label.

We control \emph{attribute correlation} by varying the conditional probability matrix associated with each edge in the Bayesian network. Specifically, we let the diagonal elements to be a specific value $p$ and set the remaining ones uniformly. Intuitively, the larger the $p$ is, the higher dependencies the attributes possess. 
For example, in an extreme case that $p=1$, each attribute (except the first one) deterministically depends on its previous attribute in the network. In such a manner, we consider two degrees of attribute correlation by setting $p=0.5$ and $p=0.9$ respectively. Moreover, similar to \dssda datasets, we also consider $\term{balanced}$ and $\term{skew}$ settings for label skewness on these datasets.

\vspace{-1em}

\subsection{Evaluation Framework} \label{sec:exp-eval}

We implement our GAN-based relational data synthesis framework, as shown in Figure~\ref{fig:archi}, using PyTorch~\cite{pytorch}

To evaluate the performance of the data synthesis framework, we split a dataset into training set $\T_{\tt train}$, validation set $\T_{\tt valid}$ and test set $\T_{\tt test}$ with ratio of $4$:$1$:$1$ respectively, following the existing works for relational data synthesis.
Next, we train a data synthesizer realized by our GAN-based framework on the training set $\T_{\tt train}$ to obtain the optimized parameters of discriminator and generator as follows. We first perform \emph{hyper-parameter search}, which will be described later, to determine the hyper-parameters of the model. Then, we run a training algorithm for parameter optimization. We divide the training iterations in the algorithm evenly into $10$ \emph{epochs} and evaluate the performance of the model snapshot after each epoch on the validation set $\T_{\tt valid}$. We select the model snapshot with the best performance and generate a synthetic relational table $\Tfake$.

After obtaining $\Tfake$, we compare it with the original table $\T_{\tt train}$ on both data utility and privacy protection.

\noindent\textbf{Evaluation on data utility for classification.}
We train a classifier $f^{\prime}$ on the fake table $\Tfake$, while also training a classifier $f$ on the training set $\T_{\tt train}$. In our experiments, we consider the following four types of classifiers for evaluation.
(1) Decision Tree (DT):
We adopt 2 decision trees with max depth 10 and 30 respectively. 
(2) Random Forest (RF):
We adopt two random forests with max depth 10 and 20 respectively. 
(3) AdaBoost (AB): It uses an iterative algorithm to train different classifiers (weak classifiers), then gathers them to form a stronger final classifier for classification. 
(4) Logical Regression (LR): A generalized linear regression model which uses gradient descent method to optimize the classifier for classification.

We evaluate the performance of a trained classifier $f^{\prime}$ on the test set $\T_{\tt test}$.
We use the F1 score, which is the harmonic average of precision and recall, as the evaluation metric for the classifier. In particular, for binary classifier, we measure the F1 score of the positive label, which is much fewer but more important than the negative label.
For the multi-class classifier, we measure the F1 score of the rare label, which is more difficult to predict than others.
%
%
We evaluate the performance of a data synthesizer by measuring the difference $\diff$ of the F1 scores between $f^{\prime}$ and $f$, as defined in Section~\ref{subsec:problem}.
%
%
The smaller the difference is, the better $\Tfake$ is for training.
Note that we also consider area under the receiver operating characteristic curve (AUC) as evaluation, and obtain similar trends with that of F1 score.


\noindent \textbf{Evaluation on data utility for clustering.}
We evaluate the performance of the well-known clustering algorithm K-Means on both $\T_{\tt train}$ and $\Tfake$.
Note that we exclude the label attribute from the features fed into K-Means and instead use it as the gold-standard. We use \emph{normalized mutual information (NMI)} to evaluate the clustering performance. NMI measures the mutual information, i.e., reduction in the entropy of gold-standard labels that we get if we know the clusters, and a larger NMI indicates better clustering performance. After obtaining NMI scores from both the clustering results on $\T_{\tt train}$ and $\Tfake$, we compute the absolute difference of the scores as $\diff_{\tt CST}$, which is defined in Section~\ref{subsec:problem}, and use $\diff_{\tt CST}$ to measure the utility of $\Tfake$ for clustering.

%

\noindent \textbf{Evaluation on data utility for AQP.}
We use the fake table $\Tfake$ to answer a given workload of aggregation queries. We follow the query generation method in~\cite{DBLP:journals/tkde/LiZLTY19} to generate $1,000$ queries with aggregate functions (i.e., $\term{count}$, $\term{avg}$ and $\term{sum}$), selection conditions and groupings. We also run the same queries on the original table $\T_{\tt train}$. For each query, we measure the relative error $e^\prime$ of the result obtained from $\Tfake$ by comparing with that from $\T_{\tt train}$.
Meanwhile, following the method in~\cite{DBLP:journals/corr/abs-1903-10000}, we draw a fixed size random sample set ($1\%$ by default) from the original table, run the queries on this sample set, and obtain relative error $e$ for each query. To eliminate randomness, we draw the random sample sets for $10$ times and compute the averaged $e$ for each query. Then, as mentioned in Section~\ref{subsec:problem}, we compute the relative error difference $\diff_{\tt AQP}$ and average the difference for all queries in the workload, to measure the utility of $\Tfake$ for AQP. 
\noindent \textbf{Evaluation on privacy protection.}
We adopt the following two metrics, which are widely used in the existing works~\cite{DBLP:journals/tdp/ParkG14,DBLP:conf/wims/LuWY19,DBLP:conf/psd/Mateo-SanzSD04} for privacy evaluation. 

1) \emph{Hitting Rate}: It measures how many records in the original table $\T_{\tt train}$ can be hit by a synthetic record in $\Tfake$. To measure hitting rate, we first randomly sample $5000$ synthetic records from $\Tfake$. For each sampled record, we measure the proportion of records in $\T_{\tt train}$ that are \emph{similar} to this synthetic record. We regard two records are similar if and only if 1) the values of each categorical attribute are the same, and 2) the difference between values of each numerical attribute is within a threshold. In our experiment, this threshold is set as the range of the attribute divided by $30$.

2) \emph{Distance to the closest record (DCR)}: This measures whether the synthetic data is weak from re-identification attacks~\cite{DBLP:journals/tdp/ParkG14,DBLP:conf/wims/LuWY19}. Given a record $\tu$ in the original table $\T_{\tt train}$, we find the synthetic record from $\Tfake$ that is closest to $\tu$ in Euclidean distance. Note that a record with DCR=0 means that $\Tfake$ leaks its real information, and the larger the DCR is, the better the privacy protection is. To measure DCR, we calculate the distance after attribute-wise normalization to make sure each attribute contributes to the distance equally. 
We sample 3000 records from the original table $\T_{\tt train}$, and find the the nearest synthetic record in $\Tfake$ for each of these records. Then, we compute the the average distance between the real record to its closest synthetic record.

\vspace{-.5em}

\subsection{Data Synthesis Methods} \label{subsec:exp-method}

\noindent \textbf{GAN-based methods.} We implement the design choices shown in Figure~\ref{fig:design-space}. We use the code provided by~\cite{DBLP:journals/pvldb/ParkMGJPK18} to implement the CNN-based model\footnote{https://github.com/mahmoodm2/tableGAN}.
We use the hyper parameters provided by the code to train the model. Moreover, the code provides three privacy settings. When evaluating the ML training utility, we choose the settings of the weakest privacy protection to achieve the best synthetic data utility. On the other hand, We implement the MLP-based and LSTM-based models by ourselves using PyTorch to enable the flexibility of adapting different transformation schemes for comprehensive evaluation.
Also, we implement the variants of training algorithms, conditional GAN and DPGAN.

\noindent \textbf{Statistical methods.}
We compare GAN with a state-of-the-art statistical data synthesis method PrivBayes (or \bsbayes for simplicity)~\cite{DBLP:conf/sigmod/ZhangCPSX14,DBLP:journals/tods/ZhangCPSX17}, using the source code downloaded here\footnote{https://sourceforge.net/projects/privbayes/}. As \bsbayes has theoretical guarantee on differential privacy~\cite{DBLP:journals/fttcs/DworkR14}, we vary the privacy parameter $\epsilon$ to examine the tradeoff between privacy protection and data utility. According to the original papers~\cite{DBLP:conf/sigmod/ZhangCPSX14,DBLP:journals/tods/ZhangCPSX17}, we run \bsbayes in multiple times and report the average result.

\noindent \textbf{Variational Autoencoder (VAE)}.
We implement variational autoencoder (VAE), which is another representative deep generative model~\cite{DBLP:journals/corr/KingmaW13,DBLP:conf/icml/RezendeMW14} for relational data synthesis.
We adopt the loss function that consists of both the reconstruction loss and the KL divergence~\cite{DBLP:journals/corr/Doersch16}. We use binary cross-entropy (BCE) loss for categorical attributes and mean squared error (MSE) loss for numerical attributes.

\vspace{-.25em}

\subsection{Hyper Parameter Search} \label{subsec:exp-param}
\vspace{-.25em}

Hyper parameter search is very important for neural networks. We adopt the method in a recent empirical study for GAN models~\cite{DBLP:conf/nips/LucicKMGB18} for hyper parameter search. Given a GAN model, we firstly generate a set of candidate hyper parameter settings.
Then, we train the model for several times, and at each time, we randomly select a hyper parameter setting and evaluate the trained model on the validation set $\T_{\tt valid}$. Based on this, we select the hyper parameter setting that results in a model with the best performance. 


\vspace{1mm}
All the experiments are conducted on a server with 2TB disk, 40 CPU cores (Intel Xeon CPU E5-2630 v4 @ 2.20GHz), one GPU (NVIDIA TITAN V) and 512GB memory, and the version of Python is 3.6.5.

\vspace{-.5em}

\section{Evaluation Results} \label{sec:result}
\vspace{-.5em}

\subsection{Evaluating GAN-based Framework} \label{subsec:result-gan}
This section explores the design space of our GAN-based framework. We focus on synthetic data utility for classification, and report privacy results and data utility for clustering and AQP in next sections.

\begin{table}[!t]
	\caption{Evaluating different neural networks of generator $\Gen$ on synthetic data utility for classification, where $\term{CLF}$ stands for classifier. For low-dimensional datasets with less attributes, LSTM with appropriate transformation achieves much less F1 differences than MLP and CNN. For high-dimensional datasets with more attributes, the performance advantage of LSTM over MLP becomes less significant.
	}\label{table:all-model} \vspace{-1em} 
	\centering
	
	\subtable[\dsadult dataset (low-dimensional).]{
		\resizebox{0.47\textwidth}{!}{%
		\begin{tabular}{|c||c|c|c|c|c|c|c|c|c|}
			\hline
			\multirow{2}{*}{\textbf{CLF}} &
			\multirow{2}{*}{\textbf{CNN}} & 
			\multicolumn{4}{c|}{\textbf{MLP}} &
			\multicolumn{4}{c|}{\textbf{LSTM}} 
			\\
			\cline{3-10}
			& &\norm/\ordinal &\norm/\onehot &\gmm/\ordinal &\gmm/\onehot
			&\norm/\ordinal &\norm/\onehot &\gmm/\ordinal &\gmm/\onehot \\  
			\hline	
			\hline DT10
			&0.475
			&0.062 &0.062 &0.056 &0.040
			&0.069 &0.113 &0.088&\textbf{0.032} \\ 
			\hline
			DT30 
			&0.485
			&0.071 &\textbf{0.049} &0.077 &0.094
			&0.059 &0.167 &0.088 &0.062 \\ 
			\hline
			RF10 
			&0.417
			&0.035 &0.038 &0.029 &0.018
			&0.136 &0.050 &0.054 &\textbf{0.015}   \\ 
			\hline
			RF20 
			&0.458
			&0.060 &0.066 &0.053 &0.034
			&0.125 &0.047 &0.051 &\textbf{0.006} \\ 
			\hline
			AB 
			&0.217
			&0.066 &0.059 &0.029 &0.042
			&0.219 &0.025 &0.064 &\textbf{0.009} \\
			\hline
			LR
			&0.047
			&0.018 &0.088 &0.018 &0.013
			&0.012 &0.009 &\textbf{0.006} &0.012 \\
			\hline		
		\end{tabular} 
		\label{table:adult-model}
}	
	}\vspace{-0.5em}
	
	\subtable[\dsct dataset (low-dimensional).]{        
		
		\resizebox{0.47\textwidth}{!}{
			\begin{tabular}{|c||c|c|c|c|c|c|c|c|}
				\hline
				\multirow{2}{*}{\textbf{CLF}} &
				\multicolumn{4}{c|}{\textbf{MLP}} &
				\multicolumn{4}{c|}{\textbf{LSTM}} 
				\\
				\cline{2-9}
				&\norm/\ordinal &\norm/\onehot &\gmm/\ordinal 	&\gmm/\onehot
				&\norm/\ordinal &\norm/\onehot &\gmm/\ordinal &\gmm/\onehot  \\  
				\hline	
				\hline DT10
				&0.190 &0.170
				&0.566 &0.241 &0.130 &0.107 &0.402 &\textbf{0.079} \\ 
				\hline
				DT30 
				&0.534 &0.327 &0.752 &0.437 &0.419 &0.606 &0.652 &\textbf{0.305} \\ 
				\hline
				RF10 
				&0.165 &0.123
				&0.455 &0.155 
				&\textbf{0.111} &0.198 &0.259 &0.113 \\ 
				\hline
				RF20 
				&0.342 &0.253 &0.648 &0.264 &0.247 &0.312 
				&0.491 &\textbf{0.197} \\ 
				\hline
				AB 
				&0.091 &0.070 
				&0.321 &\textbf{0.029} 
				&0.056 &0.036 &0.098 &0.038  \\ 
				\hline
				LR
				&0.130 &0.058
				&0.516 &0.113 
				&0.076 &0.369 &0.378 &\textbf{0.043}  \\
				\hline
				
			\end{tabular}
		} 
		
		\label{table:covertype-model}
		
	}\vspace{-0.5em}

	\subtable[\dscensus dataset (high-dimensional).]{      
			\resizebox{0.47\textwidth}{!}{
			\begin{tabular}{|c||c|c|c|c|c|c|c|c|c|}
				\hline
				\multirow{2}{*}{\textbf{CLF}} &
				\multirow{2}{*}{\textbf{CNN}} & 
				\multicolumn{4}{c|}{\textbf{MLP}} &
				\multicolumn{4}{c|}{\textbf{LSTM}} 
				\\
				\cline{3-10}
				& &\norm/\ordinal &\norm/\onehot &\gmm/\ordinal &\gmm/\onehot
				&\norm/\ordinal &\norm/\onehot &\gmm/\ordinal &\gmm/\onehot \\  
				\hline	
				\hline DT10
				&0.484
				&0.188 &0.119 &0.119 &\textbf{0.113}
				&0.211 &0.332 &0.162 &0.180 \\ 
				\hline
				DT30 
				&0.462
				&0.172 &\textbf{0.106} &0.116 &0.114
				&0.288 &0.288 &0.185 &0.157 \\ 
				\hline
				RF10 
				&0.214
				&\textbf{0.007} &0.038 &0.050 &0.028
				&0.132 &0.035 &0.058 &0.012   \\ 
				\hline
				RF20 
				&0.410
				&0.166 &\textbf{0.051} &0.053 &0.095
				&0.189 &0.244 &0.109 &0.089 \\ 
				\hline
				AB 
				&0.506
				&0.215 &0.107 &0.127 &\textbf{0.063}
				&0.144 &0.239 &0.082 &0.113 \\
				\hline
				LR
				&0.494
				&0.133 &\textbf{0.047} &0.085 &0.069
				&0.358 &0.085 &0.250 &0.053 \\
				\hline		
			\end{tabular}
		}
			\label{table:census-model}
	}\vspace{-0.5em}

	\subtable[\dssat dataset (high-dimensional).]{      
			\resizebox{0.43\textwidth}{!}{
			\begin{tabular}{|c||c|c|c|c|c|}
				\hline
				\multirow{2}{*}{~~~~~\textbf{CLF}}~~~~~ &
				\multicolumn{2}{c|}{\textbf{MLP}} &
				\multicolumn{2}{c|}{\textbf{LSTM}} 
				\\
				\cline{2-5}
				&~~~~~\norm~~~~~ &~~~~~\gmm~~~~~ &~~~~~\norm~~~~~ &~~~~~\gmm~~~~~  \\  
				\hline		
				\hline DT10
				&0.098 &0.048 &0.047 &\textbf{0.042}   \\ 
				\hline
				DT30 
				&0.063 &0.090 &0.041 &\textbf{0.041}   \\  
				\hline
				RF10 
				&0.155 &0.100 &0.149 &\textbf{0.051}   \\ 
				\hline
				RF20 
				&0.181 &0.108 &0.157 &\textbf{0.093}   \\  
				\hline
				AB 
				&\textbf{0.017} &0.065 &0.040 &0.182    \\ 
				\hline 
				LR 
				&\textbf{0.009} &0.014 &0.018 &0.017  \\
				\hline
				
			\end{tabular}
		} 
	\label{table:SAT-model}
	} \vspace{-2em}
\end{table}

\begin{figure}[!t]
	\begin{center}\hspace{-1mm}
		\subfigure[{\small LSTM-based $\Gen$ (\dsadult).}]{
			\label{exp:search-1}
			\epsfig{figure=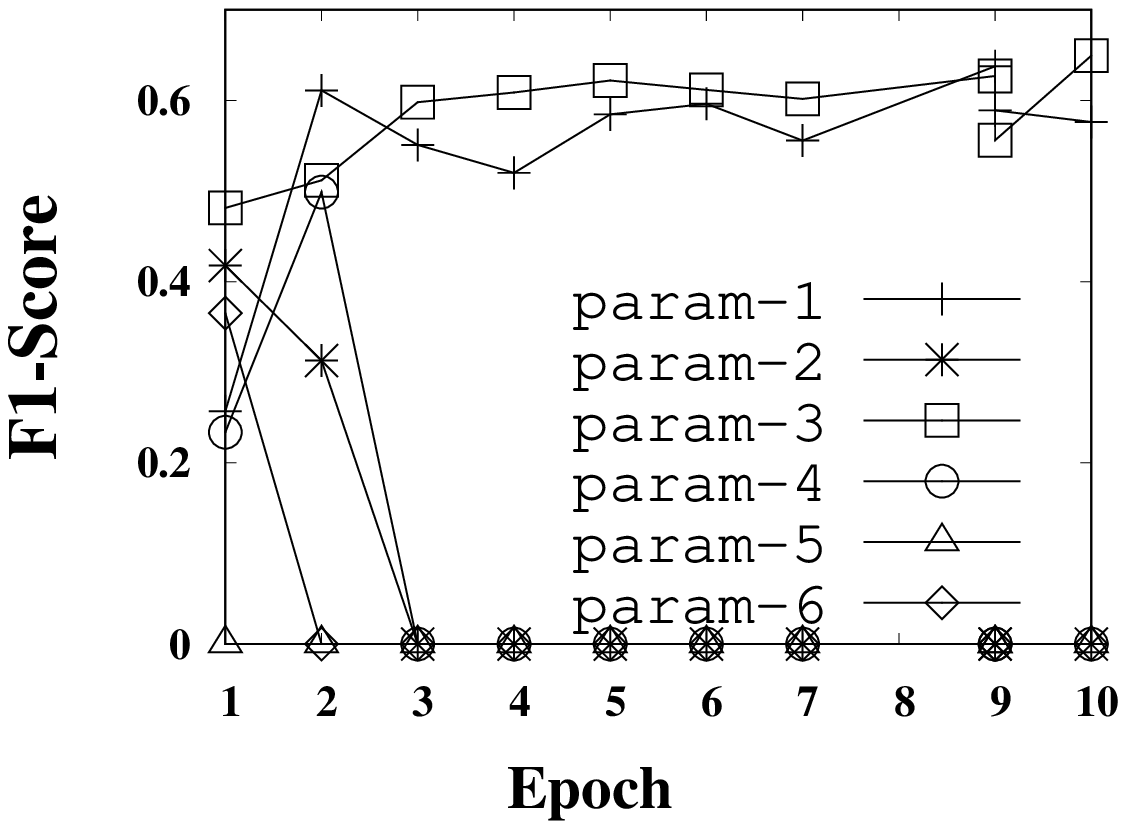,width=0.22\textwidth}
		}
		\subfigure[{\small LSTM-based $\Gen$ (\dsct).}]{
			\epsfig{figure=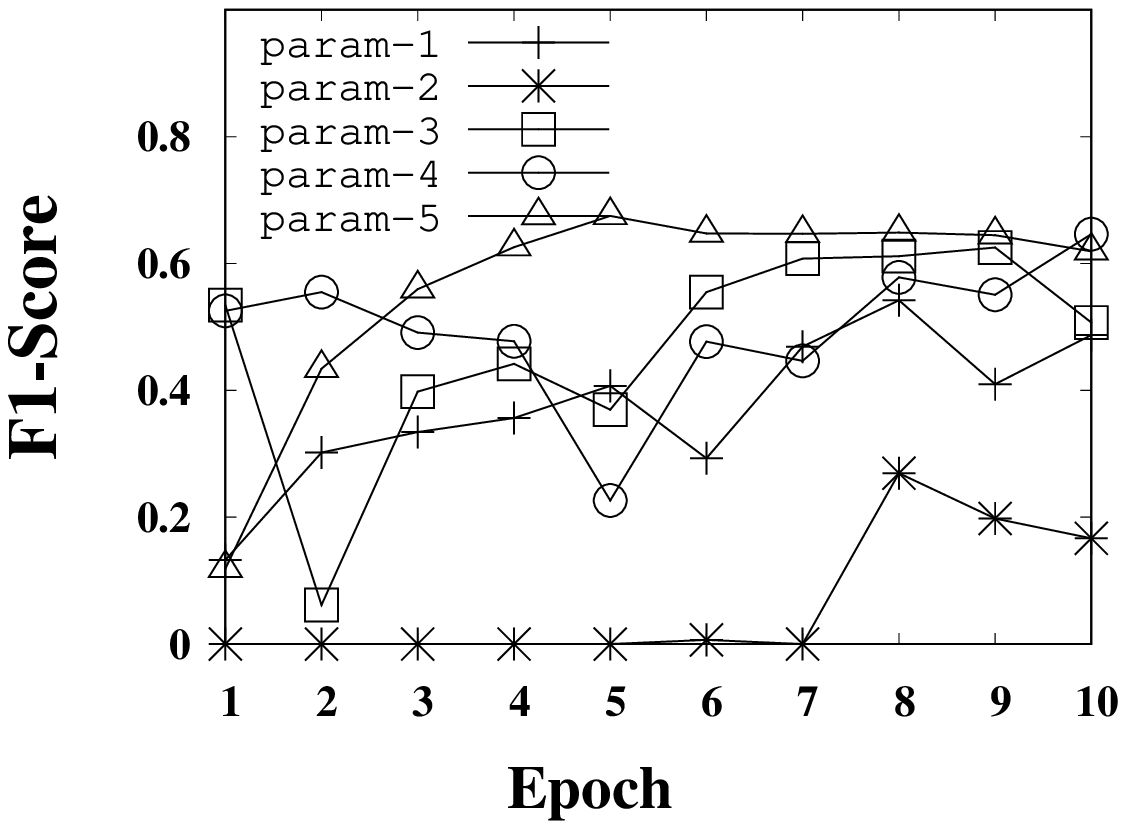,width=0.22\textwidth}
		}\vspace{-1em}
	\subfigure[{\small MLP-based $\Gen$ (\dsadult).}]{
		\label{exp:search-2}
		\epsfig{figure=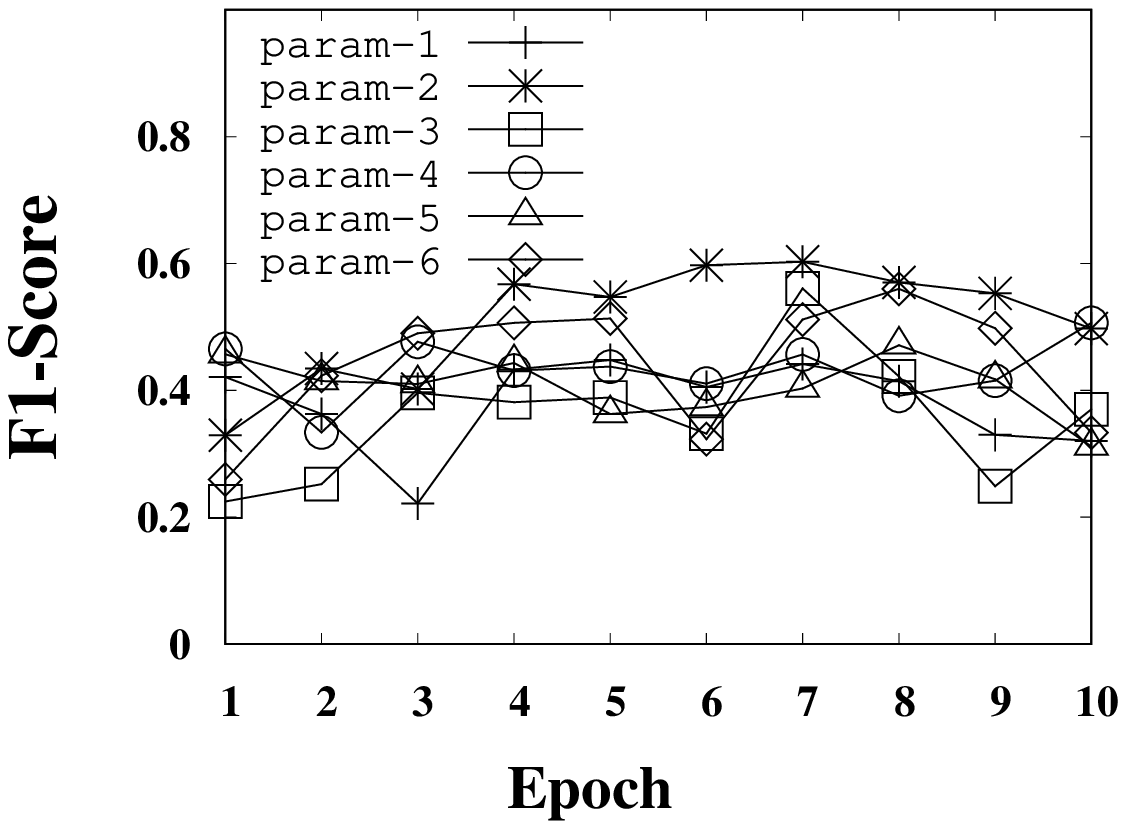,width=0.22\textwidth}
	}
		\subfigure[{\small MLP-based $\Gen$ (\dsct).}]{
			\epsfig{figure=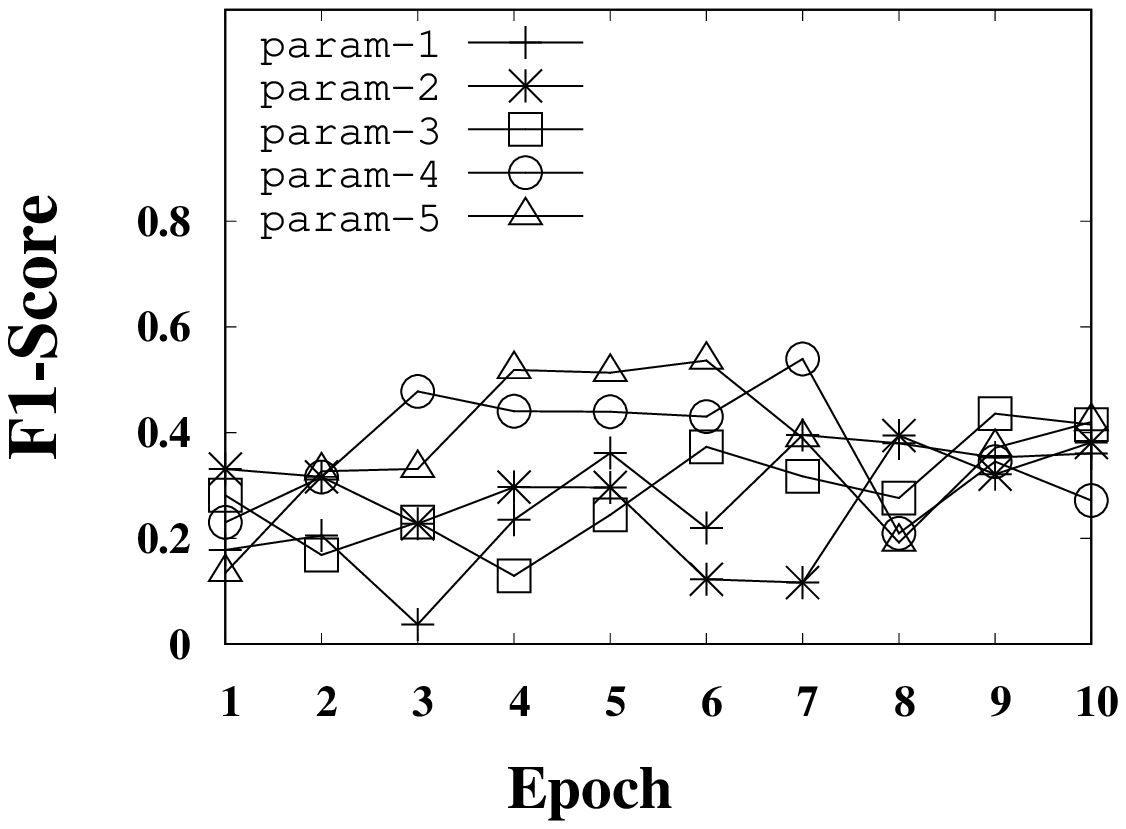,width=0.22 \textwidth}
		}
	\end{center}\vspace{-2em}
	\caption{Evaluating GAN model training on various hyper-parameter settings. MLP-based generator is more robust against various hyper parameters, while LSTM is likely to result in mode collapse.} \label{exp:search}
	\vspace{-2em}
\end{figure}

\begin{figure*}[!t]\vspace{-1em}
	\begin{center}\hspace{-1mm}
		\hspace{-3mm}
		\subfigure[{\small \dsadult dataset.}]{
			\epsfig{figure=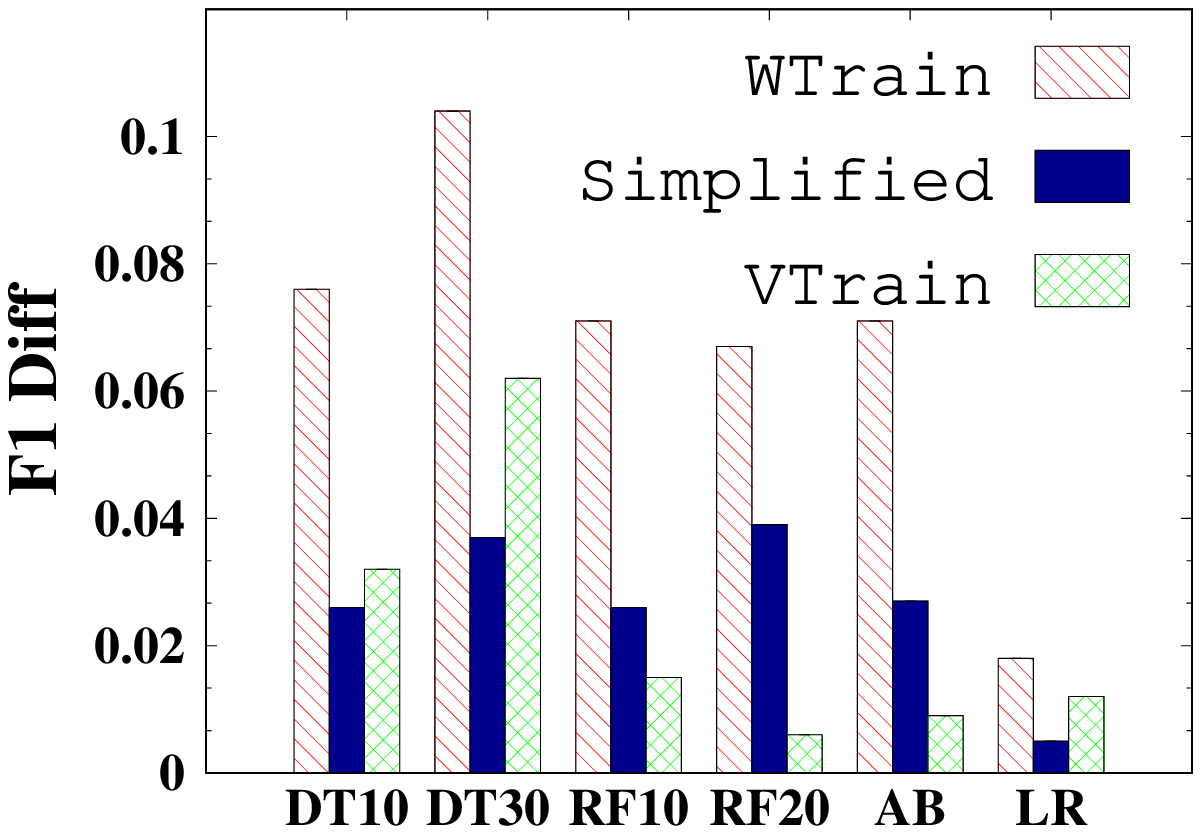,width=0.22\textwidth}
		}\hspace{-1mm}
		\subfigure[{\small \dsct dataset.}]{
			\epsfig{figure=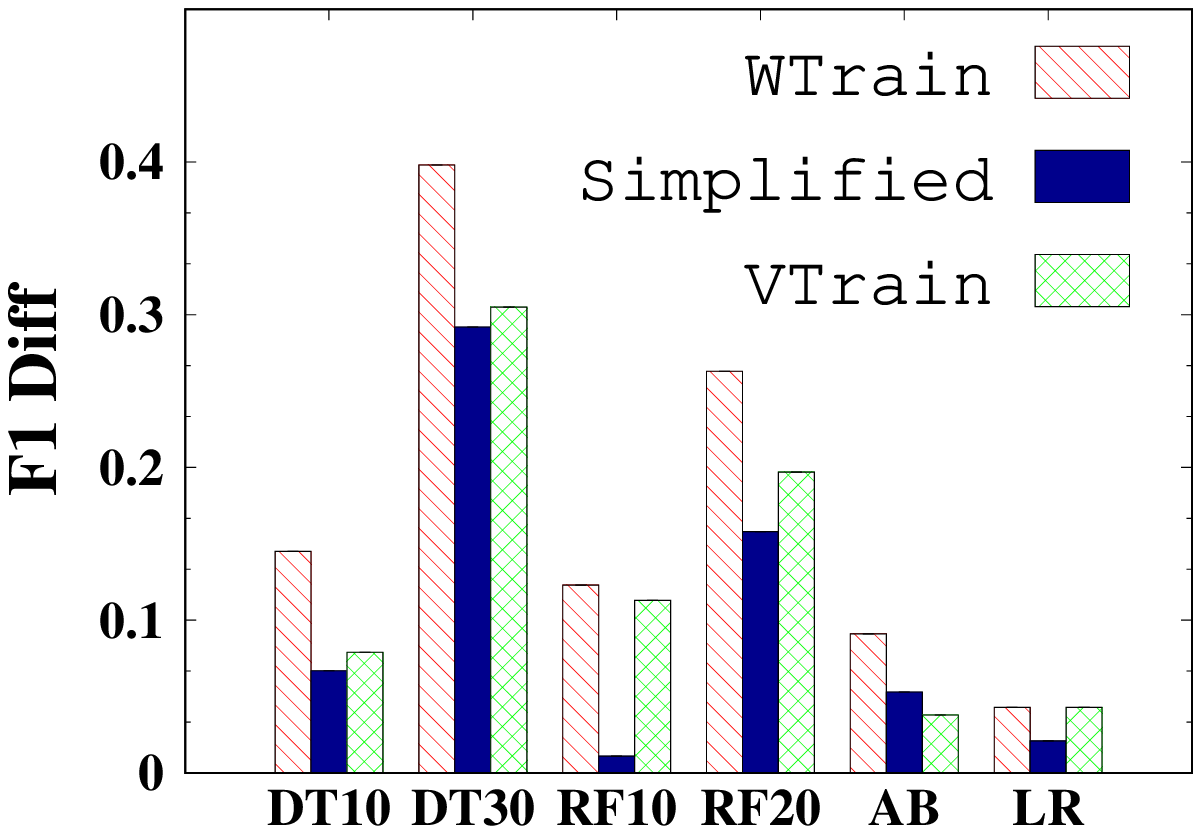,width=0.22\textwidth}
		}\hspace{-1mm}
		\subfigure[{\small \dssat dataset.}]{
			\epsfig{figure=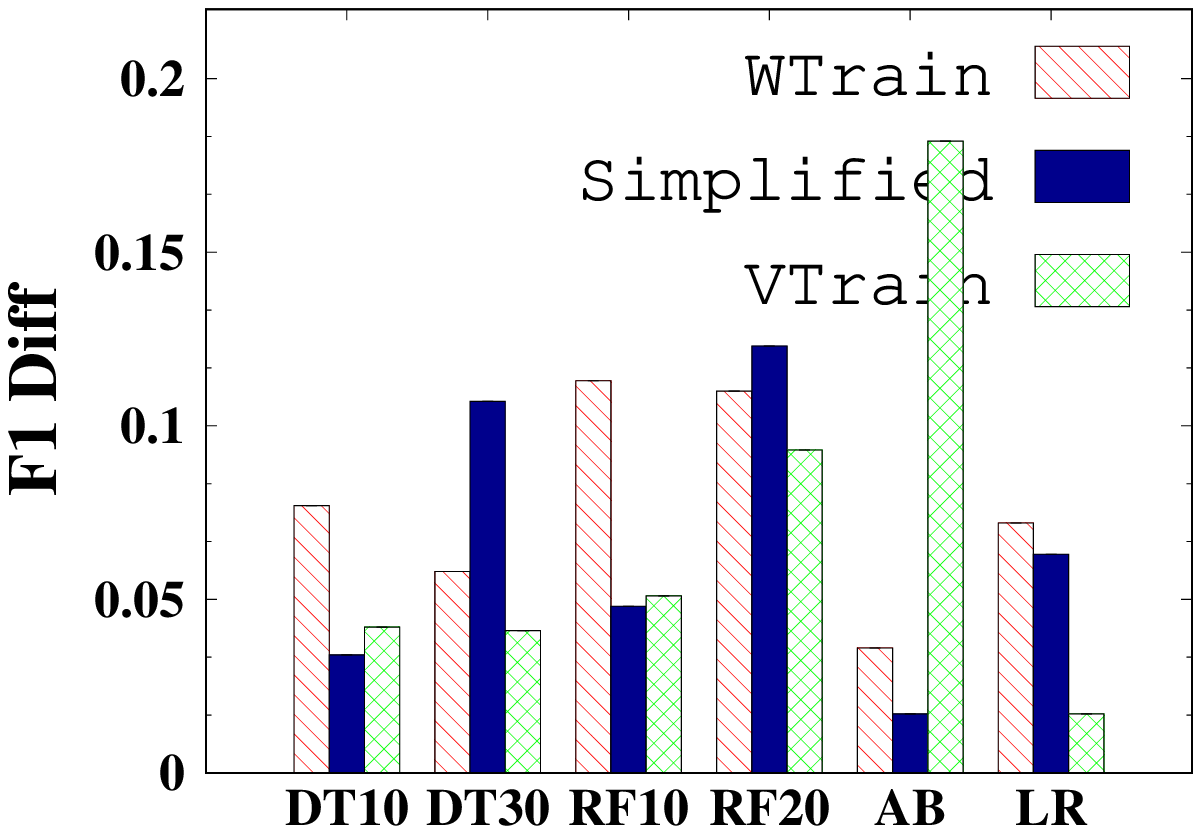,width=0.22\textwidth}
		}
		\subfigure[{\small \dscensus dataset.}]{
			\epsfig{figure=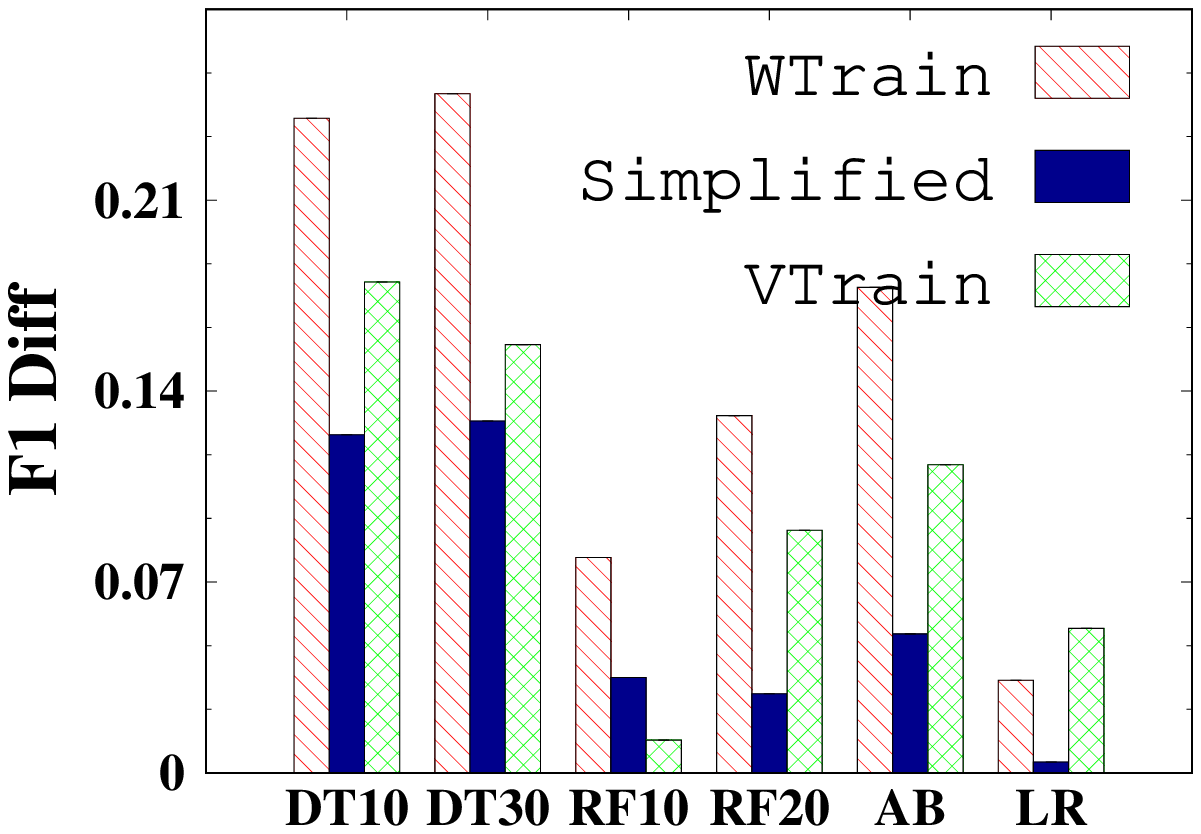,width=0.22\textwidth}
		}
	\end{center}\vspace{-2.5em}
	\caption{Comparison of strategies that are used to avoid mode collapse. 
	} \label{exp:wtrain}
\end{figure*}

%

%

\vspace{-.5em}
\subsubsection{Evaluation on Neural Networks} \label{subsec:result-nn}
We evaluate the neural networks, CNN, MLP and LSTM that realize the generator $\Gen$ in our framework. For MLP and LSTM, we fix the discriminator $\Dis$ as MLP. We also evaluate the LSTM-based discriminator and obtain inferior result (the result is included in our technical report~\cite{daisy}).

We first evaluate the data utility for classification model training.
Due to the space limit, we report the results on two low-dimensional ($\term{\#Attr}\le20$) datasets \dsadult and \dsct, and two high-dimensional ones \dssat and \dscensus, and we find similar results on other datasets.
Tables~\ref{table:adult-model}, \ref{table:covertype-model}, \ref{table:census-model} and \ref{table:SAT-model} report the experimental results on data utility for classification model training, where \norm, \gmm, \ordinal, and \onehot respectively denote simple normalization, GMM-based normalization, ordinal encoding and one-hot encoding. Note that CNN is not evaluated on \dsct and \dssat, as the original code in~\cite{DBLP:journals/pvldb/ParkMGJPK18} is not designed for multi-class classification.

On the datasets \dsadult and \dsct with less attributes, LSTM achieves the best performance in most of the cases, i.e., achieving $7\% - 90\%$ less F1 difference than the second best model MLP.
This suggests the \emph{sequence generation} mechanism in LSTM, which generates a record attribute by attribute, is more adequate for relational data synthesis. First, each attribute is generated from a separated noise $\noise$, which avoids the disturbance among different attributes. Second, LSTM does not generate an attribute from scratch. Instead, it generates an attribute based on the ``understanding'' of previous attributes, i.e., the hidden state $\bm{h}$ and previous output $\bm{f}$, and thus it would be capable of capturing column correlation.
%
%
Nevertheless, on datasets \dscensus and \dssat with more attributes, the performance advantage of LSTM is less significant. The reason is that, with more attributes, it becomes more difficult for LSTM to capture correlation among attributes, which implies that more effective models should be invented for data synthesis.

CNN achieves the inferior performance in data synthesis, which is different from image synthesis~\cite{DBLP:journals/corr/RadfordMC15}. This is because matrix input of CNN is only compatible with simple normalization and ordinal encoding, which is not effective for relational data.
Moreover, convolution/deconvolution operation in CNN is usually effective for data with \emph{feature locality}. For example, features, which are locally close to each other in the matrix of an image, may also be semantically correlated.
However, relational data does not have such locality.

\vspace{1mm}
\noindent \textbf{Finding 1: LSTM with appropriate transformation schemes generates the best synthetic data utility for classification. Nevertheless, with the increase of the number of attributes, the performance advantage achieved by LSTM becomes less significant.}

%
%

\vspace{1mm}
For ease of presentation, we use LSTM with one-hot encoding and GMM-based normalization as default setting.

\subsubsection{Evaluation on GAN Training}~\label{subsubsec:exp-training}
We evaluate the \emph{robustness} of MLP-based and LSTM-based generator wrt. hyper parameters.
Given a setting of parameters, we divide the training iterations evenly into $10$ \emph{epochs} and generate a snapshot of synthetic table after each epoch. Then, we evaluate the F1 score of a classifier trained on each synthetic table snapshot. Figure~\ref{exp:search} shows the results on datasets \dsadult and \dsct. Note that we find similar trends on other datasets, and include result in~\cite{daisy}. 
We have a surprising observation that the LSTM-based generator performs badly in some hyper parameter settings. 
For example, on the \dsadult dataset, the F1 score drops sharply to $0$ after the few early epochs in $4$ out of $6$ hyper parameter settings.
After sampling records from inferior synthetic table snapshots, we find the reason is \emph{mode collapse}: generator $\Gen$ only produces nearly duplicated samples, rather than outputting diverse synthetic records.
MLP-based generator is robust against various hyper parameter settings, and it achieves moderate results on F1 score, although its best case is worse than that of LSTM-base generator.

%
%

\noindent \textbf{Finding 2: MLP is more robust against hyper parameter settings and achieves the moderate results, while LSTM is more likely to result in mode collapse if its hyper parameters are not well tuned.}

We also examine the following training strategies to alleviate mode collapse: (i) \AlgoVTrain (with KL divergence), (ii) Wasserstein GAN training (\AlgoWTrain) and (iii) \AlgoVTrain with simplified discriminator $\Dis$ (\AlgoSimD).
As shown in Figure~\ref{exp:wtrain}, Wasserstein GAN does not have advantage over vanilla GAN training, which is different from the image synthesis scenarios, and \AlgoSimD achieves better performance than \AlgoVTrain. For example, on the \dsadult dataset, \AlgoSimD reduces F1 difference compared with \AlgoVTrain on most classifiers. We also report a result of \AlgoSimD against various hyper-parameters in~\cite{daisy} and find it more robust in avoiding mode collapse. The reason is that \AlgoSimD makes $\Dis$ not trained too well, and thus avoids the chance of gradient disappearance of generator $\Gen$.


\vspace{1mm}
\noindent \textbf{Finding 3: Vanilla GAN training with simplified discriminator is shown effective to alleviate mode collapse, and outperforms Wasserstein GAN training in preserving data utility.}


%
%

\vspace{-.5em}
\subsubsection{Evaluation on Conditional GAN}
\begin{figure*}[!t]\vspace{-1.5em}
\vspace{-.5em}
	\begin{center}
		\subfigure[{\dsadult dataset.}]{
			\label{exp:condt-1}
			\epsfig{figure=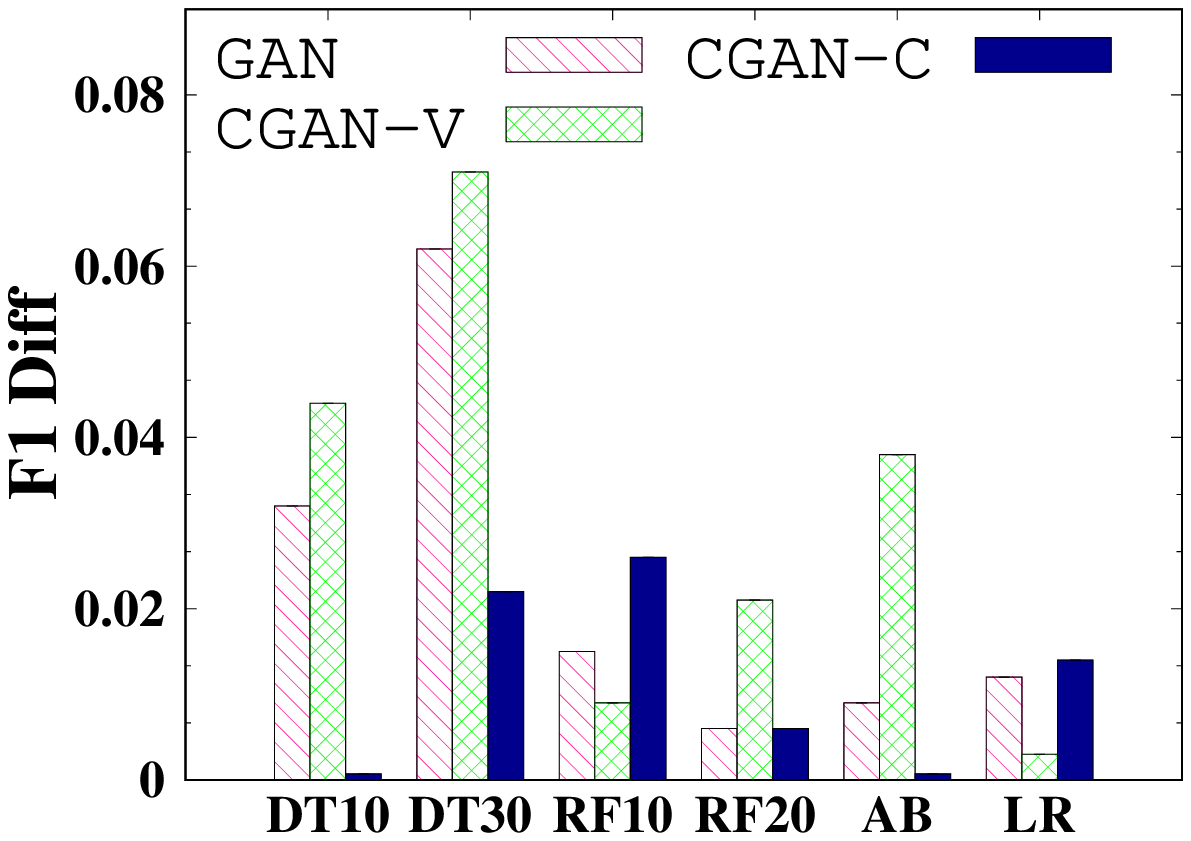,width=0.22\textwidth}
		}
		\subfigure[{\dsct dataset.}]{
			\epsfig{figure=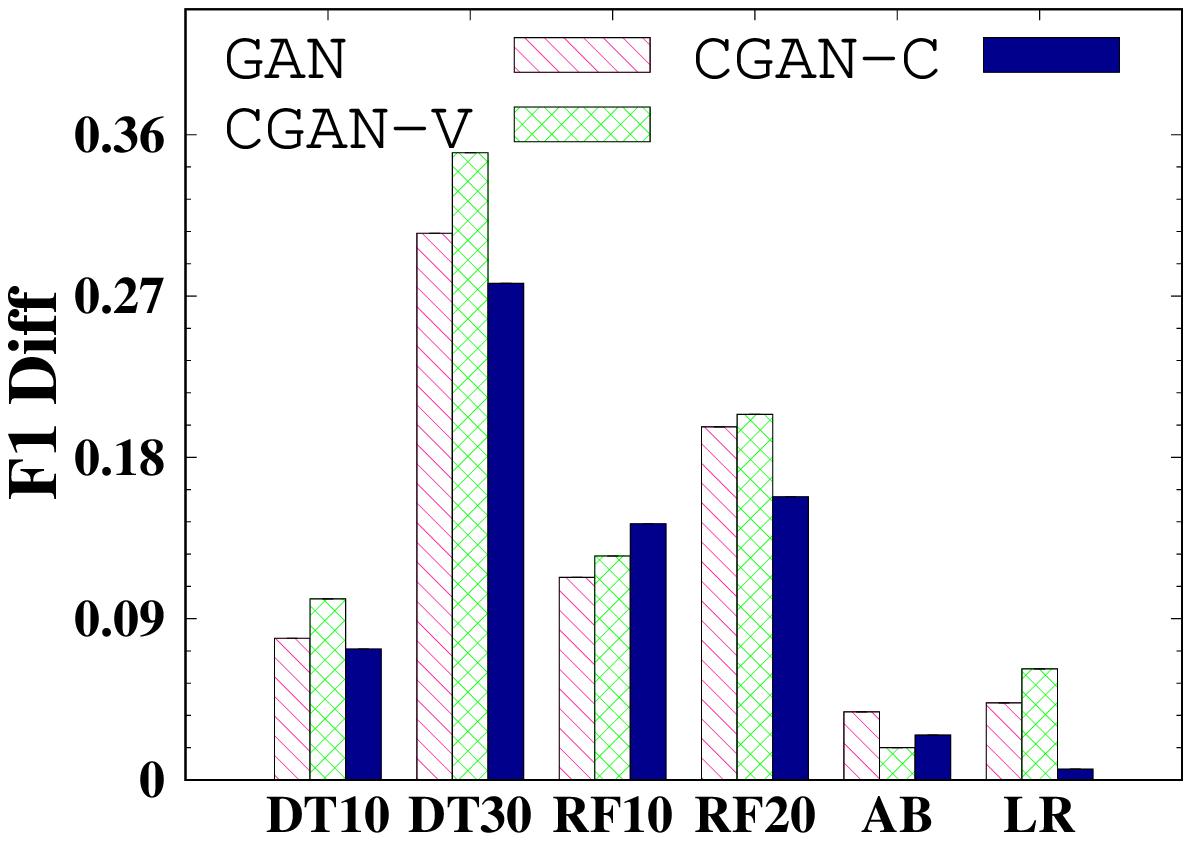,width=0.22\textwidth}
		}
		\subfigure[{\dscensus dataset.}]{
			\epsfig{figure=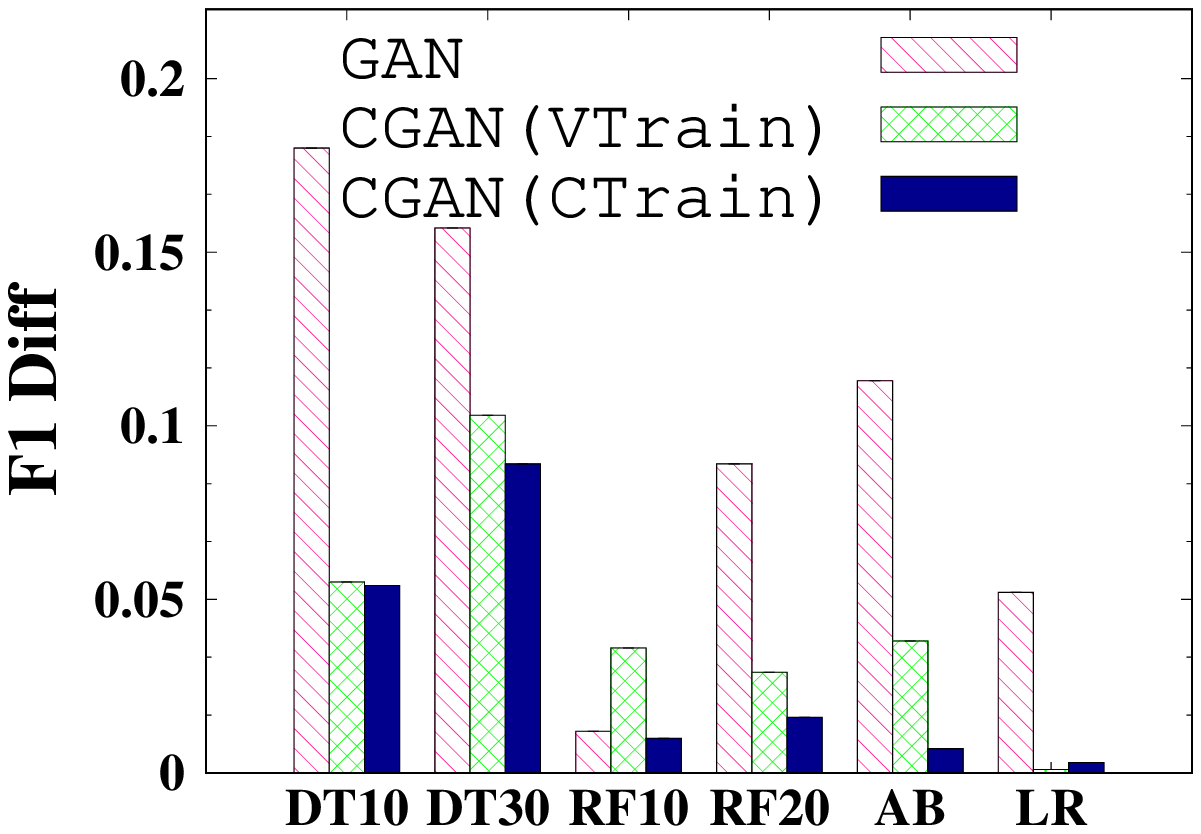,width=0.22\textwidth}
		}
		\subfigure[{\dsanuran dataset.}]{
			\epsfig{figure=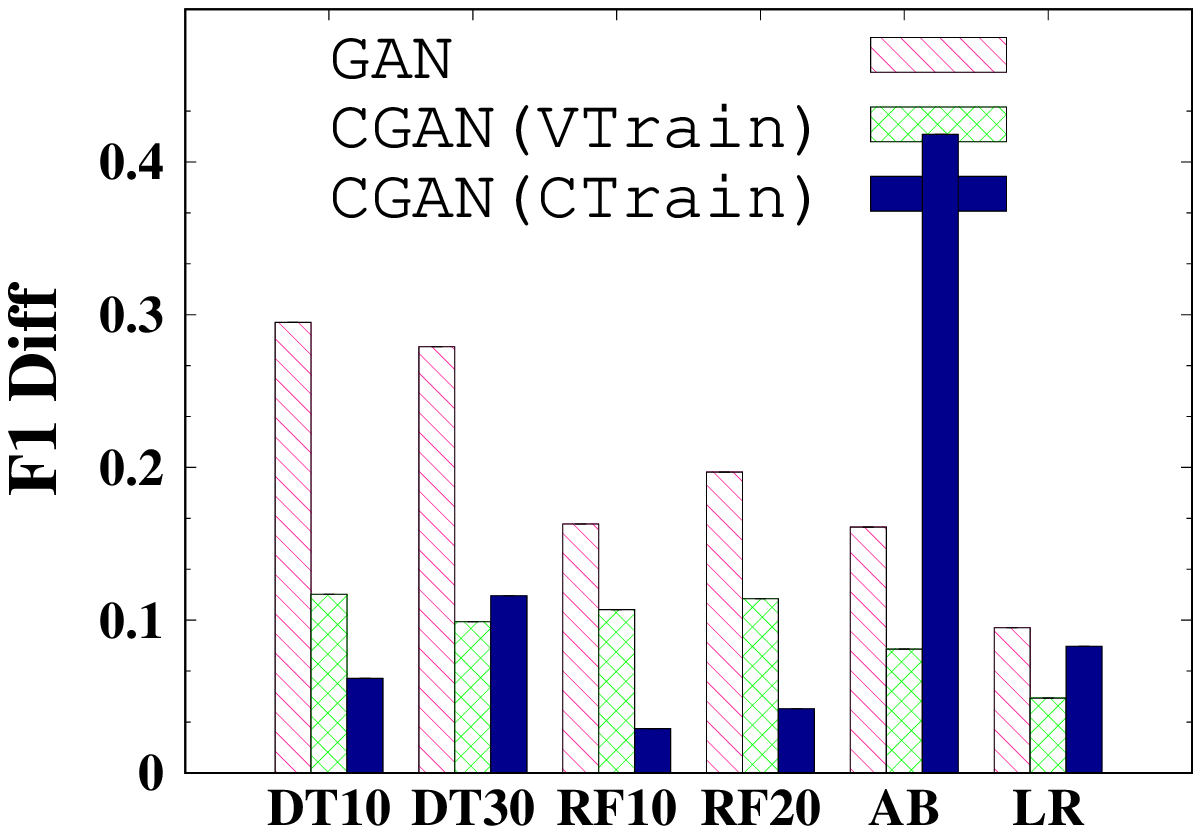,width=0.22\textwidth}
		}
	
	\end{center}\vspace{-2.5em}
	\caption{Evaluating conditional GAN on synthetic data utility for classification.} \label{exp:condt}
	\vspace{-1em}
\end{figure*}

This section investigates if conditional GAN is helpful to address the challenge of imbalance label distribution. We compare the original GAN, conditional GAN trained by random data sampling and conditional GAN trained by label-aware data sampling, which are denoted by \bsvgan, \bscgan-${\tt V}$ and \bscgan-${\tt C}$ respectively, on the skew datasets \dsadult, \dsct, \dscensus and \dsanuran.
%
As shown in Figure~\ref{exp:condt}, \bscgan-${\tt V}$ gains very limited improvements over \bsvgan, and sometimes it performs worse than \bsvgan. This is because that \bsvtrain uses the random strategy to sample each minibatch of real records. Due to the label imbalance, records with minority labels
may have less chances to be sampled, leading to insufficient training opportunities for the minority labels.
On the contrary, \bscgan-${\tt C}$ solves this problem by sampling records conditioned on given labels. This label-aware sampling method can provide fair training opportunities for data with different labels.

\vspace{1mm}
\noindent \textbf{Finding 4: Conditional GAN plus label-aware data sampling is helpful to address imbalance label distribution and improves the utility of synthetic data.}

\begin{figure*}[!t]\vspace{-.5em}
	\begin{center}
		\subfigure[{\dsadult dataset.}]{
			\epsfig{figure=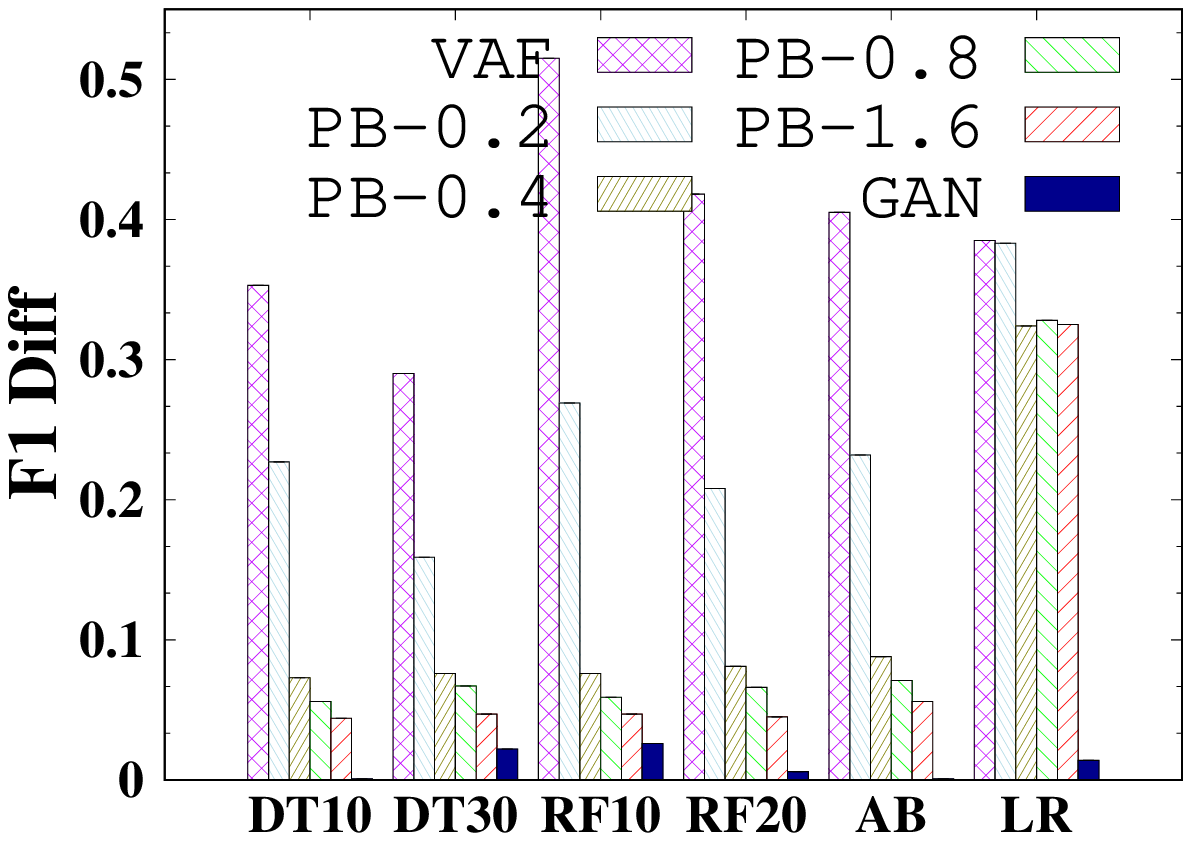,width=0.23\textwidth}
		}
		\subfigure[{\dsct dataset.}]{
			\epsfig{figure=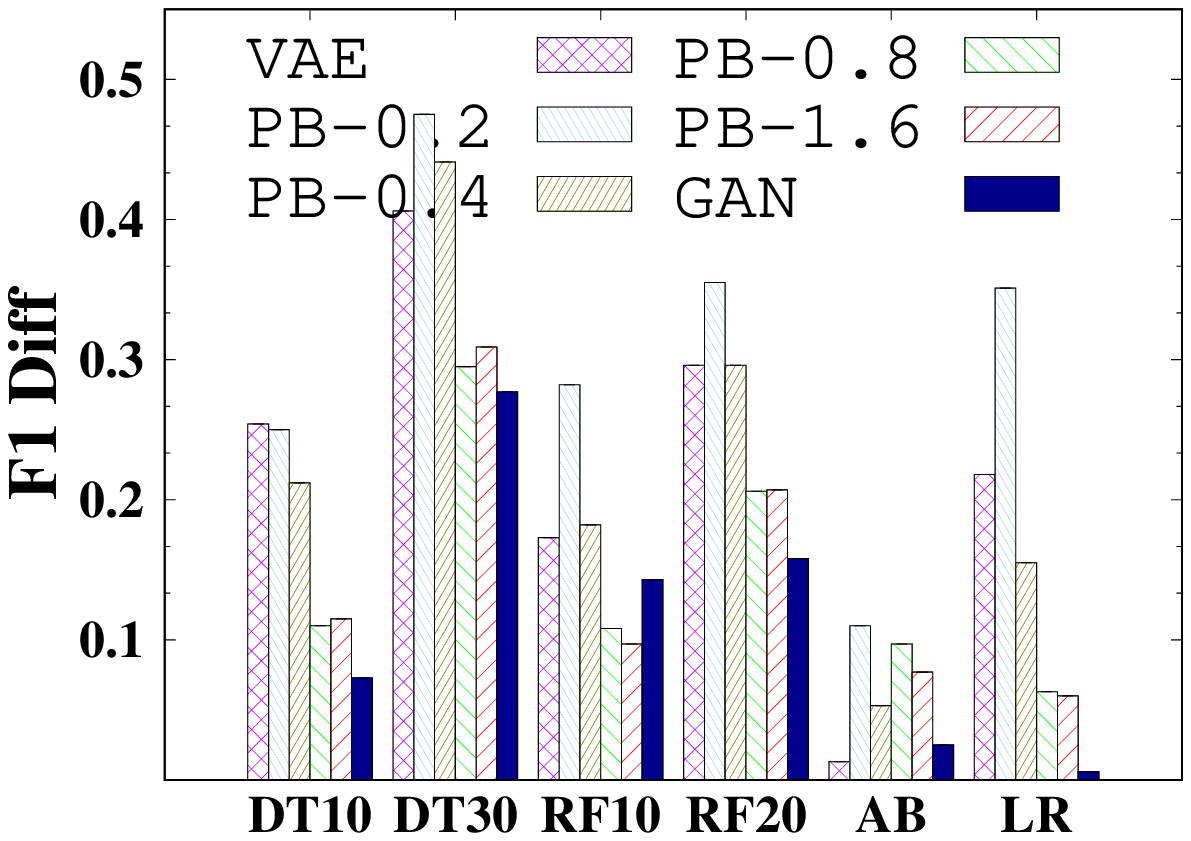,width=0.23\textwidth}
		}
		\subfigure[{\dscensus dataset.}]{
			\epsfig{figure=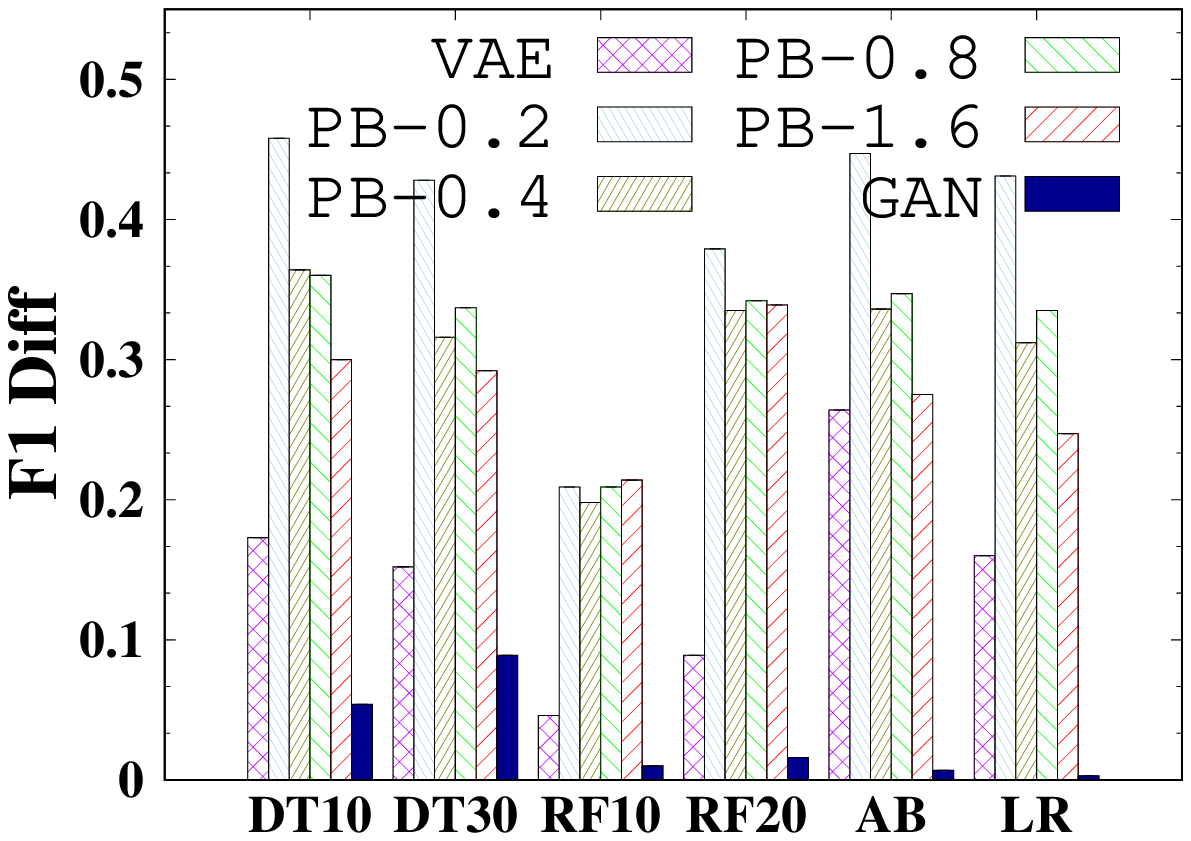,width=0.23\textwidth}
		}
		\subfigure[{\small \dssat dataset.}]{
			\epsfig{figure=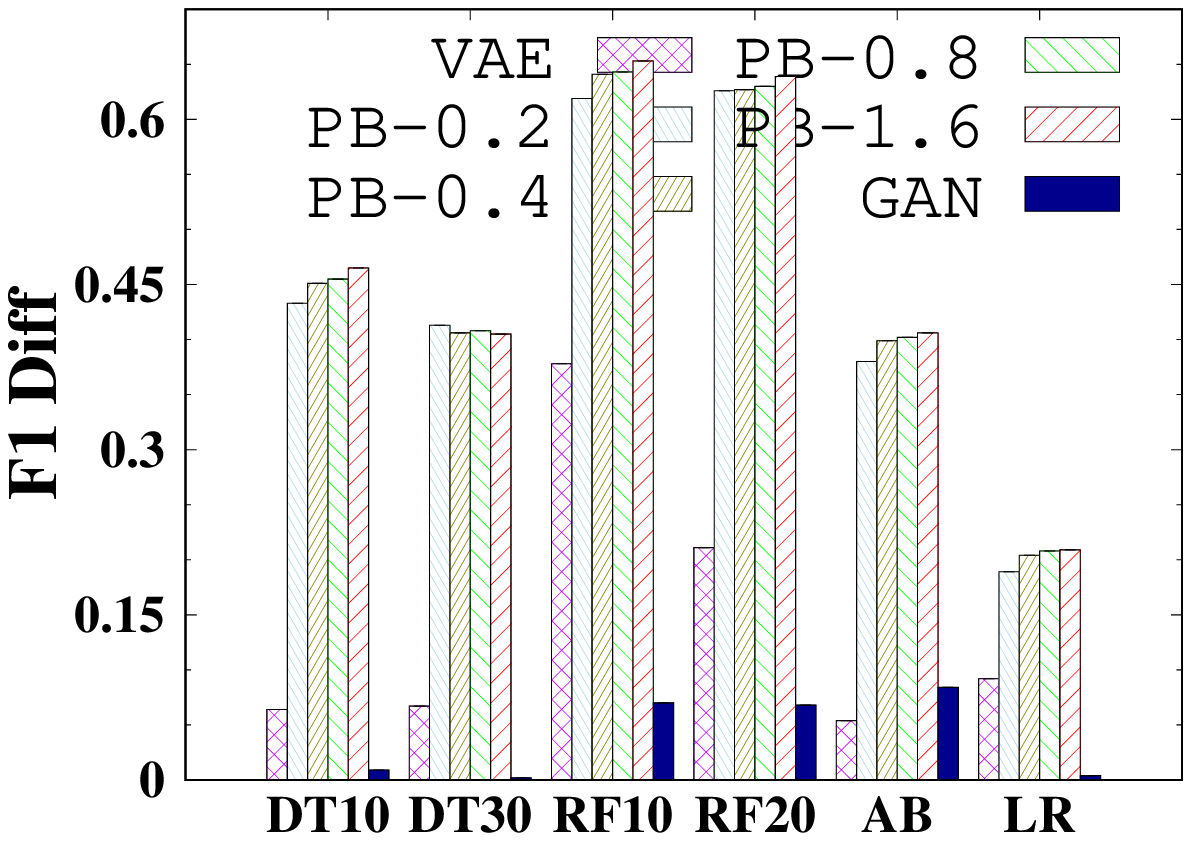,width=0.23\textwidth}
		}
	\end{center}\vspace{-2.5em}
	\caption{Comparison of different approaches to relational data synthesis on data utility for classification.} \label{exp:compa-1}
	\vspace{-1em}
\end{figure*}

\subsubsection{Effect of Sample Size for Synthetic Data}\label{subsubsec:sample-size}
This section evaluates whether the sample size $|\Tfake|$ of synthetic table
would affect the utility.
Table~\ref{table:ratio-result} reports the F1 difference $\diff$ when varying the ratio between sizes of synthetic $\Tfake$ and real $\T_{\tt train}$ tables.
We observe that, with the increase of sample size, the performance of classifier is improved, as more samples can be used for training the classifier.
However, the improvement is not very significant due to the fact that increasing synthetic data size does not actually inject more \emph{information}: synthetic tables with varying sizes are from a generator $\Gen$ with the same set of parameters.

\begin{table}[!t]
	\centering
	\caption{Effect of size ratio between synthetic and original tables (using DT10 as classifier).}
	\vspace{-1mm}
	\label{table:ratio-result}
	\resizebox{0.42\textwidth}{!}{%
	\begin{tabular}{|c||c|c|c|c|} 
		\hline 
		\multirow{2}{*}{\textbf{Dataset}} &
		\multicolumn{4}{c|}{\textbf{Size ratio: $|\Tfake| / |\T_{\tt train}|$}} \\
			\cline{2-5}
			& ~~$50\%$~~ & ~~$100\%$~~ & ~~$150\%$~~ & ~~$200\%$~~  \\  
			\hline		
			\hline \dsadult & 0.073 & 0.032 & 0.028 & \textbf{0.024} \\
			\hline \dsct    & 0.088 & 0.079 & 0.117 & \textbf{0.064} \\
			\hline \dssda   & 0.007 & 0.002 & 0.003 & \textbf{0.001} \\
			\hline \dssdb   & 0.029 & \textbf{0.013} & 0.018 & 0.016 \\
			\hline
	\end{tabular}
	\vspace{-3.5em}
	}
\end{table}

\subsection{Comparing Data Synthesis Methods} \label{subsec:result-compa}

\begin{table}[!t]
	\centering
	\caption{Comparison of GAN and \bsbayes on privacy.}
	\label{table:privacy}
	\resizebox{0.4\textwidth}{!}{%
	\begin{tabular}{|c||c|c|c|c|} 
		\hline
		\multirow{2}{*}{\textbf{Method}} &
			\multicolumn{2}{c|}{\textbf{Hitting Rate ($\%$)}} &
			\multicolumn{2}{c|}{\textbf{DCR}} \\
			\cline{2-5} & \dsadult & \dsct & \dsadult & \dsct \\
			\hline \hline
			\bsbayes-0.1 & 0.49   & 0.002     & 0.164      & 0.106 \\ \hline
			\bsbayes-0.2 & 0.88   & 0.006     & 0.147      & 0.094 \\ \hline
			\bsbayes-0.4 & 2.16   & 0.022     & 0.123      & 0.082 \\ \hline
			\bsbayes-0.8 & 4.40   & 0.056     & 0.112      & 0.073 \\ \hline
			\bsbayes-1.6 & 4.64   & 0.070     & 0.110      & 0.069 \\ \hline \hline
			GAN       & 0.30   & 0.500     & 0.113      & 0.072 \\ \hline
	\end{tabular}	}
\end{table}

This section compares GAN with \bsvae and \bsbayes, which are described in Section~\ref{subsec:exp-method}. Note that we use the conditional GAN as the default setting of GAN.

\vspace{-.25em}
\subsubsection{Evaluation on Synthetic Data Utility}
\vspace{-.25em}

Figure~\ref{exp:compa-1} shows the experimental results on synthetic data utility on our real datasets.
%
First, with the increase of privacy parameter $\epsilon$, the result of \bsbayes becomes better. This is because $\epsilon$ is used to control the privacy level: the larger the $\epsilon$, the lower the privacy level.
Second, \bsvae achieves moderate results, but the generated synthetic data is still worse than that synthesized by GAN. 
%
This is similar to the case in image synthesis~\cite{DBLP:conf/iclr/DumoulinBPLAMC17}: the images synthesized by VAE is worse than that generated by GAN. This is because the low dimensional latent variable in VAE may not be sufficient to capture complex relational data.
%

%

\begin{figure}[!t]\vspace{-1.5em}
	\begin{center}
		\subfigure[{\small \dsadult dataset.}]{
			\epsfig{figure=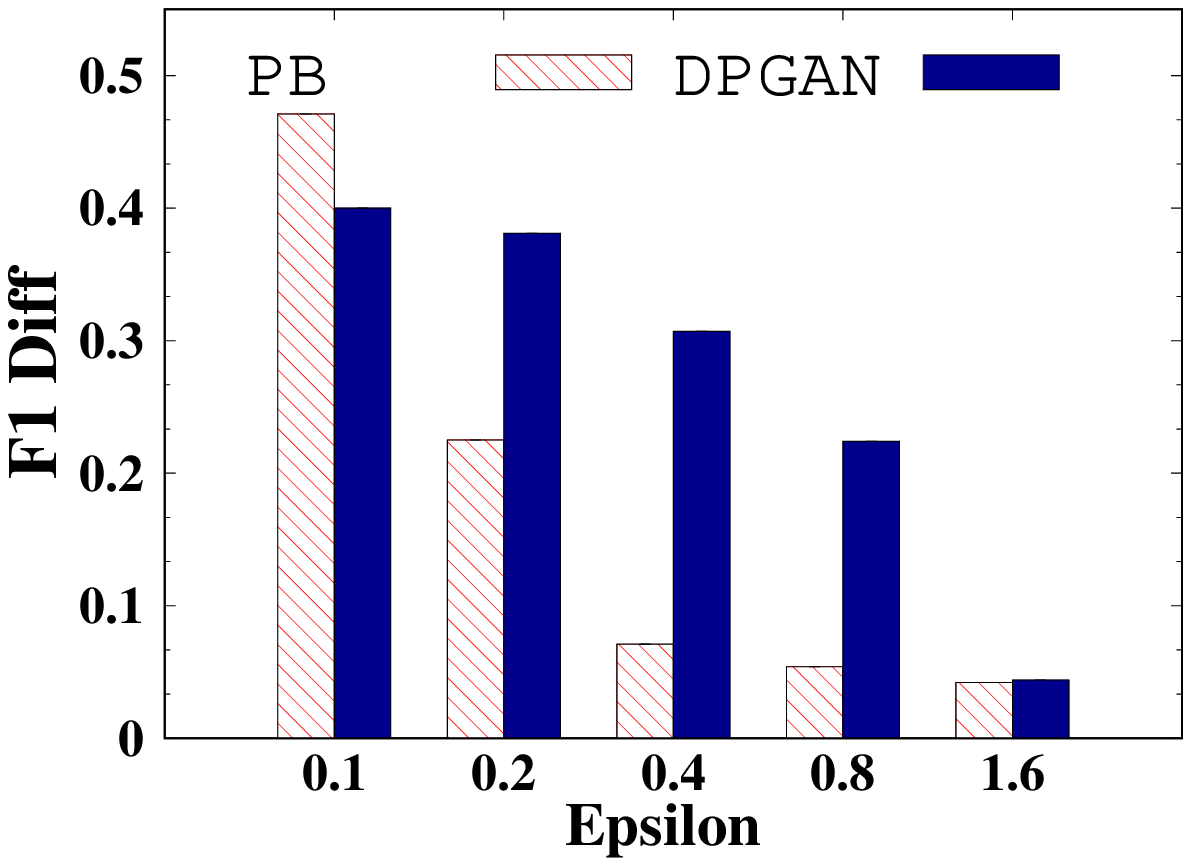,width=0.22\textwidth}
		}
		\subfigure[{\small \dsct dataset.}]{
			\epsfig{figure=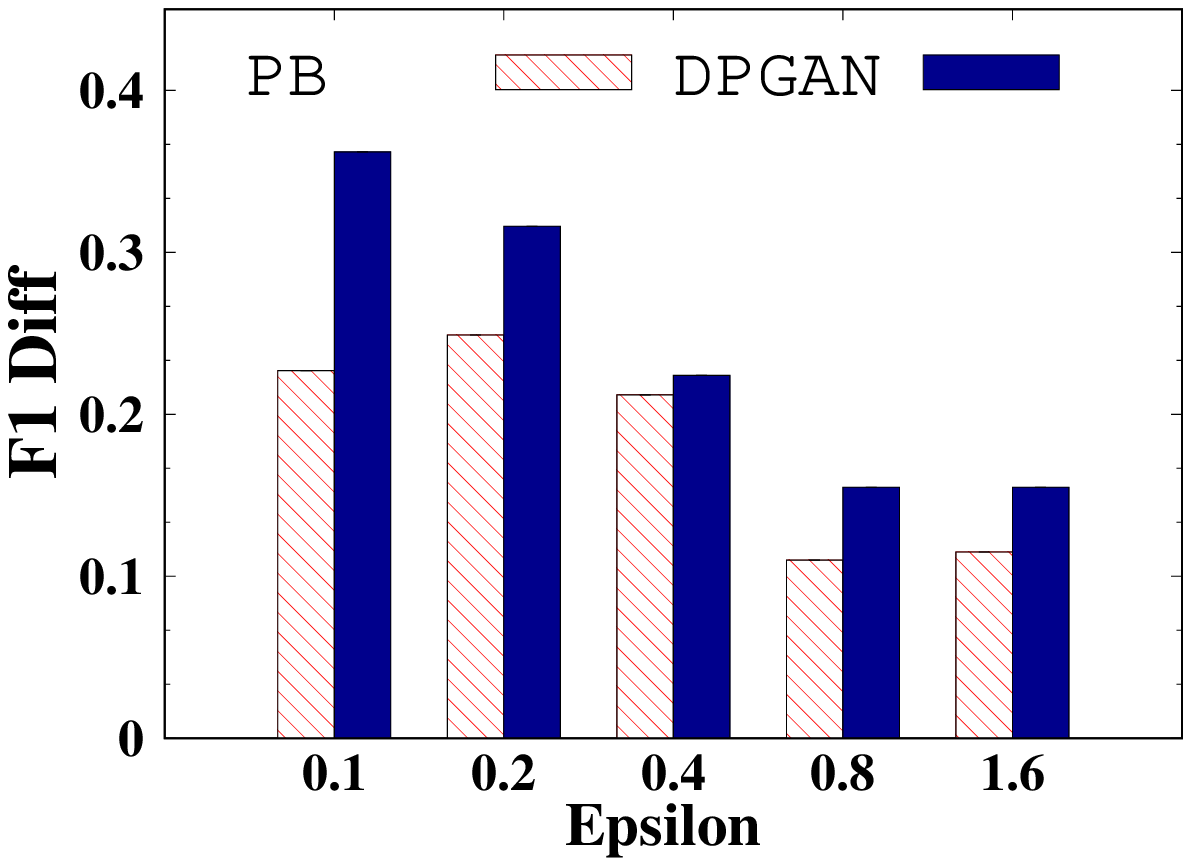,width=0.22\textwidth}
		}
	\end{center}\vspace{-2em}
	\caption{Comparing DPGAN and \bsbayes on varying privacy levels (using DT10 as classifier).} \label{exp:privacy}
	\vspace{-2em}
\end{figure}

Our GAN-based framework significantly outperforms \bsbayes and \bsvae on preserving data utility for classification. For example, the F1 difference achieved by GAN is $45-98\%$ and $10-90\%$ smaller than that achieved by \bsbayes with the lowest privacy level ($\epsilon = 1.6$) on the \dsadult and \dsct datasets respectively. This is mainly attributed to their different data synthesis mechanisms. \bsbayes aims at approximating a joint multivariate distribution of the original table, which may not perform well if the data distribution is complex. In contrast, GAN
utilizes the adversarial training mechanism to optimize generator $\Gen$. The result shows that the adversarial mechanism is useful for synthesizing relational data.

\vspace{1mm}
\noindent \textbf{Finding 5: GAN significantly outperforms \bsvae and \bsbayes on synthetic data utility. For some classifiers, the F1 difference of the synthetic data wrt. the original data achieved by GAN is smaller than that of \bsvae and \bsbayes by an order of magnitude.}

\subsubsection{Evaluation on Privacy} \label{subsec:compa-privacy}
Table~\ref{table:privacy} compares GAN with \bsbayes on protecting privacy against the risk of re-identification, measured by Hitting Rate and DCR introduced in Section~\ref{sec:exp-eval}. 
First, on the \dsadult dataset, GAN achieves lower hitting rate than \bsbayes. For example, even compared with \bsbayes with the highest privacy level $\epsilon = 0.1$, GAN reduces the hitting rate by $39\%$.
On the \dsct dataset, GAN achieves very low hitting rate $0.5\%$, i.e., only 25 out of 5000 sampled synthetic record can hit similar records in the original table. We notice that, on the \dsct dataset, the hitting rate of GAN is higher than that of \bsbayes. This is because most of the attributes on \dsct are \emph{numerical} attributes and \bsbayes discretizes the domain of each numerical attribute into a fixed number of equi-width bins~\cite{DBLP:conf/sigmod/ZhangCPSX14,DBLP:journals/tods/ZhangCPSX17}, and thus a synthetic numerical value is seldom similar to the original one.
Second, considering the metric DCR, GAN provides comparable overall performance to \bsbayes, and even outperforms \bsbayes with moderate privacy levels ($\epsilon=0.8$ or $1.6$).
The results validate our claim that GAN can reduce the risk of re-identification as there is no \emph{one-to-one} relationship between real and synthetic records.

\vspace{1mm}
\noindent \textbf{Finding 6: Empirically, the GAN-based data synthesis framework shows better tradeoff between synthetic data utility and protecting privacy against the risk of re-identification, as there is no one-to-one relationship between original and synthetic records.}
\vspace{1mm}

We evaluate the current solution DPGAN for GAN with differential privacy (DP) guarantee. Figure~\ref{exp:privacy} reports the experimental results on varying privacy level $\epsilon$. We can see that DPGAN cannot beat \bsbayes at almost all privacy levels on the \dsadult and \dsct datasets. This is because DPGAN adds noise to the gradients for updating parameters of $\Dis$ and then uses $\Dis$ to update parameters of $\Gen$. This process may make the adversarial training ineffective, as $\Dis$ now has limited ability to differentiate real/fake samples. The experimental result also implies better solutions for DP preserving GAN need to be invented.

\vspace{1mm}
\noindent \textbf{Finding 7: The current solution for differential privacy (DP) preserving GAN cannot beat traditional data synthesis methods with DP guarantees.}

\subsection{Evaluation on Simulated Datasets}\label{subsec:result-synthetic-data}

\begin{table}[!t]
	\centering
	\caption{Effect of attribute correlation on data synthesis performance (using DT30 as classifier).}\label{table:sdata-compare} \vspace{-1mm}
	\resizebox{0.48\textwidth}{!}{
		\begin{tabular}{|c||c|c|c|c|c|c|}
			\hline
			\multirow{2}{*}{\textbf{Dataset}} &
			\multicolumn{3}{c|}{\textbf{F1 Diff}} &
			\multicolumn{3}{c|}{\textbf{Synthesis Time (Min)}} \\
			\cline{2-7}
			& \textbf{CNN} & \textbf{MLP} & \textbf{LSTM} & 
			\textbf{CNN} & \textbf{MLP} & \textbf{LSTM} \\
			\hline \hline
			\dssda-$0.5$ & 0.385 & 0.010 & 0.005 & 10 & 31 & 67 \\ 
			\hline
			\dssda-$0.9$ & 0.486 & 0.047 & 0.020 & 10 & 35 & 60 \\
			\hline \hline
			\dssdb-$0.5$ & 0.200 & 0.023 & 0.014 & 6 & 27 & 60 \\
			\hline
			\dssdb-$0.9$ & 0.752 & 0.019 & 0.012 & 6 & 27 & 50 \\
			\hline
	\end{tabular}
} 
\vspace{-1em}
\end{table}

This section evaluates the effect of attribute correlation by using the simulated datasets, and the results on performance and efficiency are reported in Table~\ref{table:sdata-compare}.
We can see that LSTM-based generator achieves the best performance on datasets with various degrees of attribute correlation. This shows that the sequence generation mechanism in LSTM is effective and outperforms the other models.
In contrast, 
LSTM is less efficient than MLP and CNN, as LSTM uses a more complicated neural network structure to generate each record attribute by attribute.

\begin{figure*}[!t]\vspace{-1.5em}
	\vspace{-.5em}
	\begin{center}
		\subfigure[{\dssda-balance dataset.}]{
			\label{exp:condt-5}
			\epsfig{figure=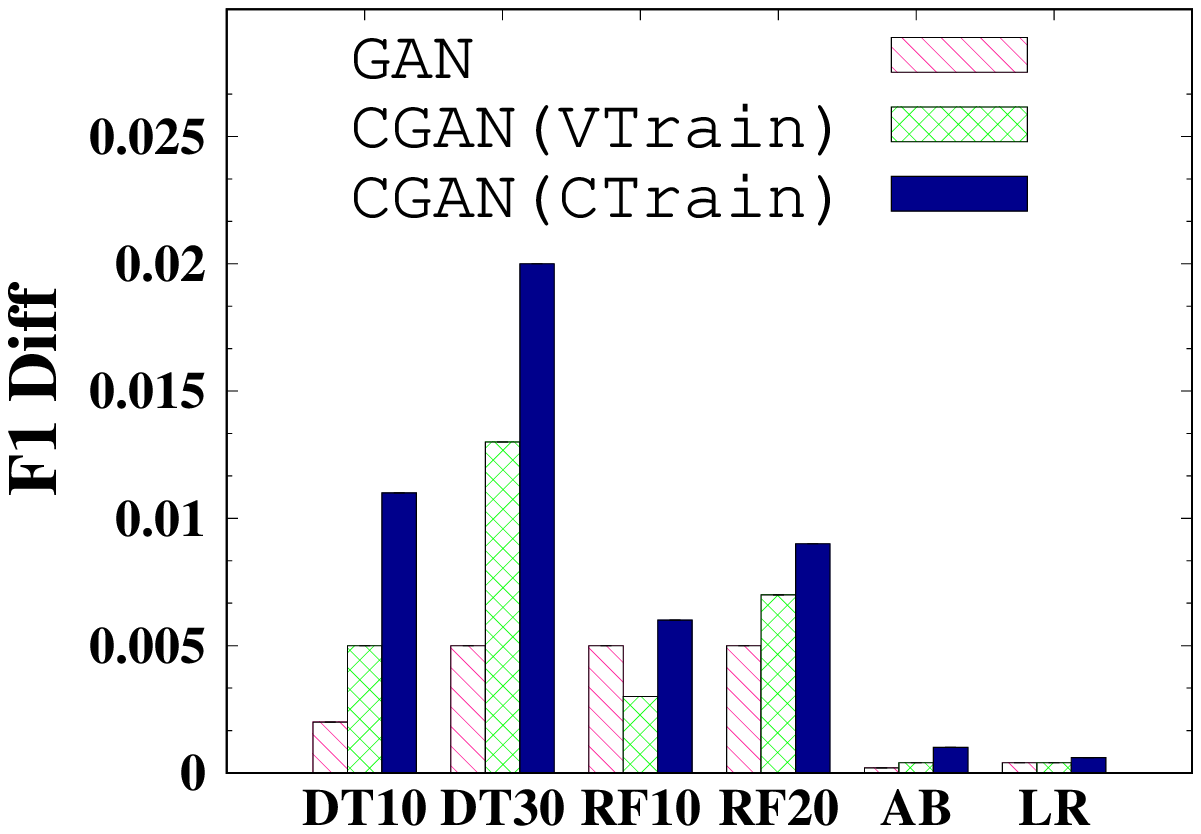,width=0.23\textwidth}
		}
		\subfigure[{\dssda-skew dataset.}]{
			\label{exp:condt-sdata-b}
			\epsfig{figure=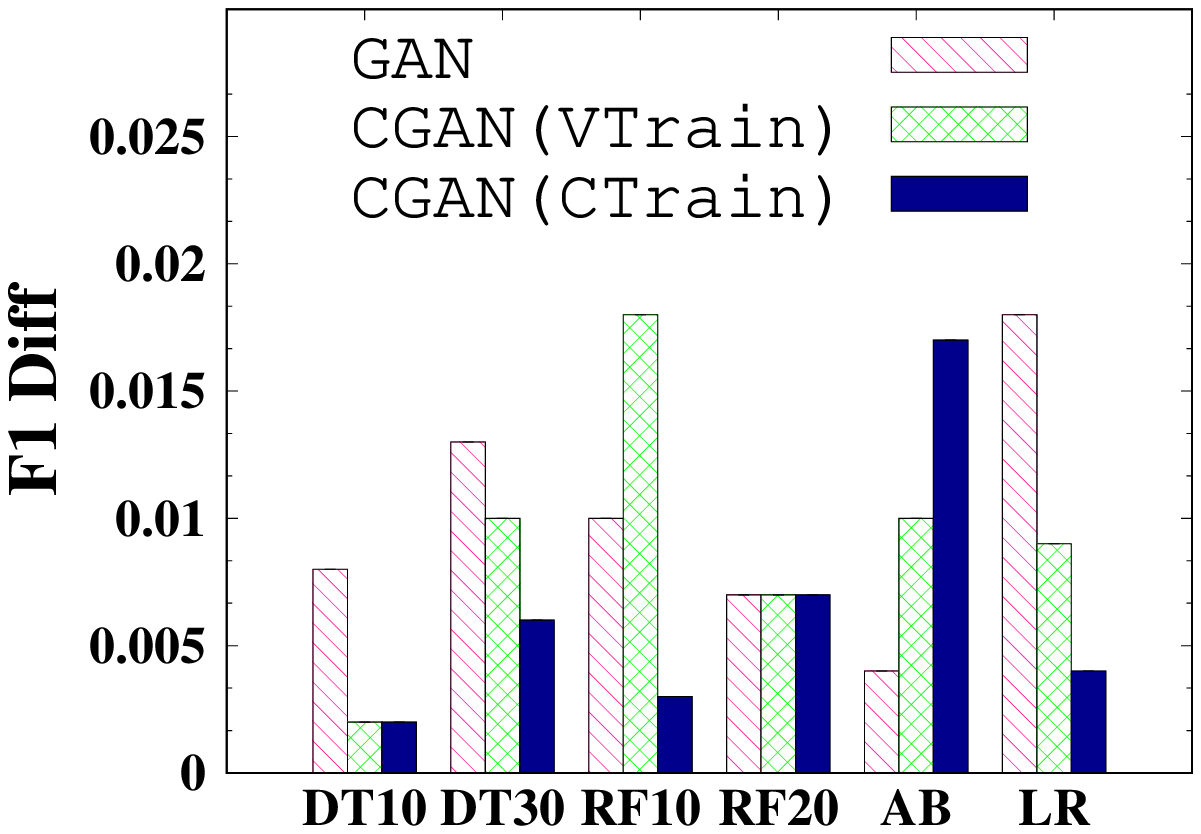,width=0.23\textwidth}
		}
		\subfigure[{\dssdb-balance dataset.}]{
			\label{exp:condt-sdata-c}
			\epsfig{figure=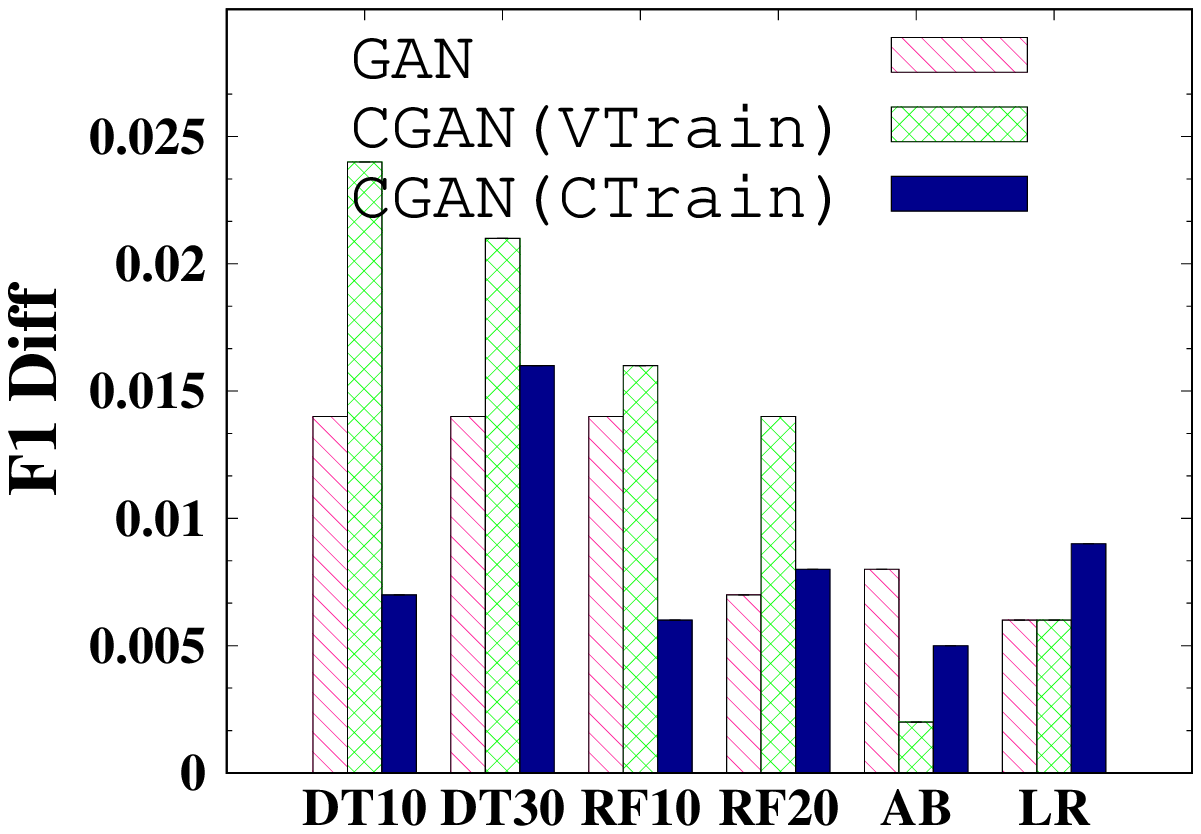,width=0.23\textwidth}
		}
		\subfigure[{\dssdb-skew dataset.}]{
			\label{exp:condt-sdata-d}
			\epsfig{figure=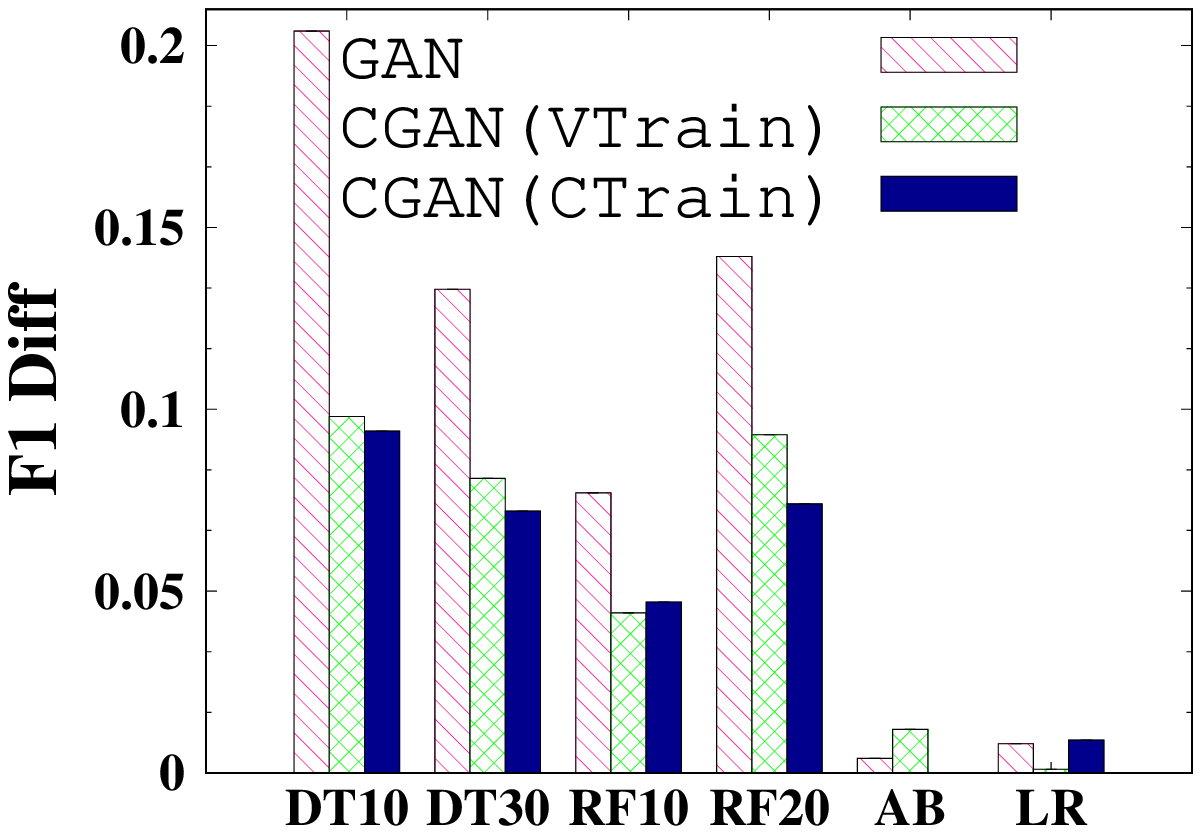,width=0.23\textwidth}
		}
	\end{center}\vspace{-2.5em}
	\caption{Evaluating conditional GAN on synthetic data utility for classification (Simulated data).} \label{exp:condt-sdata}
	\vspace{-1.5em}
\end{figure*}

We also evaluate the effect of label skewness. We set the correlation degree as $0.5$ for both \dssda and \dssdb, and consider their $\term{balanced}$ and $\term{skew}$ settings. As shown in Figure~\ref{exp:condt-sdata},
conditional GAN does not improve the data utility, and it sometimes even achieves inferior performance (e.g., on the \dssda-$\term{balanced}$ dataset) if label distribution is balanced. In contrast, if label distribution is skew, conditional GAN is helpful for improving the performance.



\subsection{Evaluation on Additional Applications} \label{subsec:result-ca}

\begin{table}[!t]
	\centering
	\caption{Evaluating neural networks of generator $\Gen$ on synthetic data utility $\diff_{\tt CST}$ for clustering.}\label{table:clustering-model} \vspace{-1mm}
		\resizebox{0.47\textwidth}{!}{%
	\begin{tabular}{|c||c|c|c|c|c|}
		\hline
		\multirow{2}{*}{\textbf{Dataset}} &
		\multirow{2}{*}{\textbf{CNN}} & 
		\multicolumn{2}{c|}{\textbf{MLP}} &
		\multicolumn{2}{c|}{\textbf{LSTM}} 
		\\
		\cline{3-6}
		&& \norm/\onehot &\gmm/\onehot
		&\norm/\onehot &\gmm/\onehot \\  
		\hline
		\hline 
		\dshtru
		&0.3614
		&0.0009 &0.0019
		&\textbf{0.0003} &0.0035 \\	
		\hline 
		\dsadult
		&0.1157
		&0.0882 &0.0124
		&0.3336 &\textbf{0.0024} \\ 
		\hline
		\dsct
		&-
		&0.0047 &0.0015
		&0.0022 &\textbf{0.0008} \\  
		\hline
		\dspendigits
		&-
		&0.0010 &0.0021
		&0.0015 &\textbf{0.0008}  \\ 
		\hline
		\dsanuran
		&-
		&0.0021 &\textbf{0.0015}
		&0.0142 &0.0645 \\
		\hline
		\dscensus
		&0.0768
		&0.0217 &0.0153
		&0.0412 &\textbf{0.0004} \\ 
		\hline	
		\dssat 
		&-
		&\textbf{0.0007} &0.0076
		&0.0007 &0.0029 \\ 
		\hline	
	\end{tabular}
}
	\vspace{-0.1em}
\end{table}

\begin{table}[!t]
	\centering
	\caption{Evaluating neural networks of $\Gen$ on synthetic data utility $\diff_{\tt AQP}$ for AQP.
	}\label{table:AQP-model} \vspace{-1mm}
	\resizebox{0.44\textwidth}{!}{%
	\begin{tabular}{|c||c|c|c|c|c|}
		\hline
		\multirow{2}{*}{\textbf{Dataset}} &
		\multirow{2}{*}{\textbf{CNN}} & 
		\multicolumn{2}{c|}{\textbf{MLP}} &
		\multicolumn{2}{c|}{\textbf{LSTM}}  \\
		\cline{3-6}
		& &\norm/\onehot & \gmm/\onehot
		&\norm/\onehot &\gmm/\onehot \\  
		\hline	
		\hline 
		\dsct
		&-
		&0.295 &0.400
		&0.609 &\textbf{0.053} \\  
		\hline
		\dscensus
		&3.499
		&0.170 &\textbf{0.167}
		&0.271 &0.204 \\ 
		\hline	
	\end{tabular}
}
	\vspace{-1em}
\end{table}

We investigate the design choices of neural networks and report the results in Table~\ref{table:clustering-model} for clustering and Table~\ref{table:AQP-model} for AQP.
For evaluating AQP, we select the datasets \dsct and \dscensus with more than 100,000 records. Observing from the tables, we find a similar result to that of data utility for classification. The results show that LSTM is effective on capturing the underlying data distribution for the original table, which is also beneficial for clustering and AQP.

We also compare GAN with \bsvae and \bsbayes on data utility for clustering and AQP.
The result on data utility for clustering is reported in Table~\ref{table:clustering-compare}. We can see that GAN outperforms the baselines by 1-2 orders of magnitude. The results show that GAN is very promising in preserving the clustering structure of the original data, e.g., synthesizing similar attributes to the records within in the same group.
For AQP, as observed from Table~\ref{table:AQP-compare-0.01}, GAN achieves less relative error difference than \bsvae and \bsbayes on preserving data utility. This is because that GAN, if effectively trained, is more capable of generating synthetic data that well preserves the statistical properties of the original table. Thus, the synthetic data could answer the query workload with less errors.
We also notice that, on the AQP benchmarking dataset \dsbing, \bsvae achieves comparable results with GAN, i.e., $0.632$ vs. $0.422$ on relative error difference. The results show that \bsvae may also be promising for supporting AQP, considering it may be more easy and efficient to train than GAN. Some existing work~\cite{DBLP:journals/corr/abs-1903-10000} studies more sophisticated techniques to optimize \bsvae, such as partitioning the data and using multiple \bsvae models, adding rejection criteria for data sampling, etc. We will leave a more thorough comparison with such new techniques in the future work.

\begin{table}[!t]
	\centering
	\caption{Comparison of approaches to relational data synthesis on data utility $\diff_{\tt CST}$ for clustering.}
	\vspace{-1mm}
	\label{table:clustering-compare}
	\resizebox{0.48\textwidth}{!}{%
	\begin{tabular}{|c||c|c|c|c|c|c|} 
		\hline 
		\multirow{2}{*}{\textbf{Dataset}} &
		\multicolumn{6}{c|}{\textbf{Approaches}} \\
			\cline{2-7}
			& \bsvae & \bsbayes-0.2 & \bsbayes-0.4 & \bsbayes-0.8 & \bsbayes-1.6 & GAN \\  
			\hline		
			\hline \dshtru & 0.0160 & 0.1769 & 0.13904 & 0.0594 & 0.0331 & \textbf{0.0007}	\\
			\hline \dsct & 0.0089 & 0.0227 & 0.0121 & 0.0071 & 0.0031 & \textbf{0.0018}   \\
			\hline \dsadult & 0.0891 & 0.0892 & 0.0959 & 0.0729 & 0.0494 & \textbf{0.0015}   \\
			\hline \dspendigits & 0.0425 & 0.2025 & 0.1839 & 0.1749 & 0.1545 & \textbf{0.0008}   \\
			\hline \dsanuran & 0.2184 & 0.2989 & 0.2170 & 0.1505 & 0.1617 & \textbf{0.0020}   \\
			\hline \dscensus &0.0010 &0.0189 &0.0101 &0.0011 &0.0112 &\textbf{0.0004}   \\
			\hline \dssat & 0.4891 & 0.2451 & 0.2277 & 0.2289 & 0.2279 & \textbf{0.0007}   \\
			\hline
	\end{tabular}
	\vspace{-3.5em}
}
\end{table}

\begin{table}[!t]
	\centering
	\caption{Comparison of approaches to relational data synthesis on data utility $\diff_{\tt AQP}$ for AQP. 
		}
	\vspace{-1mm}
	\label{table:AQP-compare-0.01}
	\resizebox{0.48\textwidth}{!}{%
	\begin{tabular}{|c||c|c|c|c|c|c|} 
		\hline 
		\multirow{2}{*}{\textbf{Dataset}} &
		\multicolumn{6}{c|}{\textbf{Approaches}} \\
			\cline{2-7}
			& \bsvae & \bsbayes-0.2 & \bsbayes-0.4 & \bsbayes-0.8 & \bsbayes-1.6 & GAN \\  
			\hline		
			\hline \dsct &0.251  & 0.201 & 0.113 & 0.183 & 0.108 & \textbf{0.015}	\\
			\hline \dscensus &0.469  & 2.348 & 1.262 & 0.786 & 0.767 & \textbf{0.240}   \\
			\hline \dsbing & 0.632 & 0.830 & 0.805 & 0.783 & 0.761 & \textbf{0.422}   \\
			\hline
	\end{tabular}
	}\vspace{-1.5em}
\end{table}

\vspace{1mm}
\noindent
\textbf{Finding 8: GAN is also very promising for preserving the utility of original data for supporting the applications of clustering and AQP.
}



%


\section{Conclusion \& Future Direction}\label{sec:conclusion}

In this paper, we have conducted a comprehensive experimental study for applying GAN to relational data synthesis. We introduced a unified framework and defined a design space of the solutions that realize GAN. We empirically conducted a thorough evaluation to explore the design space and compare GAN with conventional approaches to data synthesis. Based on our experimental findings, we summarize the following key insights that provide guidance to the practitioners who want to apply GAN to develop a relational data synthesizer.

\vspace{1mm}
\noindent \textbf{Overall Evaluation for GAN.}
GAN is very promising for relational data synthesis. 
It generates synthetic data with very good utility on classification, clustering and AQP (\emph{Findings 5 and 8}).
Moreover, it also achieves competitive performance on protecting privacy against the risk of re-identification (\emph{Finding 6}).
However, GAN has limitations on providing provable privacy protection: the current solution cannot produce superior data utility when preserving differential privacy (\emph{Finding 7}).

\vspace{1mm}
\noindent \textbf{Neural Network Selection.}
For ordinary users with limited knowledge on deep learning, we suggest to use MLP to realize GAN, as MLP is more robust and can achieve moderate results without parameter tuning (\emph{Finding 2}).
For expert users who want to spend sufficient efforts to finetune parameters, we recommend LSTM that can achieve the best performance (\emph{Finding 1}), given proper training strategies as discussed below, and data transformation schemes.

\vspace{1mm}
\noindent \textbf{Model Training Strategy.}
We provide guidelines to users on how to train GAN models.
To avoid mode collapse, we introduce solutions to boost model training, including adding KL divergence in the loss function for warm-up and using simplified discriminator to avoid gradient vanishing in generator (\emph{Finding 3}).
We leverage conditional GAN for datasets with imbalanced data distribution (\emph{Finding 4}).


\vspace{1mm}
\noindent \textbf{Relational Data Representation.}
Data transformation that converts original records to recognized input of GAN
does affect the overall performance, which shows that representation of relational data is important. This may imply an interesting future work that co-trains GAN and record representation through a hybrid optimization framework.

\vspace{1mm}
We also identify several future directions in GAN-based relational data synthesis that may be worthy of exploration.

\vspace{1mm}
\noindent\textbf{(1) Providing provable privacy protection.}
We have shown that GAN has limitations on providing provable privacy protection, i.e., differential privacy. Although enabling GAN to support differential privacy is a hot research topic in ML~\cite{DBLP:journals/corr/abs-1802-06739,DBLP:conf/iclr/JordonYS19a}, this problem is very challenging, because adding noises to the adversarial training in GAN may drastically affect parameter optimization in $\Gen$ and $\Dis$. Therefore, it calls for new solutions to equip GAN-based data synthesis with provable privacy protection.

\vspace{1mm}
\noindent\textbf{(2) Capturing attribute correlations.}
LSTM achieves good performance as its sequence generation mechanism can \emph{implicitly} capture attribute correlations. The DB community has long studied how to model attribute correlations \emph{explicitly} by providing solutions like functional dependency~\cite{DBLP:books/daglib/0011128,DBLP:conf/icde/BohannonFGJK07}. Despite some preliminary attempt~\cite{DBLP:conf/ijcai/ChenJLPSS19}, it still remains an unsolved question that how to combine techniques from the two communities to improve the synthetic data quality for the GAN-based framework.

\vspace{1mm}
\noindent\textbf{(3) Supporting more utility definitions.}
This paper studies synthetic data utility for training classifiers, evaluating clustering algorithms and supporting AQP.
However, relational data synthesis should support a variety of applications, including ML tasks over time-series data and data synthesis for supporting AQP with theoretical bounds.
%

\vspace{2mm}
\noindent
\textbf{Acknowledgment.}
This work is supported by NSF of China (61632016, U1911203, 61925205, U1711261), the Research Funds of Renmin University of China (18XNLG18), Huawei, and TAL Education.

\newpage
\balance
\bibliographystyle{abbrv}
\bibliography{data-synthesis,gan,synthesis-apps}


\newpage

\appendix

\begin{figure}[!t]
	\begin{center}
		\subfigure[\small{Generator.}]{
			\label{fig:cnn-g}
			\epsfig{figure=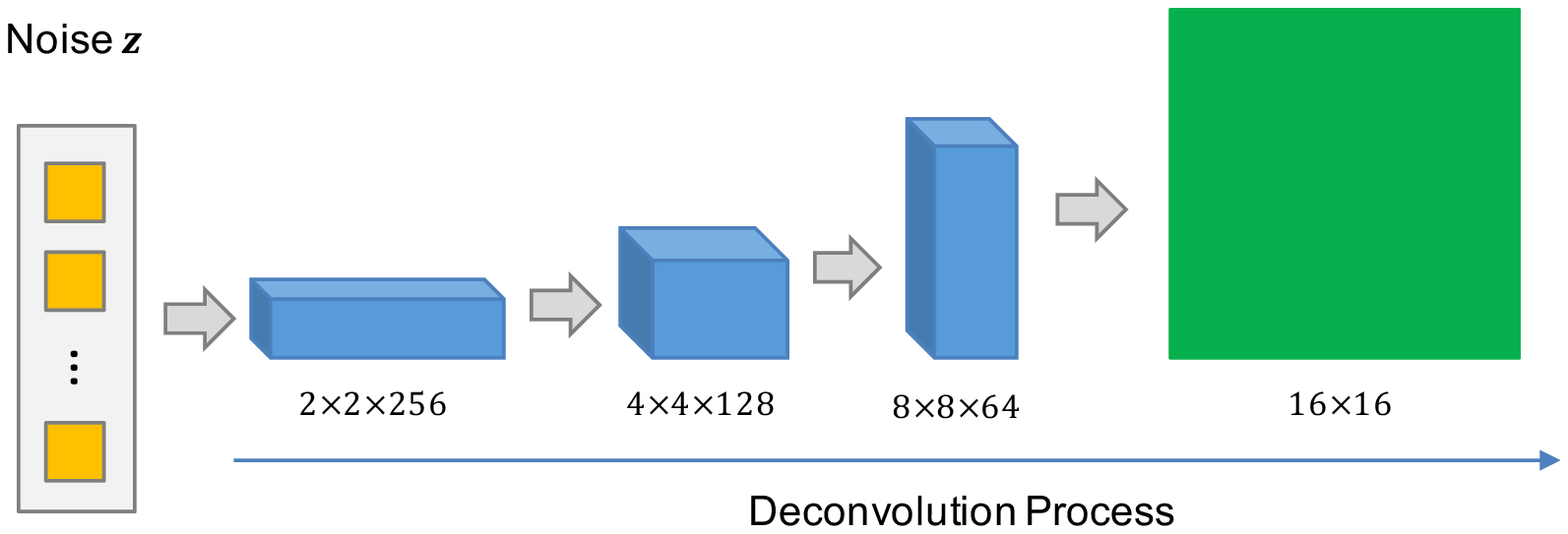,width=0.50\textwidth}
		}
		\subfigure[\small{Discriminator.}]{
			\label{fig:cnn-d}
			\epsfig{figure=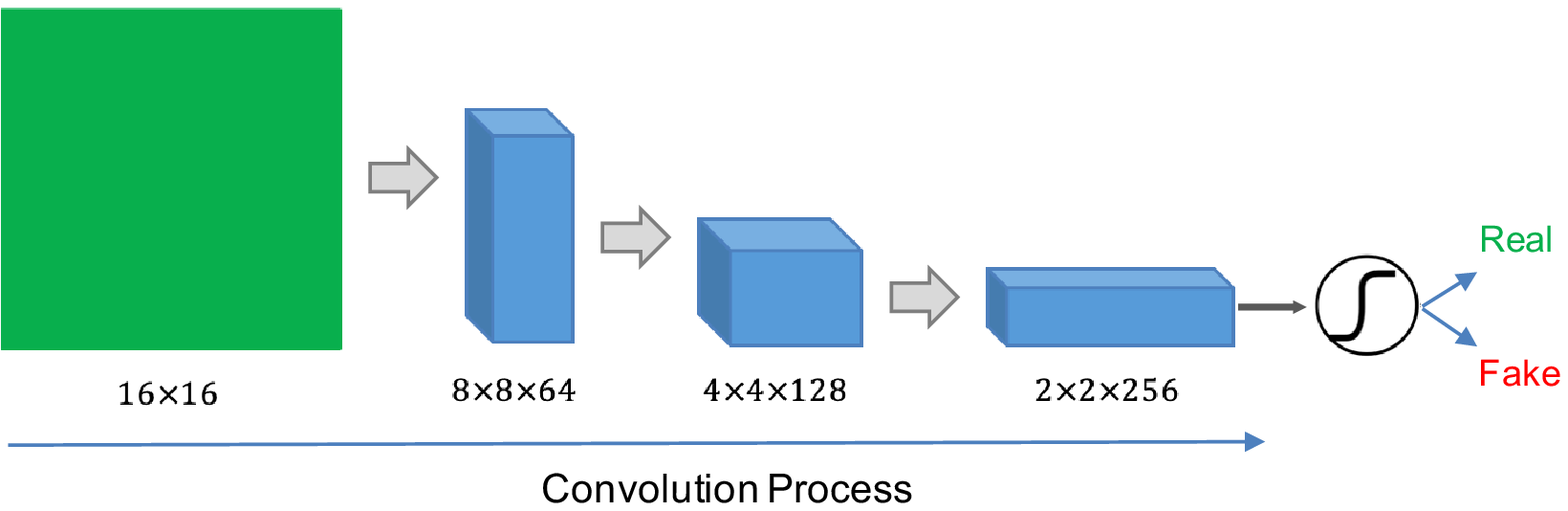,width=0.50\textwidth}
		}
	\end{center}\vspace{-2em}
	\caption{GAN module implemented by CNN.} \label{fig:cnn-gan}
	\vspace{-1em}
\end{figure}

\section{Detailed GAN Model Design}
This section presents more details of our GAN model design.
We first present the design of neural network architectures in Section~\ref{app:nn} and then provide the pseudo-codes of the training algorithms in Section~\ref{app:algos}.

\subsection{Neural Network Architectures} \label{app:nn}

\subsubsection{CNN: convolutional neural networks.}

CNN is utilized in the existing works for relational data synthesis~\cite{DBLP:conf/ijcai/ChenJLPSS19,DBLP:journals/pvldb/ParkMGJPK18}, which is inspired by the well-known DCGAN~\cite{DBLP:journals/corr/RadfordMC15}, as illustrated in Figure~\ref{fig:cnn-gan}.
One characteristic of these works is that they use a \emph{matrix} instead of a vector to represent a real/fake sample $\tuvec$. To this end, they use ordinal encoding and simple normalization to respectively preprocess categorical and numerical attributes. Then, they convert the preprocessed records into a square matrix padded with zeros. For example, consider the first record in our example table in Figure~\ref{fig:example}. The record is firstly preprocessed into $(-0.2, 0, 0, 0, 0)$ using ordinal encoding and simple normalization. Then, it is converted into a $3 \times 3$ matrix with four values are padded with zeros. Based on this, the GAN model can be trained using the converted square matrices.
Generator $\Gen$ takes as input a prior noise $\noise$, which is denoted by $\bm{h}_{g}^{0}$, and uses $L$ de-convolution layers $\{\bm{h}_{g}^{l}\}$ (i.e., fractionally strided convolution) to transform $\noise$ to a synthetic sample in the form of matrix, i.e.,  
\begin{align}
& \bm{h}_{g}^{l+1} = \relu(\bn(\deconv(\bm{h}_{g}^{l}))), \nonumber\\ 
& \tuvec = \tanhh(\deconv(\bm{h}_{g}^{L})),
\end{align}
where $\deconv$ is de-convolution function. Figure~\ref{fig:cnn-g} illustrates a de-convolution process that converts $\noise$ to a $16 \times 16$ matrix that represents a synthetic sample. 

Discriminator $\Dis$, as shown in Figure~\ref{fig:cnn-d}, takes as input a real/fake sample $\tuvec$ in matrix form, which is denoted by $\bm{h}_{d}^{0}$. It applies $L$ convolution layers $\{\bm{h}_{d}^{l}\}$ to convert $\tuvec$ to a probability indicating how likely $\tuvec$ is real, i.e., 
\begin{align}
& \bm{h}_{d}^{l+1} = \lrelu(\bn(\conv(\bm{h}_{d}^{l}))), \nonumber\\ 
& f = \sigmoid(\bn(\conv(\bm{h}_{d}^{L}))),
\end{align}
where $\conv$ is a convolution function.

\subsubsection{MLP: fully connected neural networks.}

\begin{figure}
	\begin{center}
		\subfigure[\small{Generator.}]{
			\label{fig:mlp-g}
			\epsfig{figure=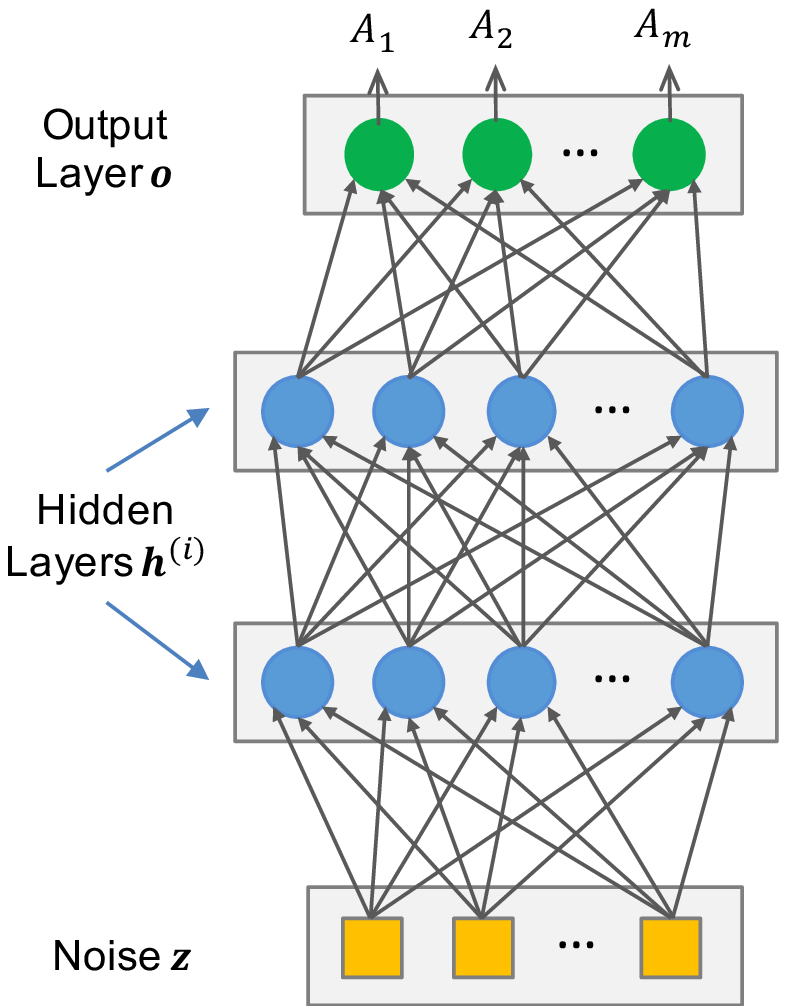,width=0.22\textwidth}
		}
		\subfigure[\small{Discriminator.}]{
			\label{fig:mlp-d}
			\epsfig{figure=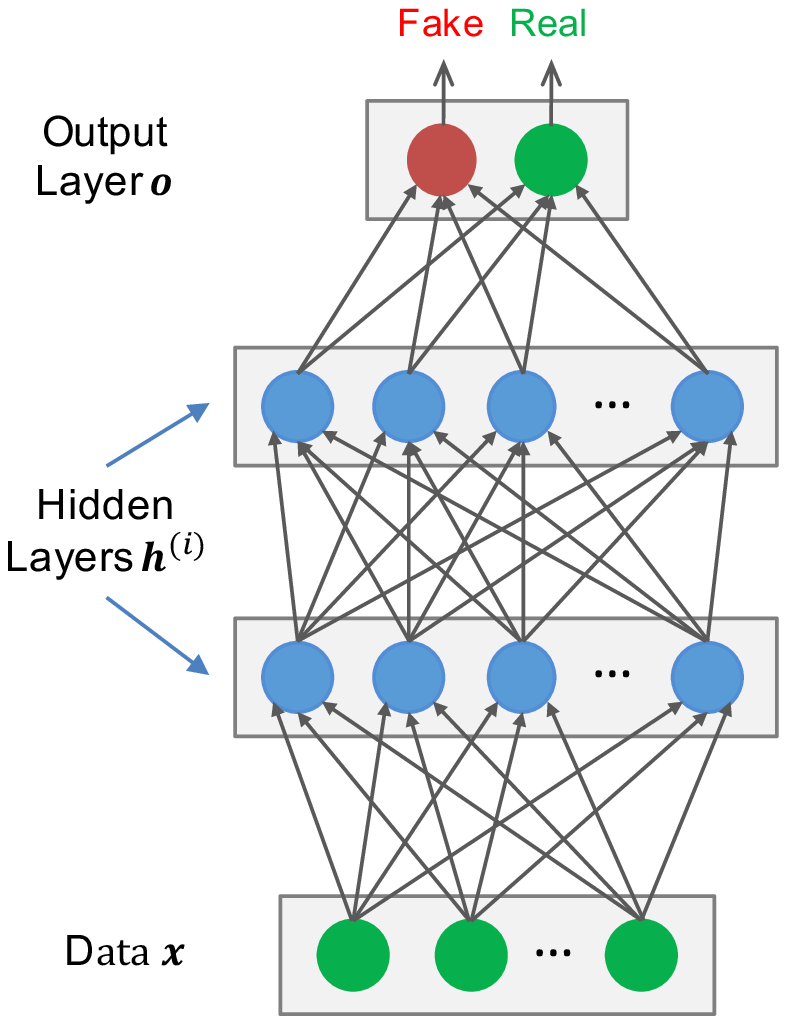,width=0.22\textwidth}
		}
	\end{center}\vspace{-2em}
	\caption{GAN module implemented by MLP.} \label{fig:mlp-gan}
	\vspace{-1em}
\end{figure}

MLP is used in the existing works for relational data synthesis~\cite{DBLP:journals/corr/ChoiBMDSS17,DBLP:journals/corr/abs-1907-00503}. 
Figure~\ref{fig:mlp-gan} provides generator $\Gen$ and discriminator $\Dis$ realized by MLP.
Specifically, $\Gen$ takes as input a prior noise $\noise$, which is also denoted by $\bm{h}^{(0)}$, and utilizes with $L$ fully-connected layers, where each layer is computed by
\begin{equation}
\bm{h}^{l+1} = \phi\big(
\bn(
\fc_{|\bm{h}^{l}| \rightarrow |\bm{h}^{l+1}|}(\bm{h}^{l})
)
\big),
\end{equation}
where $\fc_{|\bm{h}^{l}| \rightarrow |\bm{h}^{l+1}|}(\bm{h}^{l}) = \bm{W}^{l}  \bm{h}^{l} + \bm{b}^{l}$ with weights $\bm{W}^{l}$ and bias $\bm{W}^{l}$, $\phi$ is the activation function (we use $\relu$ in our experiments), and $\bn$ is the batch normalization~\cite{DBLP:conf/icml/IoffeS15}.

The challenge here is how to make the output layer in $\Gen$ \emph{attribute-aware}.
More formally, $\Gen$ needs to output a synthetic sample $\tuvec = \tuvec_{1}\oplus\tuvec_{2}\oplus \ldots \oplus \tuvec_{m}$, where $\tuvec_{j}$ corresponds to the $j$-th attribute.
Note that, for simplicity, we also use notation $\tuvec$ to represent fake samples if the context is clear.
We propose to generate each attribute vector $\tuvec_{j}$ depending on the transformation method on the corresponding attribute $\T[j]$, i.e., 
\begin{equation} \hspace{-1mm}
{\small
	\tuvec_{j} = 
	\left\{
	\begin{array}{lr}
	\tanhh(\fc_{|\bm{h}^{L}| \rightarrow 1}(\bm{h}^{L})),&  (C_1)  \\
	\tanhh(\fc_{|\bm{h}^{L}| \rightarrow 1}(\bm{h}^{L}) \oplus 
	\softmax(\fc_{|\bm{h}^{L}| \rightarrow |\tuvec_{j}|-1}(\bm{h}^{L})),  & (C_2)\\
	\softmax(\fc_{|\bm{h}^{L}| \rightarrow |\tuvec_{j}|}(\bm{h}^{L})), &  (C_3)\\
	\sigmoid(\fc_{|\bm{h}^{L}| \rightarrow 1}(\bm{h}^{L})), & (C_4)
	\end{array}\nonumber
	\right.
}
\end{equation}
where $C_1$ to $C_4$ respectively denote the cases of using simple normalization, mode-specific normalization, one-hot encoding and ordinal encoding as transformation on the attribute $\T[j]$ (see Section~\ref{sec:preproc}). For example, consider $C_2$, the GMM-based normalization, we first use $\tanhh(\fc_{|\bm{h}^{L}| \rightarrow 1}(\bm{h}^{L}) $ to generate $v_{\tt gmm}$ and then use $\softmax(\fc_{|\bm{h}^{L}| \rightarrow |\tuvec_{j}|-1}(\bm{h}^{L}))$ to generate a one-hot vector indicating which component $v_{\tt gmm}$ belongs to.
After generating $\{\tuvec_{j}\}$ for all attributes, we concatenate them to obtain $\tuvec$ as a synthetic sample. 

Figure~\ref{fig:mlp-d} shows the NN structure of our discriminator $\Dis$. 
Discriminator $\Dis$ is an MLP that takes a sample $\tuvec$ as input, and utilizes multiple fully-connected layers and a $\sigmoid$ output layer to classify whether $\tuvec$ is real or fake.

\subsubsection{LSTM: recurrent neural networks.}
%
%

\begin{figure}[!t]
	\begin{center} 
		\epsfig{figure=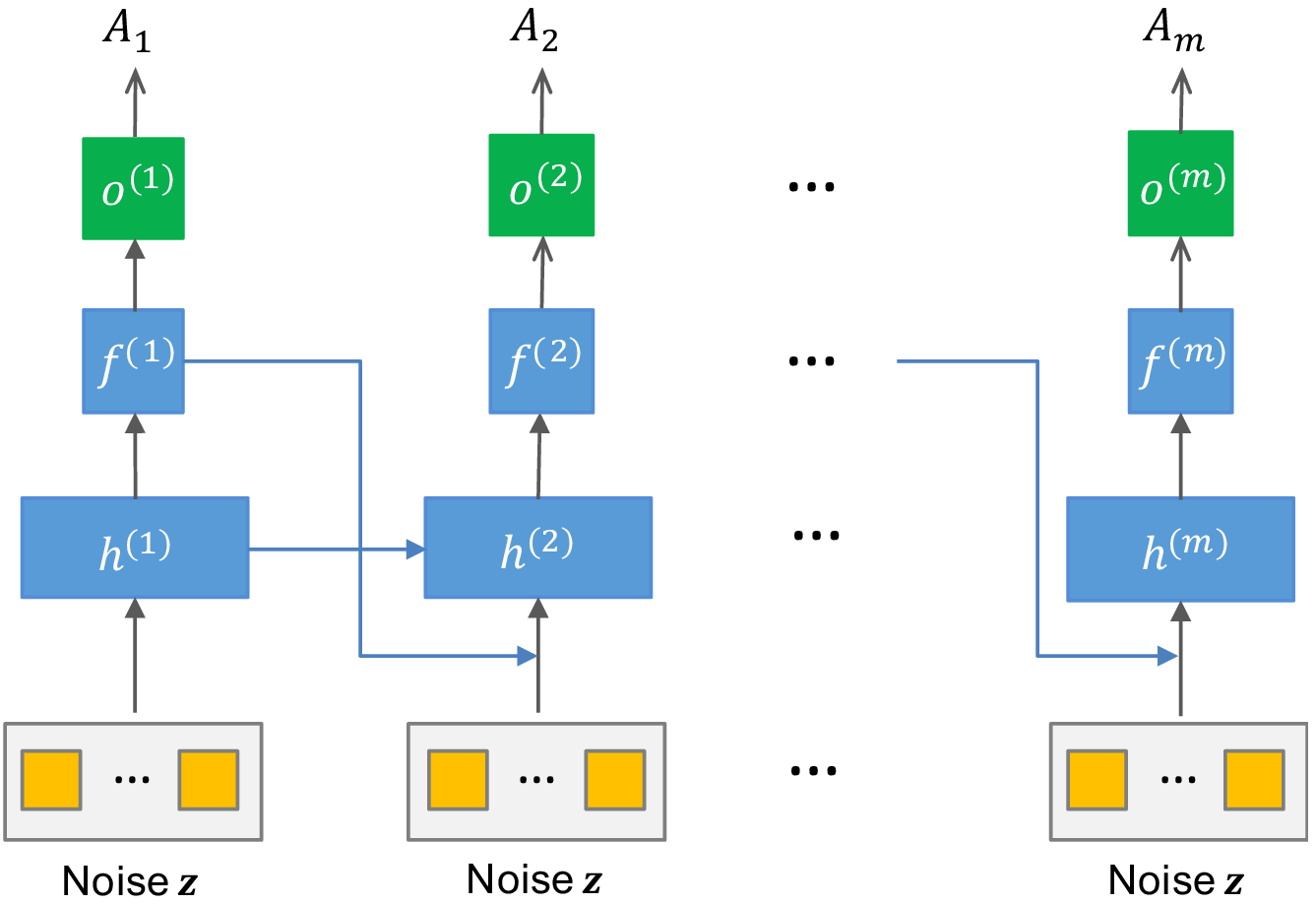,width=0.48\textwidth}
		\vspace{-1em}
		\caption{Generator implemented by LSTM.}
		\label{fig:rnn}
		\vspace{-1em}
	\end{center}
\end{figure}

Existing work also utilizes LSTM, a representative variant of RNN, to realize $\Gen$~\cite{DBLP:journals/corr/abs-1811-11264}. The basic idea is to formalize record synthesis as a \emph{sequence generation} process: it models a record $\tuvec$ as a {sequence} and each element of the sequence is an attribute $\tuvec_{j}$. It uses LSTM to generate $\tuvec$ at multiple timesteps, where the $j$-th timestep is used to generate $\tuvec_{j}$, as illustrated in Figure~\ref{fig:rnn}. 
Let $\bm{h}^{j}$ and $\bm{f}^{j}$ respectively denote the hidden state and output of the LSTM at the $j$-th timestep. Then, we have
\begin{align}
& \bm{h}^{j+1} = \lstm(\noise, \bm{f}^{j}, \bm{h}^{j}), \nonumber \\ 
& \bm{f}^{j+1} = \tanhh( \fc_{|\bm{h}^{j+1})| \rightarrow |\bm{f}^{j+1})|}(\bm{h}^{j+1})), \nonumber
\end{align}
where $\bm{h}^{0}$ and $\bm{f}^{0}$ are initialized with random values.

Next, we compute attribute $\tuvec_{j}$ by considering transformation method of the corresponding attribute. Specifically, for simple normalization, one-hot encoding and ordinal encoding, we compute $\tuvec_{j}$ as follows. 
\begin{equation}
{
	\tuvec_{j} = 
	\left\{
	\begin{array}{lr}
	\tanhh(\fc_{|\bm{f}^{j}| \rightarrow 1}(\bm{f}^{j})),&  {\rm simple~normalization}  \\
	\softmax(\fc_{|\bm{f}^{j}| \rightarrow |\tuvec_{j}|}(\bm{f}^{j})), &  {\rm one-hot~encoding} \\
	\sigmoid(\fc_{|\bm{f}^{j}| \rightarrow 1}(\bm{f}^{j})), & {\rm ordinal~encoding}
	\end{array}\nonumber
	\right.
}
\end{equation}

In the case that attribute $\tu[j]$ is transformed by GMM-based normalization, we use two timesteps to generate its sample $\tuvec_{j}$: the first timestep $j_1$ generates the normalized value $v_{\tt gmm} = \tanhh(\fc_{|\bm{f}^{j_1}| \rightarrow 1}(\bm{f}^{j_1}))$, while the second timestep $j_2$ generates a vector that indicates which GMM component $v_{\tt gmm}$ comes from, i.e., $\softmax(\fc_{|\bm{f}^{j_2}| \rightarrow |\tuvec_{j}-1|}(\bm{f}^{j_2}))$. Then, we concatenate these two parts to compose $\tuvec_{j}$. 

Note that we can use a typical \emph{sequence-to-one} LSTM~\cite{DBLP:conf/nips/SutskeverVL14} to realize the discriminator $\Dis$.

\subsection{Training Algorithms} \label{app:algos}
This section presents the pseudo-code of the training algorithms introduced in Sections~\ref{subsec:train}, \ref{subsec:cgan} and \ref{subsec:dpgan}. 

\subsubsection{Vanilla GAN Training}
\begin{figure}[t!]
{\small
	\linesnumbered \SetVline
	\begin{algorithm}[H]
		\KwIn{$m$: batch size;
			$\alpha_d$: learning rate of $\Dis$; 
			$\alpha_g$: learning rate of $\Gen$;
			$T$: number of training iterations}
		\KwOut{$\Gen$: Generator; $\Dis$: Discriminator}
		Initialize parameters $\theta_d^{(0)}$ for $\Dis$ and $\theta_g^{(0)}$ for $\Gen$ \\
		\For{training iteration $t=1,2,\ldots, T$}
		{
			\tcc{Training discriminator $\Dis$}
			Sample $m$ noise samples $\{\noise^{(i)}\}^{m}_{i=1}$ from noise prior $p_{z}(\noise)$ \\
			Sample $m$ samples $\{\tuvec^{(i)}\}^{m}_{i=1}$ from real data $p_{data}(\tuvec)$ \\
			$\bar{g}_{1} \gets \nabla_{\theta_{d}}\frac{1}{m}\sum_{i=1}^{m}[\log {\Dis(\tuvec^{(i)})}+\log{(1-\Dis(\Gen(\noise^{(i)})))}]$ \\
			$\theta_{d}^{(t)}\gets \theta_{d}^{(t-1)}+\alpha_d \cdot \adam(\theta_{d}^{(t-1)},\bar{g}_{1})$ \\
			\tcc{Training generator $\Gen$}
			Sample $m$ noise samples $\{\noise^{(i)}\}^{m}_{i=1}$ from noise prior $p_{z}(\noise)$ \\
			$\bar{g}_{2} \gets \nabla_{\theta_{g}}\frac{1}{m}\sum_{i=1}^{m}\log(1-\Dis(\Gen(\noise^{(i)})))$ \\
			$\theta_{g}^{(t)} \gets \theta_{g}^{(t-1)} - \alpha_g \cdot \adam(\theta_{g}^{(t-1)},\bar{g}_{2})$ \\
		}
		\Return $\Gen$, $\Dis$ \\
		\caption{\AlgoVTrain~($m$, $\alpha_d$, $\alpha_g$, $T$)} \label{alg:vtrain}
	\end{algorithm}
	\vspace{-1em}
}
\end{figure}

We apply the vanilla GAN training algorithm~\cite{DBLP:conf/nips/GoodfellowPMXWOCB14} (\AlgoVTrain) to iteratively optimize parameters $\theta_{d}$ in $\Dis$ and $\theta_{g}$ in $\Gen$.
\begin{align}
& \theta_{d}\gets \theta_{d}+\alpha_d \nabla_{\theta_{d}}\frac{1}{m}\sum_{i=1}^{m}[\log {\Dis(\tuvec^{(i)})}+\log{(1-\Dis(\Gen(\noise^{(i)})))}] \nonumber \\
& \theta_{g} \gets \theta_{g} - \alpha_g 
\nabla_{\theta_{g}}\frac{1}{m}\sum_{i=1}^{m}\log(1-\Dis(\Gen(\noise^{(i)}))), \nonumber
\end{align}
where $m$ is the minibatch size  and $\alpha_d$ ($\alpha_g$) is learning rate of $\Dis$ ($\Gen$).

Algorithm~\ref{alg:vtrain} presents the pseudo-code of the vanilla training algorithm~\cite{DBLP:conf/nips/GoodfellowPMXWOCB14}.
It takes as input size $m$ of minibatch, learning rates $\alpha_d$ and $\alpha_g$) of discriminator $\Dis$ and generator $\Gen$, and number $T$ of training iterations, and iteratively optimize parameters $\theta_{d}$ in $\Dis$ and $\theta_{g}$ in $\Gen$. 
In each iteration, the algorithm trains discriminator $\Dis$ and generator $\Dis$ alternately. First, it fixes $\Gen$ and trains $\Dis$ by sampling $m$ noise samples $\{\noise^{(i)}\}^{m}_{i=1} \sim p(\noise)$ and $m$ real examples $\{\tuvec^{(i)}\}^{m}_{i=1} \sim p_{data}(\tuvec)$ and updating $\theta_{d}$ with the {Adam} optimizer~\cite{DBLP:journals/corr/KingmaB14}. Second, it fixes $\Dis$ and trains $\Gen$ by sampling another set of noise samples and updating parameters $\theta_{g}$. 

\subsubsection{Wasserstein GAN Training}
\begin{figure}[t]
{\small
	\linesnumbered \SetVline
	\begin{algorithm}[H]
		\KwIn{$m$: batch size;
			$\alpha_d$: learning rate of $\Dis$; 
			$\alpha_g$: learning rate of $\Gen$;
			$T_{d}$: number of iterations for $\Dis$;
			$T_{g}$: number of iterations for $\Gen$;
			$c_{p}$, clipping parameter}
		\KwOut{$\Gen$: Generator; $\Dis$: Discriminator}
		Initialize parameters $\theta_d^{(0)}$ for $\Dis$ and $\theta_g^{(0)}$ for $\Gen$ \\
		\For{training iteration $t_1= 1, 2, \ldots, T_{g}$}
		{
			\tcc{Using $T_d$ iterations to train $\Dis$}
			\For{training iteration $t_2=1, 2, \ldots, T_{d}$}
			{
				Sample noise samples $\{\noise^{(i)}\}^{m}_{i=1}$ from noise prior $p_{z}(\noise)$ \\
				Sample samples $\{\tuvec^{(i)}\}^{m}_{i=1}$ from real data $p_{data}(\tuvec)$ \\
				$\bar{g}_{1}\gets\nabla_{\theta_{d}}\frac{1}{m}\sum_{i=1}^{m}[\Dis(\tuvec^{(i)})-\Dis(\Gen(\noise^{(i)}))]$\\
				$\theta_{d}^{(t_2)} \gets \theta_{d}^{(t_2-1)} + \alpha_d \cdot 
				\rms(\theta_{d}^{(t_2-1)},\bar{g}_{1})$\\
				$\theta_{d}^{(t_2)} \gets \clip(\theta_{d}^{(t_2)},-c_{p}, c_{p})$ \\
			}
			\tcc{Training generator $\Gen$}
			Sample noise samples $\{\noise^{(i)}\}^{m}_{i=1}$ from noise prior $p_{z}(\noise)$ \\
			$\bar{g}_{2} \gets -\nabla_{\theta_{g}}\frac{1}{m}\sum_{i=1}^{m}D(G(\noise^{(i)}))$ \\
			$\theta_{g}^{(t_1)} \gets  \theta_{g}^{(t_1-1)}-\alpha_g\cdot \rms(\theta_{g}^{(t_1-1)}, \bar{g}_{2} )$ \\
		}
		\Return $\Gen$, $\Dis$
		\caption{\AlgoWTrain~($m$, $\alpha_d$,
			$\alpha_g$, $T_{d}$, $T_{g}$, $c_{p}$)} \label{alg:wtrain}
	\end{algorithm}
	\vspace{-1em}
}
\end{figure}

We also evaluate Wasserstein GAN~\cite{DBLP:journals/corr/ArjovskyCB17} for training our data synthesizer (\AlgoWTrain). Different from the original GAN, Wasserstein GAN removes the $\sigmoid$ function of $\Dis$ and changes the gradient optimizer from $\adam$ to $\rms$. It uses the loss functions of $\Dis$ and $\Gen$ as
\begin{align} \label{eq:wtrain-loss-1}
& \mathcal{L}_{D} = -\mathbb{E}_{\tuvec \sim p_{data}(\tuvec) }[D(\tuvec)]+\mathbb{E}_{\noise \sim p(\noise)}[D(G(\noise))] \nonumber \\
& L_{G}=-\mathbb{E}_{\noise\sim p(\noise)}[D(G(\noise))].
\end{align}

Algorithm~\ref{alg:wtrain} presents the training algorithm in Wasserstein GAN~\cite{DBLP:journals/corr/ArjovskyCB17}. Wasserstein GAN removes the $\sigmoid$ function of $\Dis$ and changes the gradient optimizer from Adam to RMSProp. It uses the loss functions of $\Dis$ and $\Gen$ as shown in Equation~(\ref{eq:wtrain-loss-1}).
Algorithm~\ref{alg:wtrain} takes as input size $m$ of minibatch, learning rates $\alpha_d$ and $\alpha_g$) of discriminator $\Dis$ and generator $\Gen$, and number training iterations $T_{d}$ and $T_{g}$ and a clipping parameter $c_p$.
It uses $T_{g}$ iterations to optimize $\Gen$. In each $\Gen$'s training iteration, it first uses $T_d$ iterations to train $\Dis$, and then trains $\Gen$. In particular, it clips the parameters $\theta_{d}$ of $\Dis$ into an interval $[-c_{p},c_{p}]$ after each training iteration of $\Dis$. 

\subsubsection{Conditional GAN Training}

Algorithm~\ref{alg:ctrainplus} presents the algorithm for training conditional GAN.
Basically, it follows the framework of the vanilla GAN training in Algorithm~\ref{alg:vtrain} with a minor modification of \emph{label-aware} sampling.
The idea is to avoid that the minority label has insufficient training opportunities due to the highly imbalanced label distribution.
Specifically, in each iteration, the algorithm considers every label in the real data, and for each label, it samples records with corresponding label for the following training of $\Dis$ and $\Gen$. Using this method, we can ensure that records with different labels have ``fair'' opportunities for training. 

\subsubsection{DPGAN Training}
Algorithm~\ref{alg:dptrain} presents the training algorithm for DPGAN.
Basically, it follows the framework of Wasserstein GAN training in Algorithm~\ref{alg:wtrain} with minor modifications (\AlgoDPTrain). 
When training $\Dis$, for each sampled noise $\noise^{(i)}$ and real example $\tuvec^{(i)}$, it adds Gaussian noise $N(0,\sigma_{n}^{2}c_{g}^{2}I)$ to the gradient $\nabla_{\theta_d}[D(\tuvec^{(i)})-D(G(\noise^{(i)}))]$, where $\sigma_{n}$ is the noise scale and $c_{g}$ is a user-defined bound on the gradient of Wasserstein distance with respect to parameters $\theta_d$ (see the original paper~\cite{DBLP:journals/corr/abs-1802-06739} for details of $c_{g}$).

\begin{figure}[t!]
{\small
	\linesnumbered \SetVline
	\begin{algorithm}[H]
		\KwIn{$m$: batch size;
			$\alpha_d$: learning rate of $\Dis$;
			$\alpha_g$: learning rate of $\Gen$;
			$T$: number of training iterations;
			$\LaDom$: label domain in real data}
		\KwOut{$\Gen$: Generator; $\Dis$: Discriminator}
		Initialize parameters $\theta_d^{(0)}$ for $\Dis$ and $\theta_g^{(0)}$ for $\Gen$ \\
		\For{training iteration $t=1,2,\ldots, T$}
		{
			\For{each label $\la$ in $\LaDom$}
			{
				Encode label $\la$ as condition vector $\condt$ \\
				\tcc{Training discriminator $\Dis$}
				Sample $m$ noise samples $\{\noise^{(i)}\}^{m}_{i=1}$ from prior $p_{z}(\noise)$ \\
				Sample $m$ samples $\{\tuvec^{(i)}\}^{m}_{i=1}$ with label $\la$ from real data $p_{data}(\tuvec|\la)$ \\
				$\bar{g}_{1} \gets \nabla_{\theta_{d}}\frac{1}{m}\sum_{i=1}^{m}[\log {\Dis(\tuvec^{(i)},\condt)}+\log{(1-\Dis(\Gen(\noise^{(i)}, \condt), \condt))}]$ \\
				$\theta_{d}^{(t)}\gets \theta_{d}^{(t-1)}+\alpha_d \cdot \adam(\theta_{d}^{(t-1)},\bar{g}_{1})$ \\
				\tcc{Training generator $\Gen$}
				Sample $m$ noise samples $\{\noise^{(i)}\}^{m}_{i=1}$ from prior $p(\noise)$ \\
				$\bar{g}_{2} \gets \nabla_{\theta_{g}}\frac{1}{m}\sum_{i=1}^{m}\log(1-\Dis(\Gen(\noise^{(i)}, \condt), \condt))$ \\
				$\theta_{g}^{(t)} \gets \theta_{g}^{(t-1)} - \alpha_g \cdot \adam(\theta_{g}^{(t-1)},\bar{g}_{2})$ \\
			}
		}
		\Return $\Gen$, $\Dis$
		\caption{\AlgoCTrainPlus($m$,
			$\alpha_d$,
			$\alpha_g$,
			$T$,
			$\LaDom$)} \label{alg:ctrainplus}
	\end{algorithm}
	\vspace{-1em}
}
\end{figure}

\begin{figure}[t]
	\linesnumbered \SetVline
	\begin{algorithm}[H]
		\KwIn{$m$: batch size;
			$\alpha_d$: learning rate of $\Dis$;
			$\alpha_g$: learning rate of $\Gen$;
			$T_{d}$: number of iterations for $\Dis$;
			$T_{g}$: number of iterations for $\Gen$;
			$c_{p}$: clipping parameter;
			$c_{g}$: bound on the gradient;
			$\sigma_{n}$: noise scale}
		\KwOut{$\Gen$: Generator; $\Dis$: Discriminator}
		Initialize parameters $\theta_d^{(0)}$ for $\Dis$ and $\theta_g^{(0)}$ for $\Gen$ \\
		\For{training iteration $t_1= 1, 2, \ldots, T_{g}$}
		{
			\tcc{Using $T_d$ iterations to train $\Dis$}
			\For{training iteration $t_2=1, 2, \ldots, T_{d}$}
			{
				Sample noise samples $\{\noise^{(i)}\}^{m}_{i=1}$ from prior $p_{z}(\noise)$ \\
				Sample samples $\{\tuvec^{(i)}\}^{m}_{i=1}$ from real data $p_{data}(\tuvec)$ \\
				\For{each $i$}
				{
					$g_{1}(\tuvec^{(i)},\noise^{(i)})\gets\nabla_{\theta_d}[D(\tuvec^{(i)})-D(G(\noise^{(i)}))]$ \\
				}
				$\bar{g}_{1}\gets\frac{1}{m}(\sum_{i=1}^{m}g_{1}(\tuvec^{(i)},\noise^{(i)})+N(0,\sigma_{n}^{2}c_{g}^{2}I))$ \\
				$\theta_{d}^{(t_2)} \gets \theta_{d}^{(t_2-1)} + \alpha_d \cdot 
					\rms(\theta_{d}^{(t_2-1)},\bar{g}_{1})$\\
				$\theta_{d}^{(t_2)} \gets \clip(\theta_{d}^{(t_2)},-c_{p}, c_{p})$ \\
			}
			\tcc{Training generator $\Gen$}
			Sample noise samples $\{\noise^{(i)}\}^{m}_{i=1}$ from prior $p_{z}(\noise)$ \\
			$\bar{g}_{2} \gets -\nabla_{\theta_{g}}\frac{1}{m}\sum_{i=1}^{m}D(G(\noise^{(i)}))$ \\
			$\theta_{g}^{(t_1)} \gets  \theta_{g}^{(t_1-1)}-\alpha_g\cdot \rms(\theta_{g}^{(t_1-1)}, \bar{g}_{2} )$ \\
		}
		\Return $\Gen$, $\Dis$ \;
		\caption{\AlgoDPTrain~($m$,
			$\alpha_d$,
			$\alpha_g$,
			$T_{d}$,
			$T_{g}$,
			$c_{p}$,
			$c_{g}$,
			$\sigma_{n}$)} \label{alg:dptrain}
	\end{algorithm}
	\vspace{-1em}
\end{figure}

\section{Additional Experiments} \label{app:add-exp}

\begin{figure*}[!t]
	\begin{center}
		\epsfig{figure=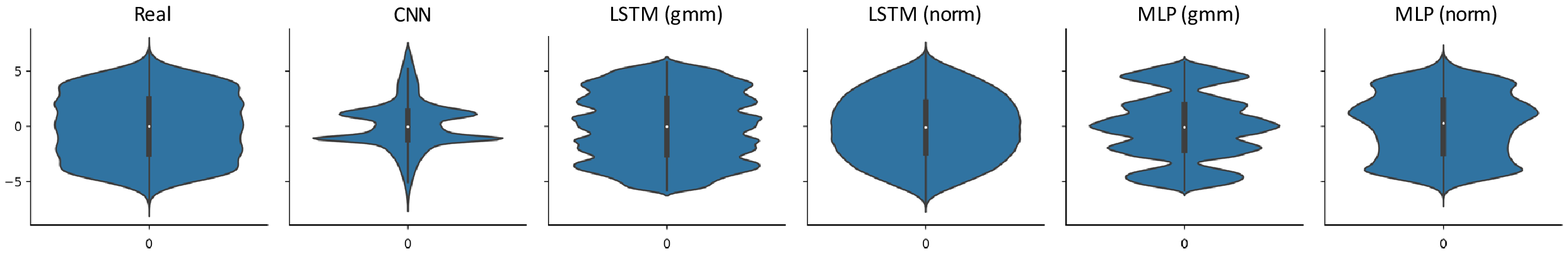,width=0.9\textwidth}
		\vspace{-1em}
		\caption{Evaluating value distribution of synthetic numerical attributes (\dssda).}
		\label{exp:dist-num}
	\end{center}
	\vspace{-2em}
\end{figure*}

\begin{figure*}[!t]
	\begin{center}
		\epsfig{figure=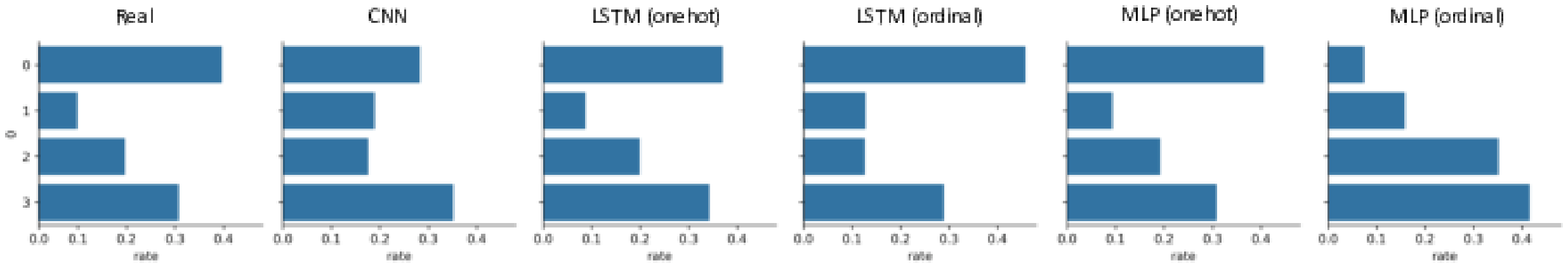,width=0.9\textwidth}
		\vspace{-1em}
		\caption{Evaluating value distribution of synthetic categorical attributes (\dssdb).}
		\label{exp:dist-cat}
	\end{center}
	\vspace{-2em}
\end{figure*}

\begin{figure}[!t]
	\begin{center}
		\subfigure[{\dssda dataset.}]{
			\epsfig{figure=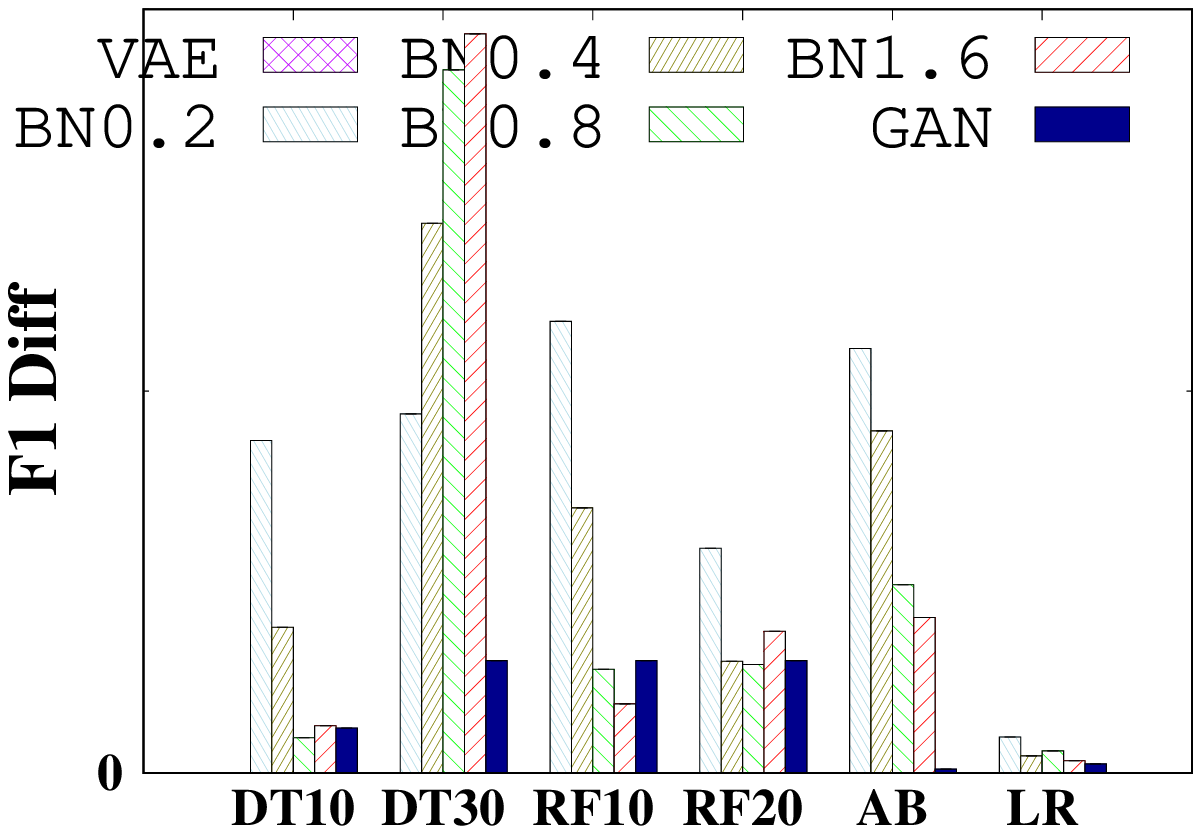,width=0.22\textwidth}
		}
		\subfigure[{\small \dssdb dataset.}]{
			\epsfig{figure=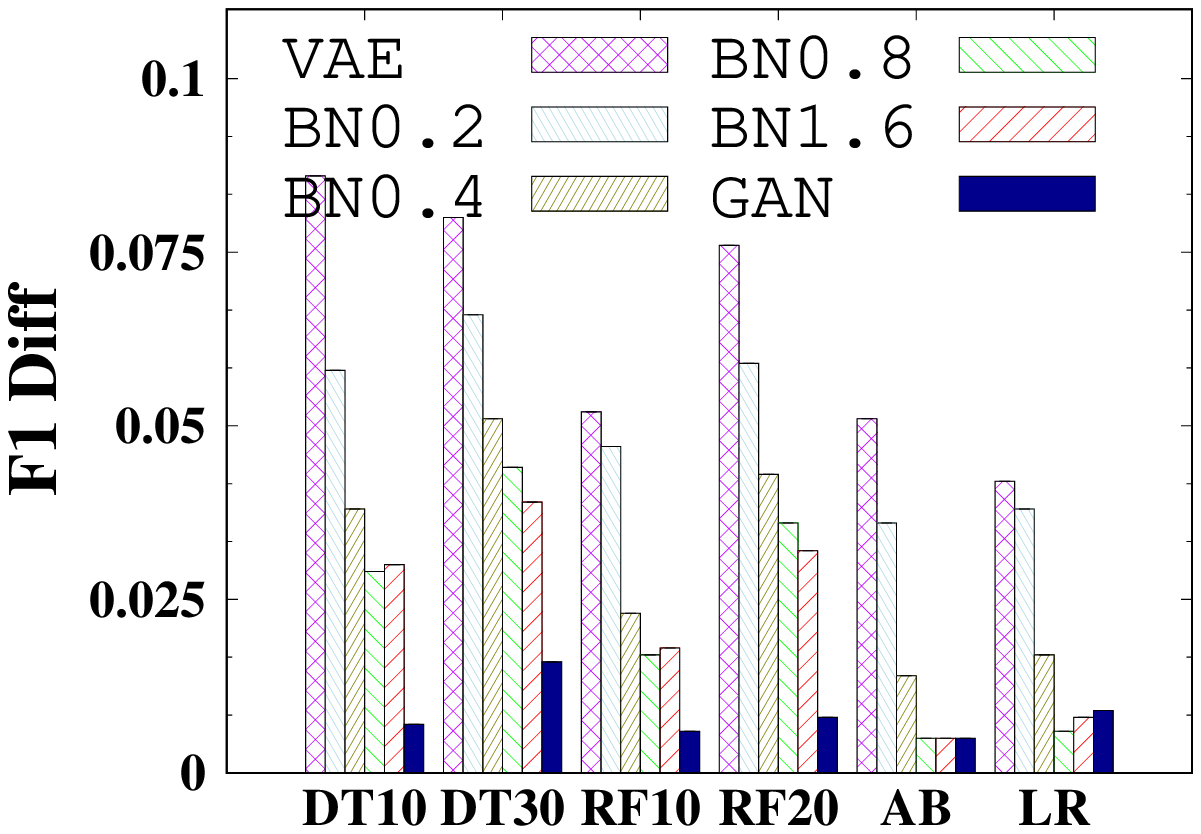,width=0.22\textwidth}
		}
	\end{center}\vspace{-2em}
	\caption{Comparison of different approaches to relational data synthesis on data utility for classification.} \label{exp:compa-synthetic}
	\vspace{-1em}
\end{figure}

\subsection{Detailed Dataset Information}
We describe the detailed information of the eight datasets we used in our experiments.

\noindent \textbf{(1) \dshtru dataset} is a physical dataset that contains $17,898$ pulsar candidates collected during the High Time Resolution Universe Survey~\cite{HTRU2}. This dataset has $8$ numerical attributes, which are statistics obtained from the integrated pulse profile and the DM-SNR curve, and a binary label (i.e., pulsar and non-pulsar). The label distribution is balanced.

\noindent  \textbf{(2) \dspendigits dataset} contains $10,992$ pen-based handwritten digits~\cite{Pendigits}. Each digit has $16$ numerical attributes collected by a pressure sensitive tablet and processed by normalization methods, and a label indicating the gold-standard number from $0-9$. The label distribution is balanced.
	%

\noindent \textbf{(3) \dsadult dataset} contains personal information of $41,292$ individuals extracted from the 1994 US census with $8$ categorical attributes, such as $\term{Workclass}$ and $\term{Education}$ and $6$ numerical attributes, such as $\term{Age}$ and $\term{Hours}$-$\term{per}$-$\term{Week}$~\cite{Adult}.
We use attribute $\term{Income}$ as label and predict whether a person has income larger than $50K$ per year (positive) or not (negative), where the label distribution is skew, i.e., the ratio between positive and negative labels is $0.34$. 

\noindent \textbf{(4) \dsct dataset} contains the information of $116,204$ forest records obtained from US Geological Survey (USGS) and US Forest Service (USFS) data~\cite{Covertype}. It includes $2$ categorical attributes, $\term{Wild}$-$\term{area}$ and $\term{Soil}$-$\term{type}$, and $10$ numerical attributes, such as $\term{Elavation}$ and $\term{Slope}$. 
We use attribute $\term{Cover}$-$\term{type}$ with $7$ distinct values as label and predict forest cover-type from other cartographic variables. The label distribution is also very skew, e.g., there are $46\%$ records with label $2$ while only $6\%$ records with label $3$. 

\noindent  \textbf{(5) \dssat dataset} consists of the multi-spectral values of pixels in 3x3 neighborhoods in a satellite image~\cite{SAT}.
	It has $36$ numerical attributes that represent the values in the four spectral bands of the 9 pixels in a neighborhood, and uses a label with $7$ unique values indicating the type of the central pixel. The label distribution is balanced in the dataset.

\noindent  \textbf{(6) \dsanuran dataset} a dataset from the life domain for anuran species recognition through their calls~\cite{Anuran}. It has $22$ numerical attributes, which are derived from the audio records belonging to specimens (individual frogs), and associates a label with $10$ unique values that indicates the corresponding species. The label distribution is very skew: there are $3,478$ records with label $2$ and $68$ with label $9$.

\noindent \textbf{(7) \dscensus dataset} contains weighted census data extracted from the 1994 and 1995 Current Population Surveys~\cite{Census}. We use demographic and employment variables, i.e., $9$ numerical and $30$ categorical attributes, as features, and $\term{total}$-$\term{person}$-$\term{income}$ as label. We remove the records containing null values and then obtain $142,522$ records with very skew label distribution, i.e., $5\%$ records with income larger than $50K$ vs. $95\%$ with income smaller than $50K$.

%
%

\noindent \textbf{(8) \dsbing dataset} is a Microsoft production workload dataset, which contains the statistics of Bing Search and is used for evaluating AQP~\cite{DBLP:journals/tkde/LiZLTY19}. We sample $500,000$ records with $23$ categorical and $7$ numerical attributes. As the dataset does not have any attribute used as label, we only use the dataset for evaluating performance of data synthesis on AQP.

\subsection{Additional Evaluation on Mode Collapse}
This section shows the results of GAN model training on various hyper-parameter settings on other datasets and the performance of \AlgoSimD strategy to avoid mode collapse.

\begin{figure*}[!t]\vspace{-1em}
	\begin{center}\hspace{-1mm}
		\subfigure[{\small LSTM-based $\Gen$ (\dssat).}]{
			\epsfig{figure=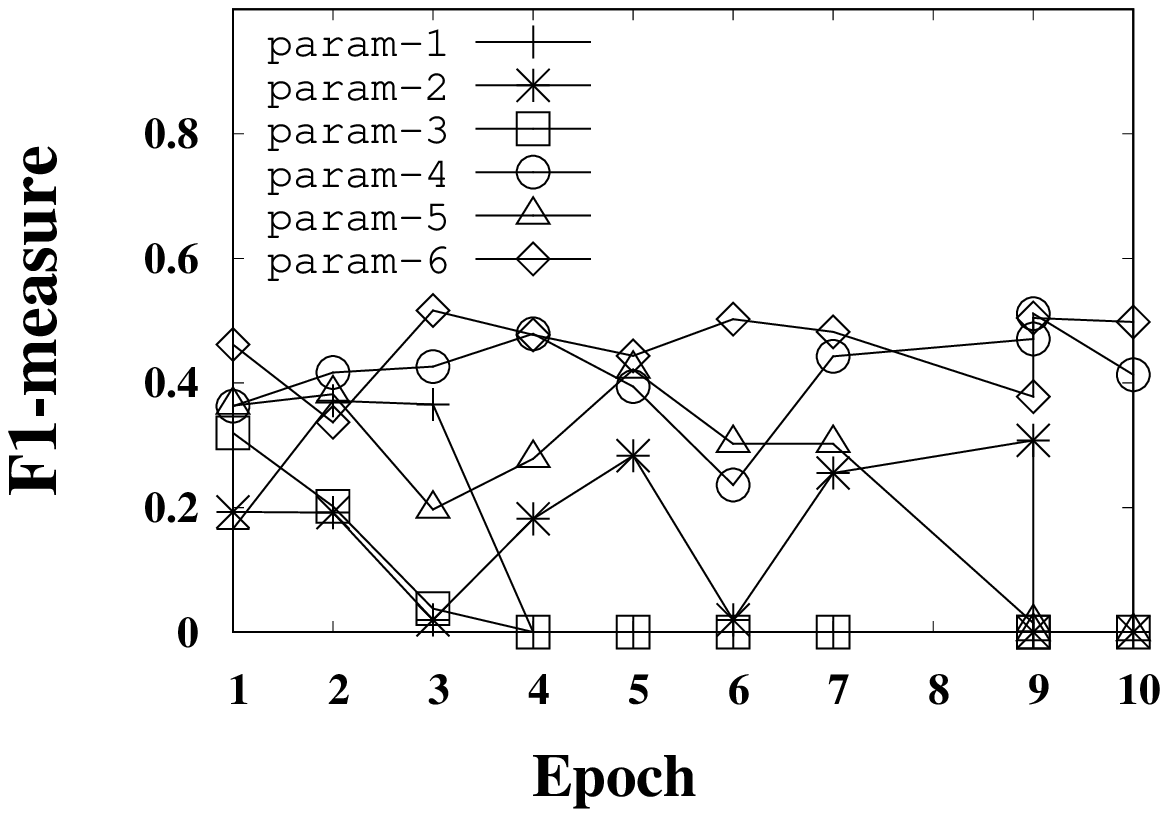,width=0.23\textwidth}
		}
		\subfigure[{\small MLP-based $\Gen$ (\dssat).}]{
			\epsfig{figure=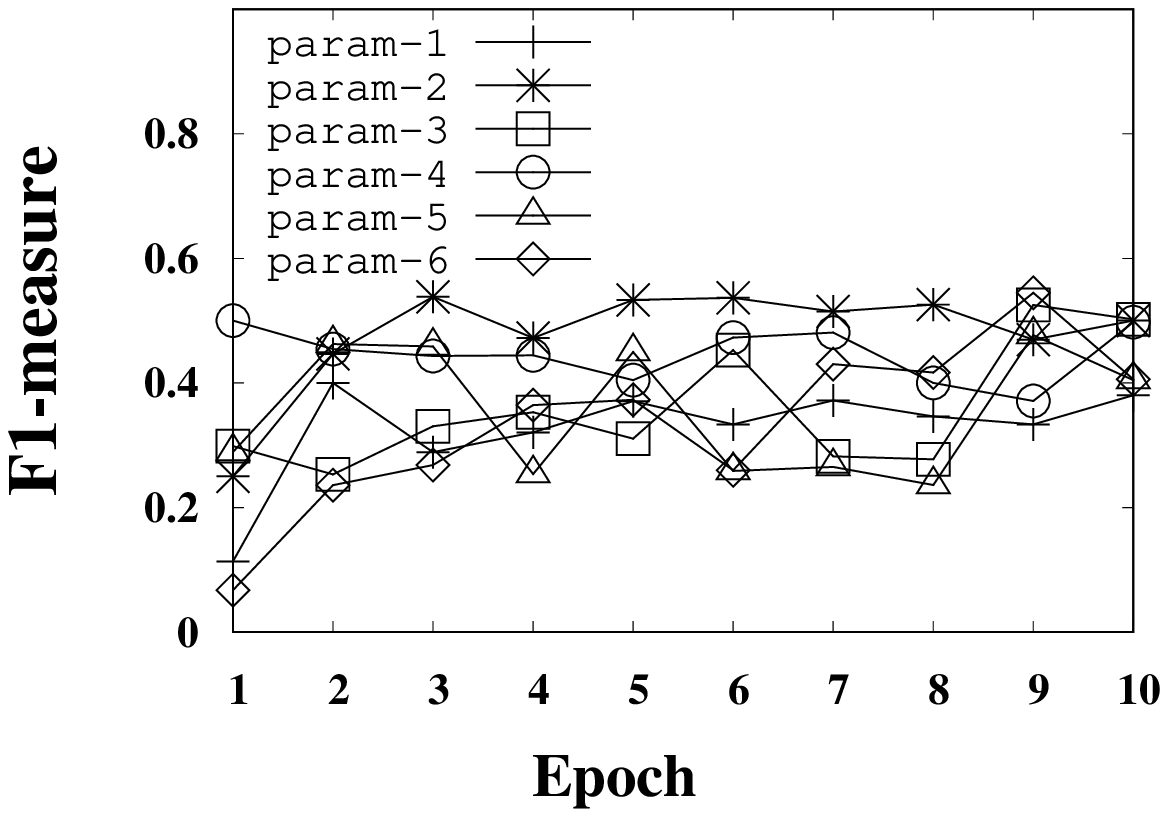,width=0.23\textwidth}
		}
		\subfigure[{\small LSTM-based $\Gen$ (\dscensus).}]{
			\epsfig{figure=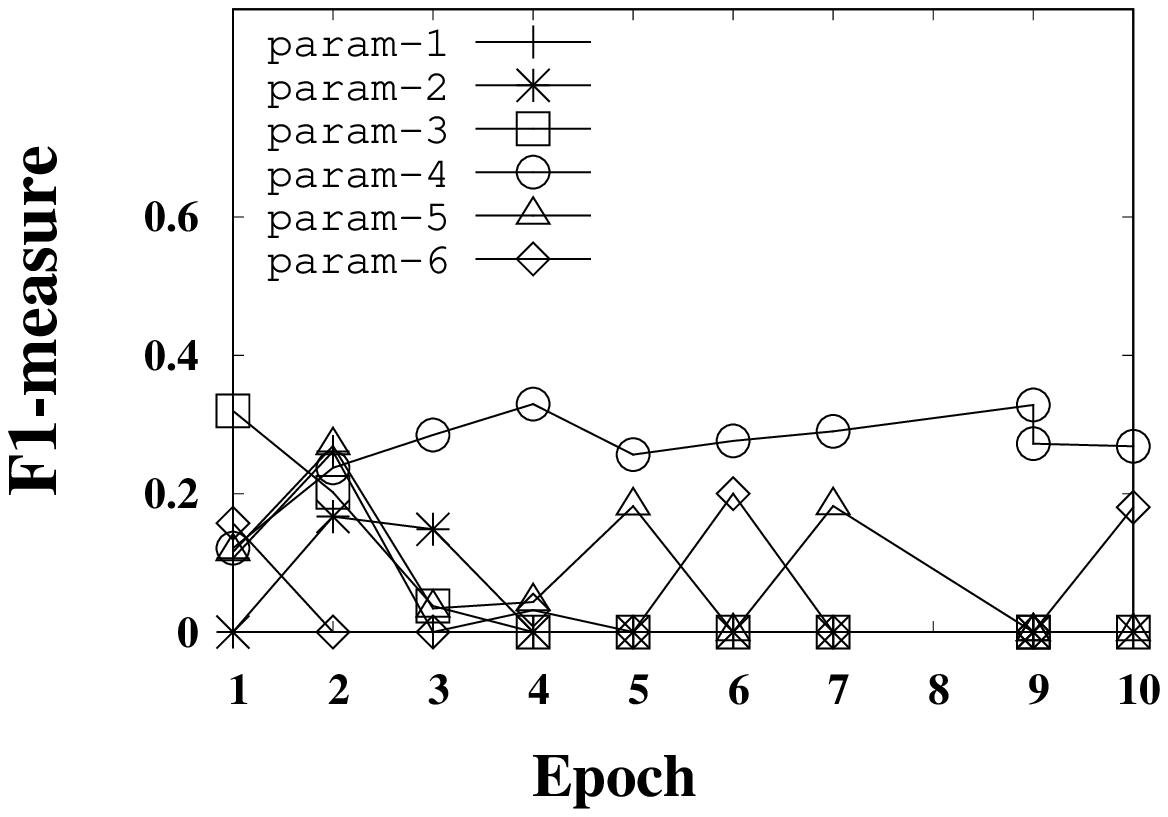,width=0.23\textwidth}
		}
		\subfigure[{\small MLP-based $\Gen$ (\dscensus).}]{
			\epsfig{figure=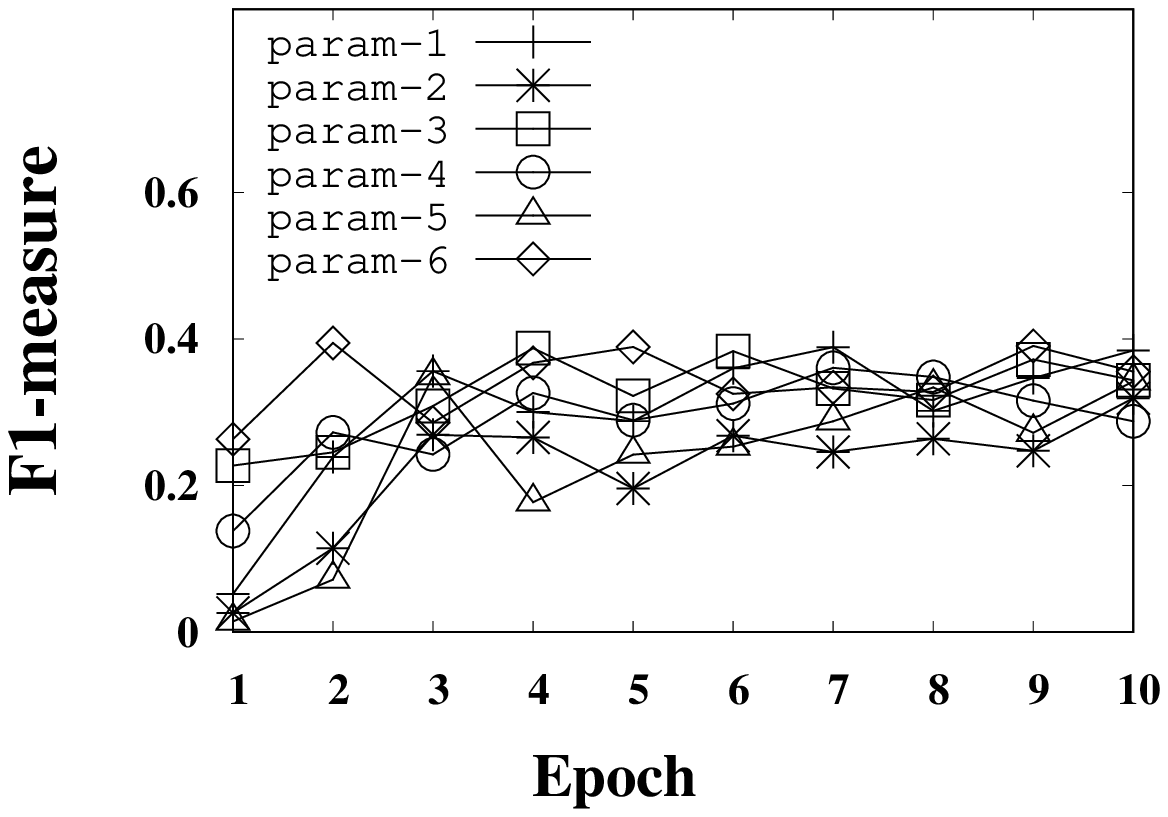,width=0.23\textwidth}
		}
	\end{center}\vspace{-1.5em}
	\caption{Evaluating GAN model training on various hyper-parameter settings (1).} \label{exp:search_2}
	\vspace{-2em}
\end{figure*}

\begin{figure*}[!t]\vspace{-1em}
	\begin{center}\hspace{-1mm}
		\subfigure[{\small Normal $\Dis$ (\dsadult).}]{
			\epsfig{figure=adult-search-kl-l.eps,width=0.23\textwidth}
		}
		\subfigure[{\small Simplified $\Dis$ (\dsadult).}]{
			\epsfig{figure=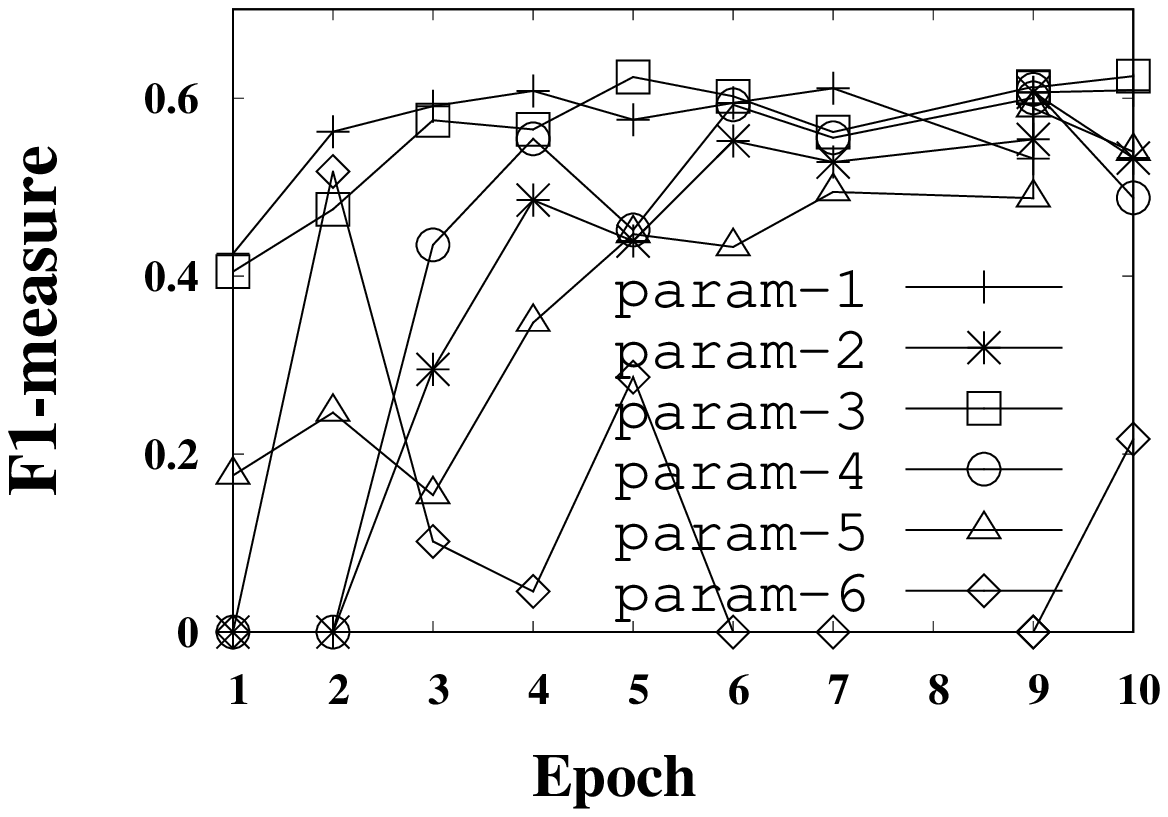,width=0.23\textwidth}
		}
		\subfigure[{\small Normal $\Dis$ (\dsct).}]{
			\epsfig{figure=covertype-search-kl-l.eps,width=0.23\textwidth}
		}
		\subfigure[{\small Simplified $\Dis$ (\dsct).}]{
			\epsfig{figure=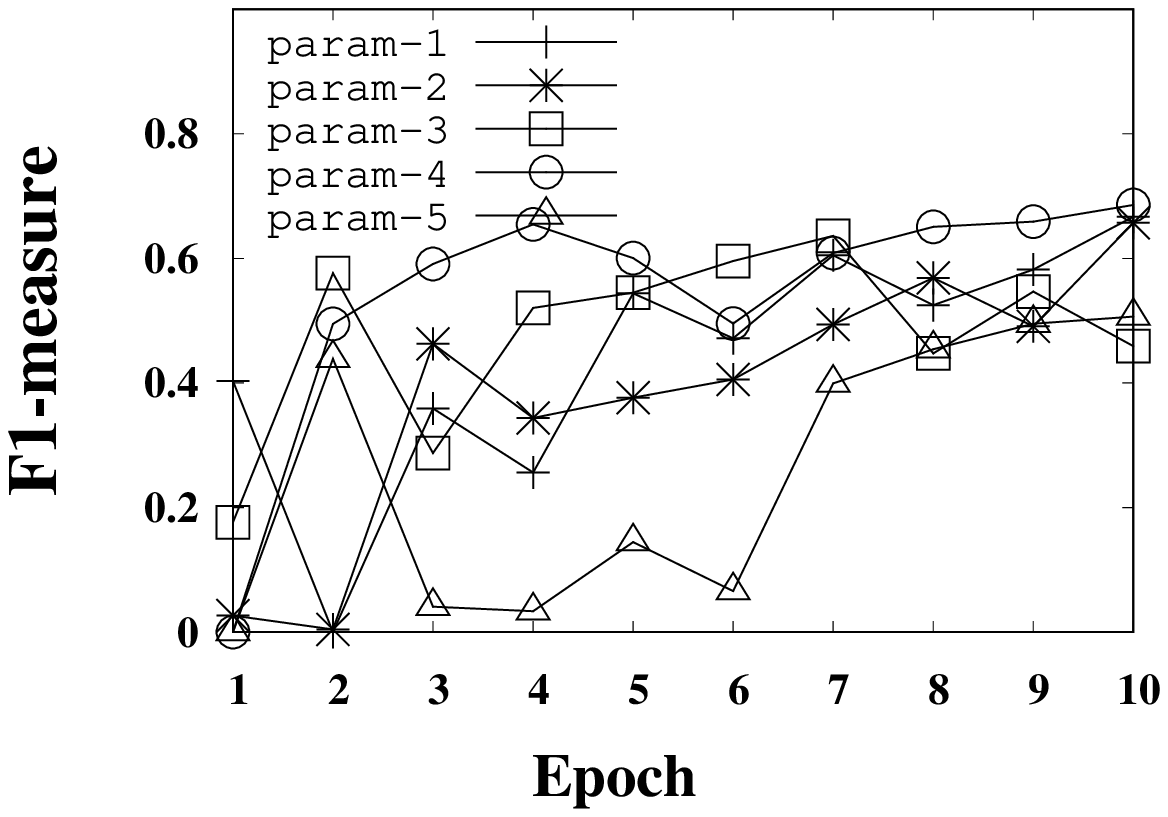,width=0.23\textwidth}
		}
	\end{center}\vspace{-1.5em}
	\caption{Evaluating GAN model training on various hyper-parameter settings (2).} \label{exp:search_3}
	\vspace{-2em}
\end{figure*}

\begin{figure*}[!t]\vspace{-1em}
	\begin{center}\hspace{-1mm}
		\subfigure[{\small Normal $\Dis$ (\dssat).}]{
			\epsfig{figure=SAT-search-kl-l.eps,width=0.23\textwidth}
		}
		\subfigure[{\small Simplified $\Dis$ (\dssat).}]{
			\epsfig{figure=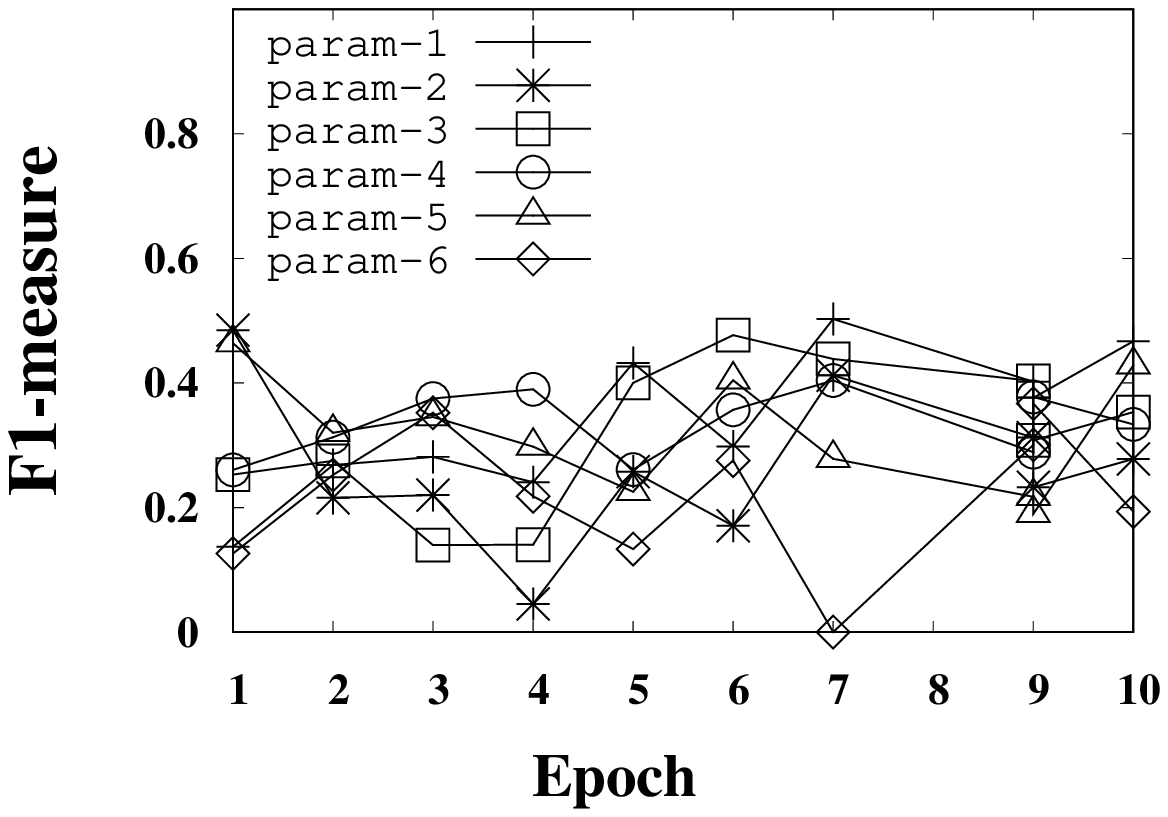,width=0.23\textwidth}
		}
		\subfigure[{\small Normal $\Dis$ (\dscensus).}]{
			\epsfig{figure=census-search-kl-l.eps,width=0.23\textwidth}
		}
		\subfigure[{\small Simplified $\Dis$ (\dscensus).}]{
			\epsfig{figure=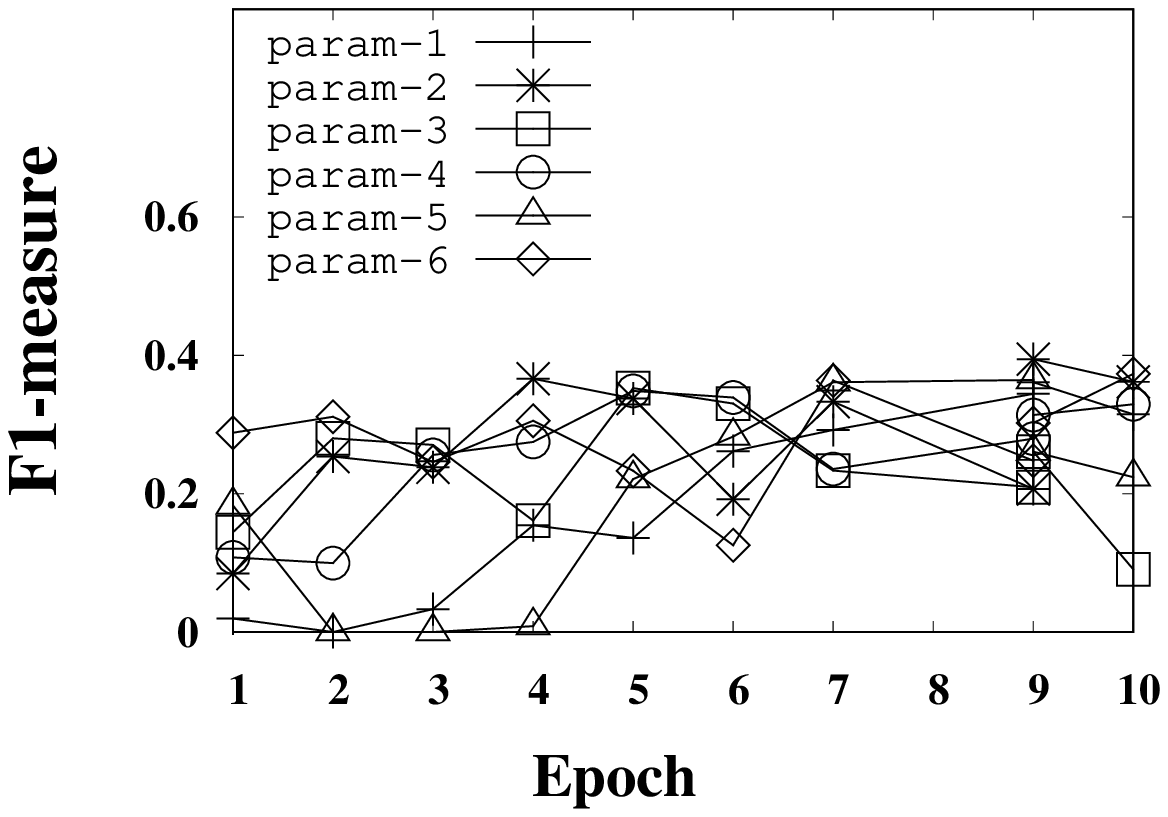,width=0.23\textwidth}
		}
	\end{center}\vspace{-1.5em}
	\caption{Evaluating GAN model training on various hyper-parameter settings (3).} \label{exp:search_4}
	\vspace{-2em}
\end{figure*}

Figure~\ref{exp:search_2} shows the results of GAN model training on various hyper-parameter settings on datasets \dscensus and \dssat, which is similar with the results in figure~\ref{exp:search}. Figure~\ref{exp:search_3} and \ref{exp:search_4} show the performance of the \AlgoSimD strategy on various hyper-parameter settings, we find that mode collapse can be effectively alleviated by replacing the simplified $\Dis$, for example, on the \dsadult dataset, \AlgoVTrain enables the LSTM-based generator to be more robust to hyper parameters and the chances of mode collapse are largely reduced. Moreover, compared with the MLP-based generator, it achieve much higher scores on F-measure.

\subsection{Additional Evaluation on Synthetic Data Utility}
\begin{figure*}[!t]
	\begin{center}
		\subfigure[{\dsanuran dataset.}]{
			\epsfig{figure=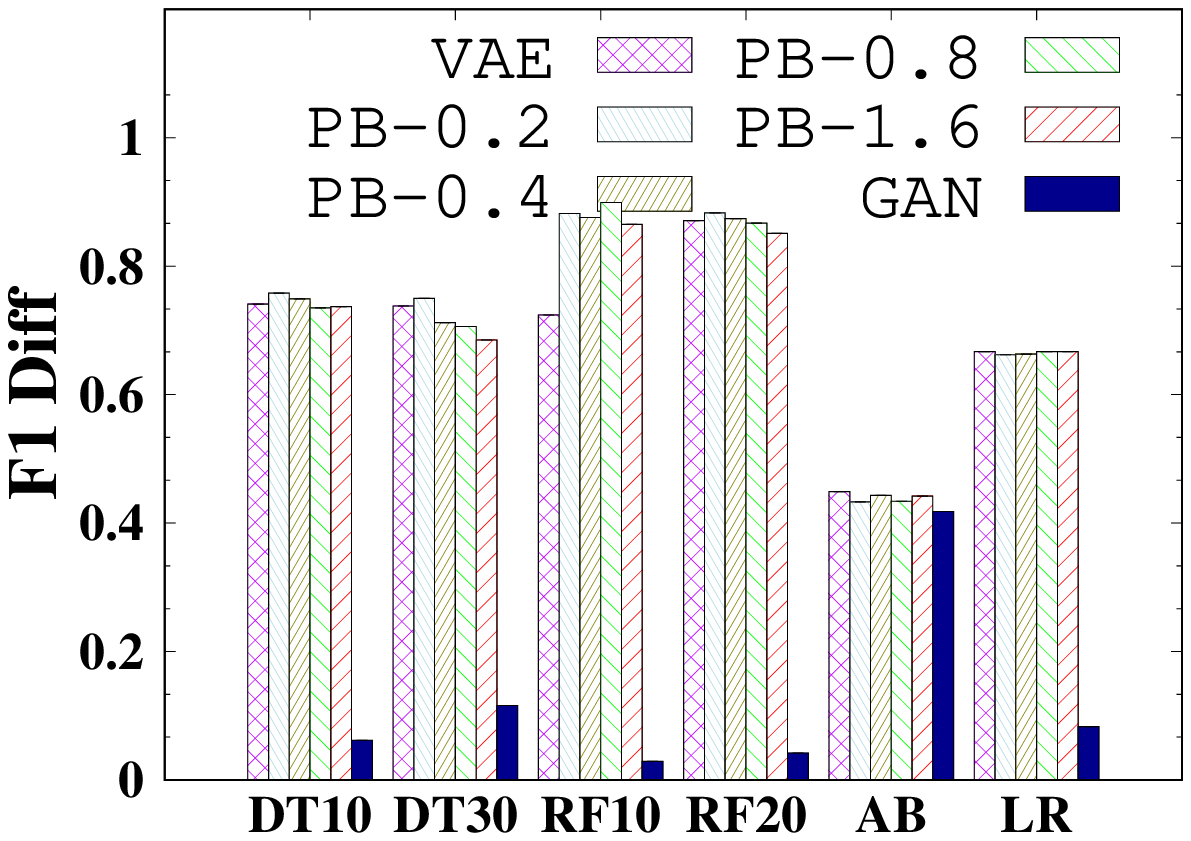,width=0.23\textwidth}
		}
		\subfigure[{\dspendigits dataset.}]{
			\epsfig{figure=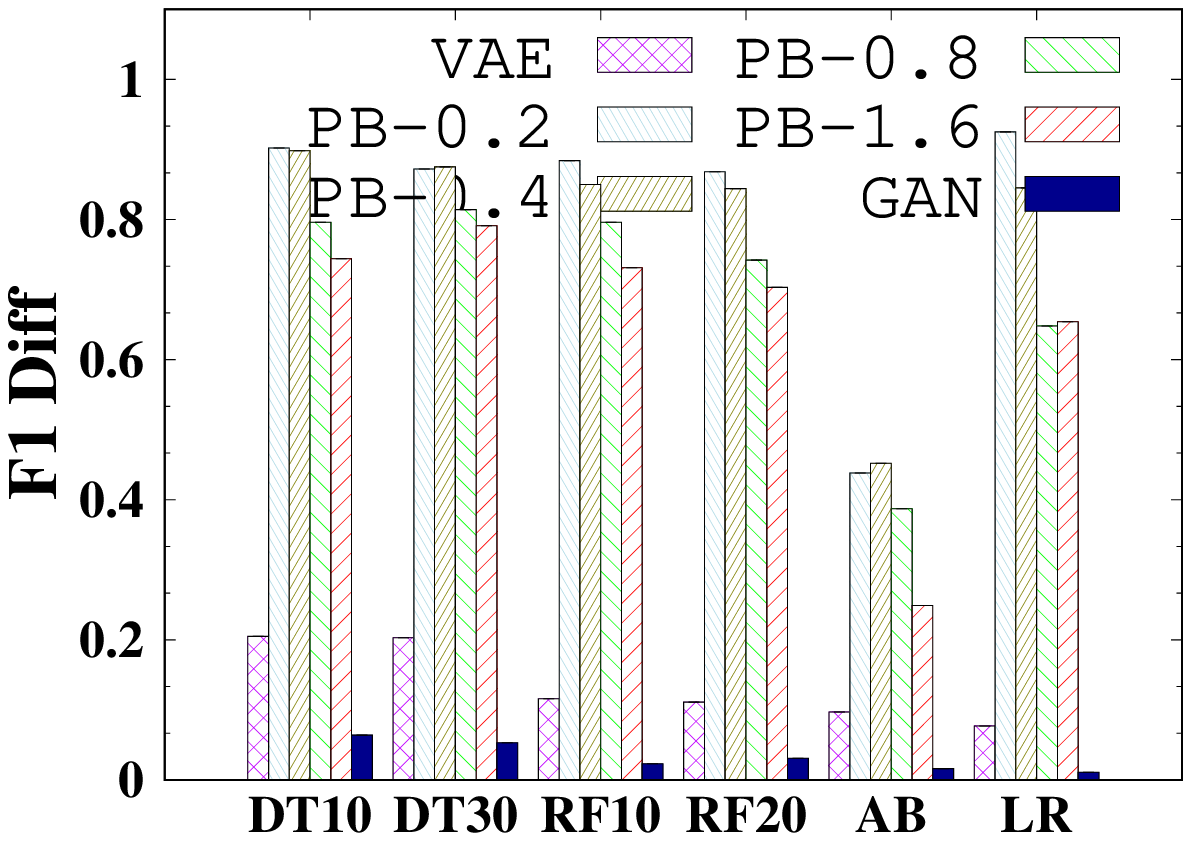,width=0.23\textwidth}
		}
		\subfigure[{\dshtru dataset.}]{
			\epsfig{figure=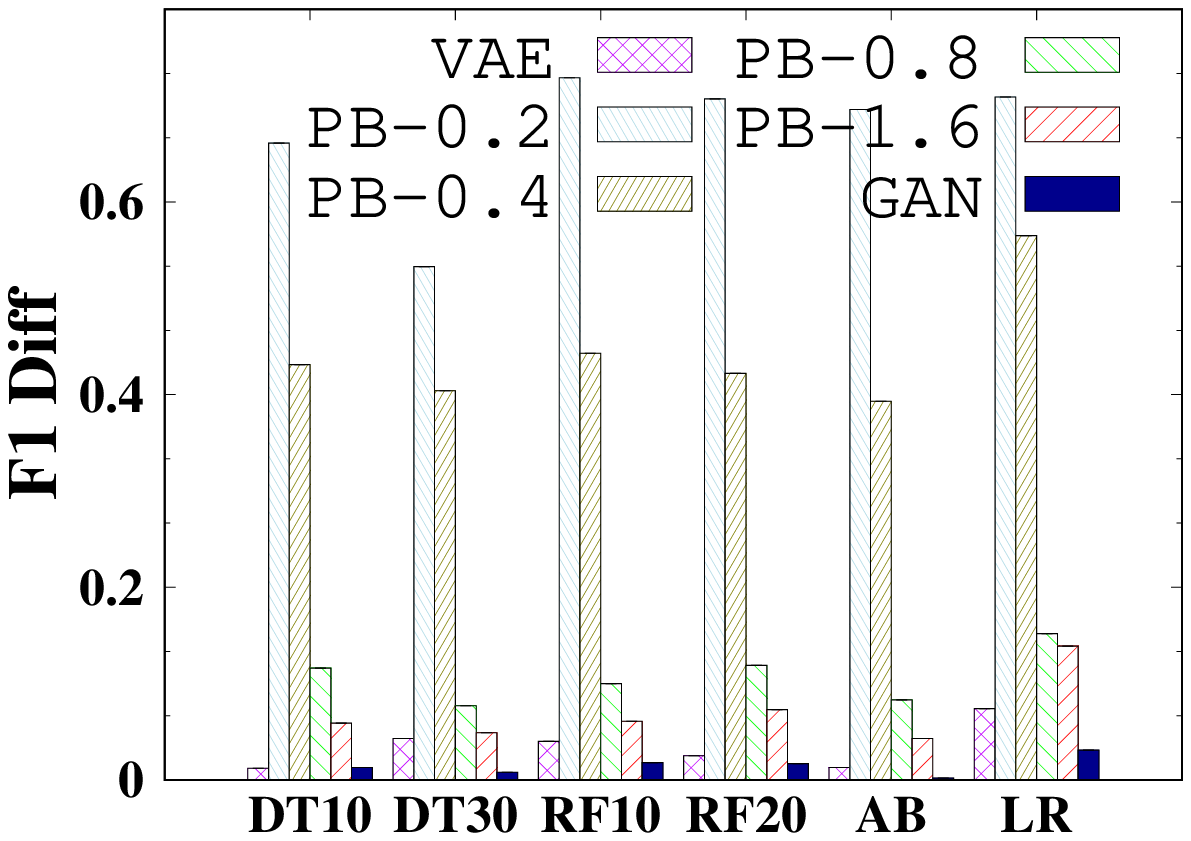,width=0.23\textwidth}
		}
	\end{center}\vspace{-2em}
	\caption{{Comparison of different approaches to relational data synthesis on data utility for classification.}} \label{exp:compa-2}
	\vspace{-1em}
\end{figure*}

	This section shows the results of evaluation on synthetic data utility on the other datasets. 
	Figure~\ref{exp:compa-2} shows the results of synthetic data utility on datasets \dsanuran, \dspendigits and \dshtru, which has a similar trend to Figure~\ref{exp:compa-1}, and we find that GAN-based framework still work well on the simulated datasets \dssda and \dssdb from the results in figure~\ref{exp:compa-synthetic}. 


\subsection{Evaluating LSTM Discriminator}

\begin{table}[!t]
	\centering
	\caption{Evaluating  LSTM-based discriminator on synthetic data utility (\dsadult dataset).}\label{table:adult-model-lstm}
	\resizebox{0.5\textwidth}{!}{%
	\begin{tabular}{|c||c|c|c|c|c|c|c|c|}
		\hline
		\multirow{2}{*}{\textbf{Classifier}} &
		\multicolumn{4}{c|}{\textbf{MLP}} &
		\multicolumn{4}{c|}{\textbf{LSTM}} 
		 \\
		\cline{2-9}
		&\norm+\ordinal &\norm+\onehot &\gmm+\ordinal &\gmm+\onehot
		&\norm+\ordinal &\norm+\onehot &\gmm+\ordinal &\gmm+\onehot  \\  
		\hline	
		\hline DT10
		&0.099 &0.151 &0.096 &0.157
		&0.125 &\textbf{0.085} &0.104 &0.165  \\ 
		\hline
		DT30 
		&0.136 &0.136 &0.079 &0.156
		&0.076 &\textbf{0.069} &0.166 &0.131 \\ 
		\hline
		RF10 
		&\textbf{0.041} &0.126 &0.125 &0.118
		&0.141 &0.071 &0.085 &0.117   \\ 
		\hline
		RF20 
		&0.107 &0.232 &0.142 &0.139
		&0.123 &\textbf{0.083} &0.118 &0.115  \\ 
		\hline
		AdaBoost 
		&0.105 &0.277 &0.126 &0.143
		&0.089 &0.126 &\textbf{0.074} &0.156\\
		\hline
		LR
		&0.093 &0.019 &\textbf{0.008} &0.265
		&0.065 &0.133 &0.013 &0.055\\
		\hline		
	\end{tabular}
	} 
	\vspace{-0.5em}
\end{table}

This section evaluates LSTM-based discriminator $\Dis$ for GAN-based relational data synthesis on the \dsadult dataset. Note that we use a typical \emph{sequence-to-one} LSTM~\cite{DBLP:conf/nips/SutskeverVL14} to realize $\Dis$.
Table~\ref{table:adult-model-lstm} reports the experimental results.
We can see that, compared with MLP-based discriminator $\Dis$ reported in Table~\ref{table:adult-model}, the F1 difference is significantly higher.
Considering classifier DT10 as an example, the F1 difference increases by $18 - 416\%$ when changing MLP to LSTM for realizing discriminator.
We also find similar results in other datasets.
Therefore, we use MLP to implement discriminator in our experiments reported in Section~\ref{sec:result}.

\subsection{Synthetic Data Distribution}

%

\subsubsection{Evaluation on Data Transformation}

We also find that data transformation in preprocessing does affect the overall utility of synthetic data.
GMM-based normalization and one-hot encoding achieve the best performance in most of the cases. To provide in-depth analysis, we further examine whether the value distribution of a synthetic attribute is similar to that of its counterpart real attribute. We report the results on \dssda and \dssdb to purely evaluate numerical and categorical attributes respectively. 

Figure~\ref{exp:dist-num} shows the distribution for numerical attributes using the violin plots. LSTM with \gmm can generate the attribute having the most approximate distribution to their counterpart real attribute, and it is remarkably effective for the attribute with \emph{multi-modal} distribution. 
This is attributed to the Gaussian Mixture model used in this method, which is more powerful to represent multi-modal attributes. Moreover, it also outperforms MLP with \gmm. This is because that LSTM uses two time steps to generate normalized value $v_{\tt gmm}$ and components probabilities $\{\pi^{(i)}\}$ separately, which is shown more effective than generating them together in MLP. 
Figure~\ref{exp:dist-cat} shows the distribution for categorical attributes. We can see that one-hot is significantly better than ordinal embedding. This is because values in a categorical attribute usually do not have ordinal relationships, and thus a single number is insufficient for attribute representation. 

\vspace{1mm}
\noindent \textbf{Finding: Data transformation does affect overall utility of synthetic data: GMM-based normalization performs better than simple normalization, especially for numerical attributes with multi-modal distribution; One-hot encoding is better than ordinal encoding for categorical attributes.}


\balance

\end{document}